\documentclass[3p,,preprint,12pt]{elsarticle}
\makeatletter\if@twocolumn\PassOptionsToPackage{switch}{lineno}\else\fi\makeatother

\usepackage{tabulary,xcolor}
\usepackage{amsfonts,amsmath,amssymb}
\usepackage[T1]{fontenc}
\makeatletter
\let\save@ps@pprintTitle\ps@pprintTitle
\def\ps@pprintTitle{\save@ps@pprintTitle\gdef\@oddfoot{\footnotesize\itshape \null\hfill\today}}
\def\hlinewd#1{%
  \noalign{\ifnum0=`}\fi\hrule \@height #1%
  \futurelet\reserved@a\@xhline}
\def\tbltoprule{\hlinewd{.8pt}\\[-12pt]}
\def\tblbottomrule{\noalign{\vspace*{6pt}}\hline\noalign{\vspace*{2pt}}}
\def\tblmidrule{\noalign{\vspace*{6pt}}\hline\noalign{\vspace*{2pt}}}
\AtBeginDocument{\ifNAT@numbers \biboptions{sort&compress}\fi}
\makeatother

\usepackage{ifluatex}
\ifluatex
\usepackage{fontspec}
\defaultfontfeatures{Ligatures=TeX}
\usepackage[]{unicode-math}
\unimathsetup{math-style=TeX}
\else 
\usepackage[utf8]{inputenc}
\fi 
\ifluatex\else\usepackage{stmaryrd}\fi

\usepackage{url,multirow,morefloats,floatflt,cancel,tfrupee}
\makeatletter

\AtBeginDocument{\@ifpackageloaded{textcomp}{}{\usepackage{textcomp}}}
\makeatother
\usepackage{colortbl}
\usepackage{xcolor}
\usepackage{pifont}
\usepackage{smartdiagram}
\usepackage{float}
\usepackage{placeins}
\usepackage{setspace}
\usepackage{ulem}
\usepackage{endnotes}
\usepackage[nointegrals]{wasysym}
\usepackage{xcolor}
\usepackage{graphicx}
\usepackage{setspace}
\usepackage{multicol}
\usepackage{etoolbox}
\usepackage{ulem}
\usepackage{graphicx}
\makeatletter

\def\mcWidth#1{\csname TY@F#1\endcsname+\tabcolsep}

\def\cAlignHack{\rightskip\@flushglue\leftskip\@flushglue\parindent\z@\parfillskip\z@skip}
\def\rAlignHack{\rightskip\z@skip\leftskip\@flushglue \parindent\z@\parfillskip\z@skip}

\@ifundefined{etal}{}{}

\usepackage{ifxetex}
\ifxetex\else\if@twocolumn\@ifpackageloaded{stfloats}{}{\usepackage{dblfloatfix}}\fi\fi

\AtBeginDocument{
\expandafter\ifx\csname eqalign\endcsname\relax
\def\eqalign#1{\null\vcenter{\def\\{\cr}\openup\jot\m@th
  \ialign{\strut$\displaystyle{##}$\hfil&$\displaystyle{{}##}$\hfil
      \crcr#1\crcr}}\,}
\fi
}

\AtBeginDocument{%
  \@ifpackageloaded{endfloat}%
   {\renewcommand\efloat@iwrite[1]{\immediate\expandafter\protected@write\csname efloat@post#1\endcsname{}}}{\newif\ifefloat@tables}%
}%

\def\BreakURLText#1{\@tfor\brk@tempa:=#1\do{\brk@tempa\hskip0pt}}
\let\lt=<
\let\gt=>
\def\processVert{\ifmmode|\else\textbar\fi}

\@ifundefined{subparagraph}{
\def\subparagraph{\@startsection{paragraph}{5}{2\parindent}{0ex plus 0.1ex minus 0.1ex}%
{0ex}{\normalfont\small\itshape}}%
}{}

\newcommand\role[1]{\unskip}
\newcommand\aucollab[1]{\unskip}
  
\@ifundefined{tsGraphicsScaleX}{\gdef\tsGraphicsScaleX{1}}{}
\@ifundefined{tsGraphicsScaleY}{\gdef\tsGraphicsScaleY{.9}}{}
\def\checkGraphicsWidth{\ifdim\Gin@nat@width>\linewidth
	\tsGraphicsScaleX\linewidth\else\Gin@nat@width\fi}

\def\checkGraphicsHeight{\ifdim\Gin@nat@height>.9\textheight
	\tsGraphicsScaleY\textheight\else\Gin@nat@height\fi}

\def\fixFloatSize#1{}
\let\ts@includegraphics\includegraphics

\def\inlinegraphic[#1]#2{{\edef\@tempa{#1}\edef\baseline@shift{\ifx\@tempa\@empty0\else#1\fi}\edef\tempZ{\the\numexpr(\numexpr(\baseline@shift*\f@size/100))}\protect\raisebox{\tempZ pt}{\ts@includegraphics{#2}}}}

\AtBeginDocument{\def\includegraphics{\@ifnextchar[{\ts@includegraphics}{\ts@includegraphics[width=\checkGraphicsWidth,height=\checkGraphicsHeight,keepaspectratio]}}}

\DeclareMathAlphabet{\mathpzc}{OT1}{pzc}{m}{it}

\def\URL#1#2{\@ifundefined{href}{#2}{\href{#1}{#2}}}

\def\UrlOrds{\do\*\do\-\do\~\do\'\do\"\do\-}%
\g@addto@macro{\UrlBreaks}{\UrlOrds}

\edef\fntEncoding{\f@encoding}

\makeatother

\newif\ifmultipleabstract\multipleabstractfalse%
\usepackage{color}
\usepackage{xcolor}
\usepackage{graphicx}
\usepackage{setspace}
\usepackage{ulem}
\usepackage{graphicx}
\usepackage[export]{adjustbox}
\usepackage{subfig}
\usepackage{scrextend}

\usepackage[symbol]{footmisc}

\usepackage{marvosym}

\usepackage{tikz,hyperref}
\DeclareGraphicsExtensions{.pdf,.png,.jpg}

\usepackage{float}

\begin{document}

\begin{frontmatter}
	

\title{Decision Making For Celebrity Branding: An Opinion Mining
Approach Based On Polarity And Sentiment Analysis Using Twitter Consumer-Generated Content (CGC)}
%
\author[af08f922b7273]{Ali Nikseresht\footnote{E-mail Address:  \url{al.nikseresht@mail.sbu.ac.ir}}$^{,}$\footref{refnotee}, Mohammad Hosein Raeisi\footnote{E-mail Address:  \url{m.raeisifardabadi@mail.sbu.ac.ir}}$^{,}$\footref{refnoteee}, Hossein Abbasian Mohammadi\footnote{E-mail Address:  \url{h.abbasian.m@aut.ac.ir}}$^{,}$}

     \address[af08f922b7273]{\scriptsize\label{refnotee}
 Department of Business Administration\unskip, Faculty of Management and Accounting\unskip, 
    Shahid Beheshti University\unskip, Tehran\unskip, Iran}

 \address[af08f922b7273]{\scriptsize\label{refnoteee}
 Department of Business Administration\unskip, Faculty of Management\unskip, 
    University of Tehran\unskip, Tehran\unskip, Iran}

 \address[af08f922b7273]{\scriptsize\label{refnoteeee}
 Electrical Engineering Department\unskip, Amirkabir University of Technology\unskip, Tehran\unskip, Iran}

\begin{abstract}
\small
The volume of discussions concerning brands within social media provides digital marketers with great opportunities for tracking and analyzing the feelings and views of consumers toward brands, products, influencers, services, and ad campaigns in the consumer-generated content (CGC). The aim of the present study is to assess and compare the performance of firms and celebrities (i.e., influencers that with the experience of being in an ad campaign of those companies) with the automated sentiment analysis that was employed for CGC at social media while exploring the feeling of the consumers toward them to observe which influencer (of two for each company) had a closer effect with the corresponding corporation on consumer minds.
For this purpose, a number of consumer tweets from the pages of brands and influencers were utilized to make a comparison of machine learning and lexicon-based approaches to the sentiment analysis through the Naïve algorithm (lexicon-based) and Naïve Bayes algorithm (machine learning method) and obtain the desired results to assess the campaigns. The findings suggested that the approaches were dissimilar in terms of accuracy; the machine learning method (Naïve Bayes algorithm) yielded higher accuracy. Finally, the results showed which influencer was more appropriate according to their existence in previous campaigns and helped choose the right influencer in the future for our company and have a better, more appropriate, and more efficient ad campaign subsequently.
It is required to conduct further studies on the accuracy improvement of the sentiment classification. This approach should be employed for other social media CGC types, e.g., Instagram feeds. The results revealed decision-making for which sentiment analysis methods (or their combinations) are the best approaches for the analysis of social media. It was also found that companies should be aware of their consumers' sentiments and choose the right person every time they think of a campaign.
\textbf{Paper type: }Research paper
\end{abstract}
\end{frontmatter}
\textbf{}\\
\section{Introduction}
Imperial endorsements were employed as a celebrity branding technique for the promotion of products in the 1760s. The first celebrity endorsement utilizing product was in the 1760s, when Josiah Wedgwood and Sons, who made ceramics and chinaware, made use of regal endorsements as an advertising gadget to exhibit an incentive in their organization and advance their products \cite{KalyaniVemuri2004}. In 1875–1900s exchange cards were presented. It is the place where an image of a celebrity along with a photograph of the product would exist. Ordinarily, such exchange cards would be delivered to purchasers with the product or be embedded in the product bundling. This would highlight famous people (e.g., actors or game stars. Cigarette brands turned out to be tremendously associated with celebrity branding. 'Kodas' cigarettes brought the cards of baseball player into the cigarette parcels as a client dedication conspire component. This encouraged shoppers to purchase a larger number of cigarettes to increase the entire baseball player cards due to the celebrity endorsement attribute of the cards. Competitors were the greatest inclining endorsers in the early 1930s. However, the pattern changed until 1945, with motion picture stars becoming the next enormous superstar endorsers. Color TV became aware of the marketing in 1965, and a prominent rising interest began to happen.  At this time, TV characters and performers turned into the celebrity endorsements of communication products and services. Organizations and corporations decided to make products around celebrities in the 1980s\cite{KalyaniVemuri2004}.\\ A case of this is in 1984 when the organization Nike saw a skilled furthermore, a youthful basketball player called Michael Jordan. Michael Jordan at that point became a superstar brand minister of Nike for the game industry. Nike incredibly relied on the societal position of Michael Jordan to universally introduce the brand. Since organizations began to make items around superstars and celebrities became brand envoys for the organization, athletes and entertainment celebrities began arranging pay rates and payouts for those who spoke to the organization due to the challenge of different firms. Since pay rates continued to expand due to the requests of celebrities, for the most part, the organization’s interactions that were advanced by the celebrities would rise. Celebrity endorsements shifted to another level in the late 1900s; instead of exploiting celebrity pictures to advance brands, organizations began holding celebrity question and answer sessions celebrities that declared unique deals. This implied that celebrities had begun to be representatives for corporations. As a larger number of organizations had superstar representatives to hold question and answer sessions and report unique interactions, sales were substantially expanding for the brands, with a larger number of deals being introduced to the market. In the 2000s, research demonstrated that a celebrity minister considerably increased interactions for a corporation. For example, marking Tiger Woods, Nike predicted a \$50million elevation in golf ball sales by 2002 in 1996\cite{KalyaniVemuri2004}.\\
Considering a large number of side-by-side brands in brand competition, firms attempt to establish brand images in order to be differentiated from their rivals. In this respect, they commonly make use of celebrity endorsers to create a significant and persuasive appeal to customers, in light of the positive image relating to celebrities. Celebrities that serve as endorsers may be actors, entertainers, singers, or athletes whose personalities are highly known in society. Exploiting a celebrity for commercial purposes, however, does not represent a one-way procedure; to the same as with a corporate or a consumer brand, a celebrity has their value in their audiences’ minds and, therefore, becomes a brand in their manner\cite{Arai2013}\cite{Seno2007}. A celebrity is an individual that enjoys public recognition through a great share of a specific set of individuals, while attributes such as an extraordinary lifestyle, specific skills, and attractiveness skills are merely examples, and particular common characteristics are not observable. One can say that celebrities are generally different from social norms in a related social group and have a high public awareness level. Also, celebrities are selected based their credibility. Highly credible Endorsers with high credibility are expected to generate a larger number of positive changes in the attitude toward the advocated position advocated and trigger a larger number of behavioral changes than sources with lower credibility. Advertisers generally believe the brand correspondence messages that are conveyed through celebrities or famous characters to generate a greater intrigue, review, and consideration than those that are performed by ordinary individuals. In today’s exceptional and aggressive condition, the snappy message-reach and impact are of fundamental importance. Therefore, communication instruments in marketing have helped maintain and raise market share. The cost, however, is substantially high.\\ 
Regardless of the positive endorsement aspects, the utilization of celebrities in campaigns might possibly be an effective approach\cite{Misra1990}. Famous people have been employed to positively impact the behavior of purchasers toward products\cite{Tripp1994}. Also, advertising campaigns connect products and the celebrities and induce either a positive or negative implying move toward the products \cite{Till1998}. In order to choose a suitable celebrity, among various elements, marketers or advertisers need to take into account the campaign lifespan, celebrity acknowledgment, and celebrity significance, the sentiment of the targeted consumer toward the use of celebrities for correspondence, and the targeted shopper receptivity while relating the celebrity to the products or the brand. Different factors are considered when selecting a famous endorser or brand support in an endeavor to create campaigns for brands. Such factors include: \\
distinction: extensive  acknowledgment among shoppers or particular groups. \\
adjustment: the mixing or matching of the attributes of the brand and those of the celebrity\\
attributes concerning the perceptions of consumers. \\
financial highlights: the expenses and returns f utilizing celebrity endorsers.\\
roles: different approaches to celebrity utilization for communications in marketing\cite{Pringle2005}. \\
One can sustain the efficacy of celebrity endorsements through credibility, attractiveness, and expertise\cite{Ohanian1991}. Credibility alludes to the confidence in the celebrity to influence the public. Attractiveness is concerned with appearance, thinking nature, and elegance. Finally, expertise deals with the experience and information of the endorser on a given subject. When purchasers admit that their romanticized self-picture and self-idea are reflected by the endorser, the commercial’s evaluation is positive and creates intentions to purchase the product\cite{Choi2012}. Despite such a positive assessment, this endorsement makes such customers exhibit brand or product loyalty. Likewise, research indicates the exploitation of celebrities in campaigns to be associated with various strategies. \cite{Rumschisky2009} suggested that people are willing to pay an amount of up to 20\% higher for an item, depending on its endorser, yielding more significant incomes for the corporation. Those advertisements that feature celebrities generally increase the organization’s estimation at the stock trade; they also impact the perception of the endorsed firm \cite{Agrawal1995}. The effects of celebrity endorsements on product assessment, which indicates that fans are influenced by celebrities, have been increasingly studied\cite{Sliburyte2009} \cite{McNamara2009}. The research has found these effects to majorly arise results from the associations of consumers the celebrities and endorsed products\cite{Choi2012} \cite{Till1998}.\\

The present study combined two sentiment analysis approaches, demonstrating that the companies should work on these kinds of data mining seriously since the minds of consumers are very different from companies’ assumptions and achievements of their conventional market research practice.\\

The present study attempted to dealing with two objectives, including:\\
1- Based on the consumer sentiment analysis of the company, and influencer and their correlation, have companies chosen suitable celebrities for their former ad campaigns?\\
2- Based on the consumer sentiment analysis performed on the company, and influencer and their correlation, which one of the two celebrities are more suitable and effective for the company’s future ad campaigns?\\
To find answers to these questions, challenges regarding text classification techniques (lexicon-based (naïve)) were summarized for the sentiment analysis that is performed on social media data today. Then, the research methodology was outlined, empirically evaluating the machine learning (Naïve Bayes) and lexicon-based approaches through a large CGC sample of brand pages on Twitter and related hashtags and keywords.

\section{Literature}

Decision-making is an essential activity in every organization—perhaps the most important activity. Decision-making impacts the success/failure and performance of organizations its performances. It is becoming difficult to make decisions due to internal and external factors. Making proper decisions can yield very high rewards. The same case applies to losses of inappropriate decisions.
It is unfortunately difficult to make a decision. Several decision types exist, each of which needs a different approach for decision-making. For instance, McKinsey \& Company management consultants\cite{DeSmet2017} classified organizational decisions into four groups:
Big-bet, high-risk decisions.
Cross-cutting decisions, which allow for repetition but pose risk and require group work.
Ad hoc decisions, which appear episodically.
Decisions to delegated to people or small groups.\\
Moreover, market studies, which incorporate social and feeling studies, are the methodical translation and get-together of data on people or associations that adopt measurable and investigative strategies and processes of the applied sociologies to understand or improve decision-making.
The downsides of the market research include a high cost, high time-consumption,  and inexactness.\cite{proctor2007}
\subsection{Decision making for celebrity branding (endorsement)}

The use of famous individuals as a marketing communication strategy component is a regular decision-making exercise for large corporations in the support of corporate or brand imagery. This is an important decision-making process encountered and found to be difficult by corporates. Companies invest critical amounts in the comparison of brands and associations to endorser characteristics, e.g., allure, reliability, and agreeability. They believe such characteristics to work in a transferable manner and yield attractive campaign results. Nevertheless, superstar characteristics may be unseemly, insignificant, and bothersome. As a result, an important question is “at what capacity are organizations able to choose and maintain the 'right' celebrity among numerous contending options and simultaneously deal with this asset while avoiding possible entanglements?” In a marketing communication (MarCom) perspective, organizations must plan procedures to help support concentrated differential favorable positions for the items or administrations of an association. Likewise, MarCom practices back up various components in the marketing mix, e.g., branding, packaging, product design, , pricing, and place decisions (i.e., circulation channels and physical dispersion) in an endeavor to produce advantageous outcomes in the minds of customers. To accomplish this, the use of celebrity endorsers is a widely-exploited MarComs methodology. Corporations contribute vast cash amounts to adapt their brands and organizations to endorsers. These endorsers are considered to be unique and have agreeable and appealing features\cite{Atkin1983} and organizations believe these characteristics to be transferred to products through MarCom exercises \cite{Langmeyer1991}\cite{McCracken1989}\cite{Walker1992}. Furthermore, due to their notoriety, celebrities not only create and look after attention but also fulfill high recall rates for MarCom messages in the current profoundly-jumbled circumstances \cite{Croft1996}\cite{Friedman1979}\cite{Kamen1975}\cite{Kamins1989}\cite{Ohanian1991}. First, this methodology seems to be a no-chance/all-gain condition. Be that as it may, similar to any dynamic marketing communication strategy, potential dangers exist. Individuals may change, and endorsement relationships may be harsh. It can be said that the celebrity endorsement approach can become a two-edged sword. This makes it very challenging to select a celebrity endorser among several options under potential traps \cite{Erdogan1999}.\\
Advertisements are non-individual correspondences provided via the media. They are supported and paid to move forward the thoughts, administrations, or merchandise that are presented in commercials and provide a stage to create advertised product’s familiarity \cite{Ayanwale2005}. At the same time, EAs induce an enthusiastic reaction in purchasers’ minds. It similarly influences spectators in making a purchase choice, bringing problems to the brand light, and impacting their positive reaction to the promoted item \cite{Matthes2014}. Corporations spend billions of dollars to create marketing systems and commercials to influence the conduct of purchases. Be that as it may,\cite{Lewis2012} it should be noted that these venture endeavor types are lost when the ideal reactions of purchases are not gained. This may arise from the fact that purchasers have different feeling-molded practices\cite{Khuong2015}. Human feelings are described as an inclusive language\cite{Khuong2015}and a mind state formed by thinking, cognizance, and encounters with expanded sentiments in order to apply variations to social and physical exercises\cite{Ekman1992}\cite{Foo2011}\cite{Oatley2014}. They are connected to activities, e.g., stance, signals, and facial developments, on a regular basis, empowering respondents to supervise their practices\cite{Hakkak2016}\cite{Lewis2012}. Thus, behaviors can be either positive (e.g., love, joy, pride, warmth, and happiness) or negative (e.g., voracity, coerce, sadness, and anger, and fear) \cite{SaadAslam2011}\cite{Hakkak2016}. \\
Scholastic works provide insight into a classification of a mixture of feelings, e.g. , (essential) feelings, complex (optional) feelings, or a mixture of fundamental ones, complex feelings.\cite{Plutchik1997} and \cite{Zammuner1998} epitomize crucial feelings with the sub-builds of positive and negative emotions, i.e., outrage, dread, happiness, pity, love, sicken, shock, tension, sympathy, etc. \cite{Bindu2007} further enriched the dialog with an array of some other essential feelings, i.e., pride, bliss, desire, distress, rage, vengeance, anguish, disgrace, energy, and despise coupled with the essential feelings suggested by\cite{Plutchik1997} and \cite{Zammuner1998}. \cite{Bindu2007} further assure that such essential feelings can also bring forth complex feelings, which can include regret, pardoning, sentimentality, desire, cheer, offense, frustration, trust, etc. 
Be that as it may, \cite{Ekman2007} contentions are worth being referenced here, by which they mention the comprehensiveness of the essential feelings, i.e., joy, pride, pity, dread, outrage, disturb, and shock. They propose that such fundamental feelings are entirely communicated and acknowledged by those that respect time, spot, and culture due to their basic organic source. \\
Nevertheless, it has widely been discussed that people that have a place with various societies might encounter various feelings, and relevant enthusiastic improvements may likewise command the conduct of such individuals  \cite{Boiger2012}\cite{Kidwell2014}. The purchasing conduct is consolidated by conduct expectations in its differentiated dimensional outlines \cite{Shiau2012}. The classified variety in the conduct of customers causes distinct passionate reactions to inducing upgrades created by organizations for purchasers. Some analysts add to the advertising view that advertisers must attempt to comprehend the conduct of purchasers toward limited time boosts \cite{Martin2011}\cite{Chu2011} since the understanding of purchasers may yield them better advertising techniques. 
Contemplating the language individuals use to use all, the more likely comprehend their considerations and practices, which isn’t new in the sociology \cite{Krippendorff2012}. The assessment has been estimated using self-detailed information in purchaser reviews, e.g., the Michigan Consumer Sentiment Surveys. Nevertheless, the use of self-reported information has limitations, as with most of the self-provided information. According to reviews, marketing experts rely on the capacities of purchasers to accurately review their felt encounters. This can be deeply factor and difficult to be verbalized and remade \cite{Cooke2008}\cite{Nabi2009}. Analyses suggest that there are concerns that identify with incorrect conditions in which information is accumulated, and this can compel the enthusiastic reactions of shoppers \cite{Nabi2007}.\\
Social media stages are currently prominent instruments to ponder the customer assessment at a large scale and within a whiz setting \cite{Kivran-Swaine2012} owing to the considerable portion of online discussions that communicate the emotions, contemplations, and conclusions purchasers on items and brands \cite{Jansen2009}. The regular investigation of feelings in printed content is dependent on basic conclusion comment assignments in which annotators are required to decide whether a sentence is certain, unbiased, or negative \cite{Rosenthal2015}\cite{Mohammad2015}.
Considering the immense social media content volume, it is unrealistic to propose a manual supposition explanation. Augmented by most of computerized content arrangement systems, opinion examination is typically employed by advertisers as a PC-empowered, rapid, viable, and versatile method to check the slant of purchasers \cite{Murdough2009}. \\
The scholar community and industry increasingly consider computerized assumption investigation\cite{Chen2010}. Furthermore, it has found to be a key system for supervising vast online life information volumes. Mechanized supposition investigation systems are regularly employed to arrange text-based records into predefined categories reflecting the extremity of conclusion referred to within the content. As of late, \cite{Canhoto2015} have tried a relative study of computerized and manual social media discussions. Their findings demonstrate low understanding degrees in manual and mechanized investigations, which has "grave concern given the notoriety of the last in shopper explore" \cite{Canhoto2015}. 
There are a number of reasons for the mechanized characterization of communicated slant within social media discussions. In the beginning, the recognition of feelings and slants of a text-based characteristic language entails a deep comprehension of the unequivocal, certain, sporadic, linguistic, standard, and semantic rules of the language \cite{Cambria2013}. In addition, notion examination encounters problems in the utilization of natural language processing (NLP) on non-structured content, CGC, and an average of online life discussions. For instance, CGC content typically reflects the casual nature and moment of correspondence through web-based networking media \cite{Canhoto2015}. The substance is normally a free-streaming content and easily goes in its word and sentence structure use \cite{Tirunillai2014},generally incorporates shortened forms, emoticons, emojis, incorrect spellings,  and makes use of an SMS-like linguistic structure on a regular basis, which is not satisfactorily bolstered by current slant investigation strategies.\\
Additionally, particular stage highlights, similar to the 140-character limit for Twitter messages, prevent the sufficiency of current mechanized notion examination systems \cite{Kiritchenko2014}. Lastly, social media discussions’ sheer volume is a considerable test. Such a challenge is transformed into an open door by robotized advances by eliminating the need for the expensive and risk-inclined manual examination that rather utilizes an electronic methodology in order to extract experiences from social media discussions.
\subsection{Market research}
The importance of decision-making in a market strategy and marketing is dependent on the conclusions of market research. The market research procedure varies from a form to another but has a set of steps. The standard procedure of market research helps avoid errors, misunderstanding, and uncertainty risk. One can define the procedure of marketing research as
 “The scientific and systematical procedure which involves defining the problem, collecting data, and analyzing and reporting the data and results of a specific issue when the problem is encountered by the organization” \cite{Kotler2007} 
Marketing research procedure overview: the procedure of marketing research includes a step-by-step loom in which each step is to respond to the scrupulous questions \cite{Kotler2007}: 
Why is it required to perform the research?
What does research require to perform?
How can the research objectives be accomplished?

Marketing research procedure stages:\\
\textbf{}\\
Step 1: Research Purpose\\
The complete and precise research purpose definition and understanding system consistently needs endeavor.\cite{Malhotra2005}\\
Step 2: Research Objectives\\
The research objective is an accurate report required in the sequence. It more accurately and intelligibly appreciates the problem statement. The research objectives are designed such that the chore of finding the required information perfectly satisfies the research purpose.
Research objectives consist of three crucial parts: 
Research Question
The research question defines the information that is required by the organization that is accountable for decision-making. It represents that the in a sequence should be obtained in accordance with the research purpose. \cite{Kotler2007}
Hypothesis
The hypothesis represents the researcher’s perspective of the likely answer to the question of the research. The Researcher can generate = the possible research question results in one of the previous stages and perform the research to answer whether a hypothesis proposed at the beginning of the research was correct.
Research Scope
Hypothesis development helps keep the research procedure more precise and focused to contribute to the research purpose. It is another important aspect of the research to indicate the boundaries limitations of the research. \cite{Kotler2007}\\
Step 3: Value estimation of research information\\
Once the purpose, objectives, and scope of the research have been defined, it is similarly very important to gauge the estimation of the required data or that of the research issue to be conceivably answered in the research question.\\
Step 4: Research Design\\
The research design is generally the research’s structure or framework developed to conduct the research.\\
Step 5: Data Collection\\
Step 6: Data Analysis\\
Step 7: Reporting Results and Presentation. 
\subsubsection{Issues in data collection during "Market Research"}
The next aspects to be discussed are those that may appear during the assortment and estimation of information and distress chiefs and decision-makers: data cannot be accessed. 
Therefore, the model is built of dependent on probably-mistaken evaluations. It may be costly to acquire information. The information can be sufficiently precise or exact. The estimation of information is subjective regularly. Information may be unreliable. 
The important information that impacts the outcomes may be qualitative (delicate). An excessive amount of information may exist (that is, data over-burden). The results (or outcomes) might take place during an inclusive period. Therefore, costs, profits, and incomes are recorded at various focuses on time. In order to tackle this difficulty, one can adopt a present-esteem approach when the outcomes cannot be quantified. Future information is expected to be similar to historical data.\\
In case this is not the circumstance, the change’s nature should be predicted and included in the investigation. After finishing the preliminary investigation, it is possible to decide whether an issue really exists, where it is identified, and how important it is. 

A key aspect is whether the data framework reports an issue or solely an issue’s side effects. For instance, if the reports suggest that sales have reduced, there is a problem; however, the situation undoubtedly indicates symptoms of the problem. The actual problem must be known. It sometimes may be a perception problem, organizational procedures, or incentive mismatch, instead of a low-performance decision model.\cite{RameshShardaDursunDelen2019}
\subsection{Social Media Mining}
Before we go towards social media mining, first we have to comprehend social media. Social media is characterized as any platform that encourages user-generated content and ongoing shared communication. In previous articles many of the researchers explicitly characterize social media as a set of web and mobile instruments and applications that stimulate interpersonal communication and opinion sharing, and the creation and dissemination of user-generated content. All the world's significant brands are presently active via social media\cite{Harrigan2020} and so many papers investigated that in many ways including the usage analysis of social media for competitive business outcomes \cite{Cao2018}. Social media data have recently attracted significant attention as a rising voice of the client as it has quickly become a channel for trading and storing customer-generated, huge amount, and unregulated voices about different aspects of business\cite{Jeong2019}\cite{DelVecchio2020}. 
The noticeable advancements in social in the past decade, as well as digital channels’ profusion, e.g., social network websites (such as Facebook), microblogs (such as Twitter) and media sharing (such as Instagram or YouTube), have brought a revolution not only in the communications of brands with their customers but also in the customers’ roles within the marketing procedure like eWOM \cite{Abedi2019}. For example, social media provides consumers with a voice equal to or even higher than brands, thereby disrupting marketing procedures and bringing immense challenges and dilemmas to marketers \cite{Constantinides2008}.\\ Managers in branding are no longer able to ignore the essential online voice of their consumers \cite{Gensler2013}. They also enjoy new opportunities to exploit the unfettered consumer-generated content (CGC) that are available on platforms in social media. The term social media refers to the strengthening advances in social interactions among people in which information, opinions, and ideas are shared and exchanged within virtual networks and communities. Social media consists of a set of internet-based programming applications manufacturing on Web 2.0 innovative and ideological establishments and allowing user-generated content to be generated and traded\cite{Kaplan2010}.\\
Since digital marketing is currently considered to be a “many-to-many conversation” between companies and customers as well as between customers \cite{Lusch2010}, the classic one-way business-to-customer transmissions have begun becoming outdated.
The tracking and analysis of the feelings and views of customers on particular brands, services, or products associated with the CGC on social media are a recent trend in the environment of digital marketing analytics. The aim is to categorize positive and negative CGC, commonly text-based (naïve) content, in accordance with a number of manual or automated categorization methods (i.e., naïve Bayes). Marketers, for instance, can receive timely customer feedback on a newly-offered service or product by assessing customer sentiment stated in the comments of a Facebook post or those of tweets with a particular hashtag associated with the service/product.\\
Considering the immense CGC volume, which is typically known as “Big Data” and has expanded along with the employment of social media platforms, it is no longer feasible to qualitatively and manually analyze the sentiment of consumers that is expressed in online brand-associated content. For example, Twitter produces more than 500 million tweets every day, and Facebook hosts 4.75 billion content pieces each day. This adds to the need for developing automated instruments to identify and analyze text-expressed customer sentiment \cite{Wang2012}. There are two eminent approaches to the automated analysis of sentiment. It is a widely used approach to perform classification via a set of weighted words \cite{Taboada2011} as an approach to the analysis of sentiment within the community of marketing research \cite{Bolat2017}, since it does not need the classifier to be pre-processed or trained.\\
On the other hand, the use of machine learning for sentiment analysis, which is also considered as a supervised learning method, is mostly reported to have accuracy \cite{Chaovalit2005} Also, it has been employed in marketing research \cite{Pathak2017}. The machine learning approach, however, needs a training phase, which is performed by either the researchers or the provider of the sentiment software. Since each one of these methods offers its specific advantages and limitations, researchers and marketers should verify classification accuracy to avoid any action on inaccurate outcomes of data analysis \cite{Canhoto2015}. In addition, considering the wide platform range in social media and their particularities on content type that can be created by consumers (e.g. Twitter tweets, Facebook comments, emoticons, hashtags, emojis, abbreviations, and slang language), the available sentiment analysis techniques, which is typically examined on English texts, need to be carefully validated before they are put in use on social media data by marketers.\\
SMM makes different contributions to market research, such as almost freeness, quick accessibility, and including no information bias, which may be analyzed using various algorithms.
It also helps make decisions in many manners, and it is mostly the market research outcome based on SMM, Eventually, SMM can contribute to celebrity branding.

\section{Methodology}
\FloatBarrier
\begin{figure}[H]
\centering
\ignorespaces
\includegraphics[max size={\textwidth}{\textheight}, keepaspectratio]{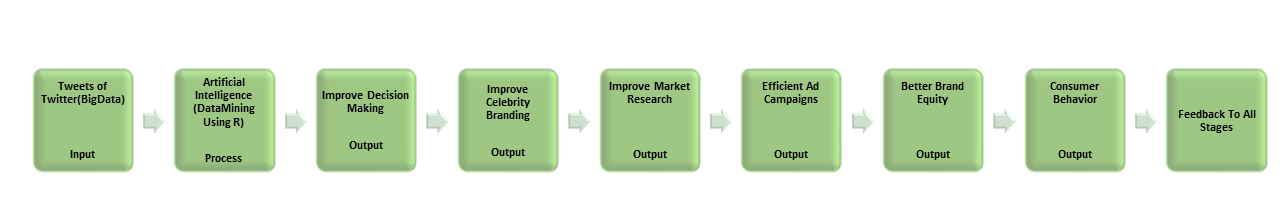}
\caption{{Flowchart of the all processes of the Research(Compiled By The Author)}}
\label{caption.conceptual model}
\end{figure}    
The current data age is known for the rapid development in the measure of data and information that are electrically collected, stored, provided. A considerable portion of business information is stored in basically-unstructured text documents. According to Merrill Lynch and Gartner, 85\% of a corporate datum is captured and stored in an unstructured format. The equivalent study also suggested that the size of such non-structurally-stored information is increasing every 18 months. Since knowledge provides power in the current business world and is derived from information and data, organizations which make effective and efficient use of their text information sources enjoy the crucial knowledge to make better choices, inducing a competitive advantage over the organizations that lag behind. Thus, the need for text analytics and mining becomes part of the all-inclusive view of the current organizations.\\ Although the all-encompassing goal of text analytics and mining is transforming the unstructured textual data format into actionable information by applying natural language processing (NLP) and investigation, such terms’ definitions They indicated "text analytics" to be a broader idea which includes the retrieval of information (e.g., the scanning and recognizing of pertinent reports for a specific key term arrangement), as data mining, information extraction, and Web mining, whereas "text mining" was indicated to fundamentally be focused on identifying new, high-value knowledge from textual data sources:\cite{RameshShardaDursunDelen2019}\\
$Text\;Analytics\;=\;Information\;Retrieval\;+\;Information\;Extraction\;+\;Data\;Mining\;+\;Web\;Mining $
\subsection{Text mining process}
The semantic extraction of the textual content of Social Media represents a further step towards better
understanding and getting insight out of the true context of a consumer’s content as a social network user\cite{Abu-Salih2018}. \\
For success in text mining, as \cite{Usai2018} remarked that in Text mining process we try to find hidden patterns in any textual content including CGCs and better understanding patterns found to be analyzed, investigations should pursue the best rehearse-dependent sound technique. A standardized process model, e.g., Cross-Industry Standard Process, is required for Data Mining (CRISP-DM) - it is the business standard for data mining practices.  
Although CRISP-DM pieces are mostly material to text mining practices, a particular data mining process model would include information preprocessing activities with significantly more details. \cite{RameshShardaDursunDelen2019}

\FloatBarrier
\begin{figure}[H]
\centering
\ignorespaces
\includegraphics[max size={4in}{4in}, keepaspectratio]{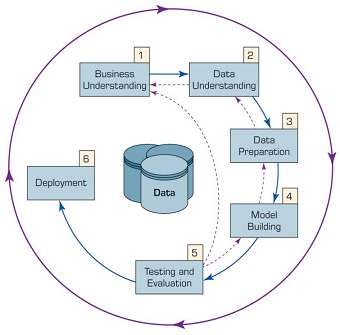}
\caption{{CRISP-DM Model}}
\label{caption.conceptual model}
\end{figure}

\subsection{Extract the knowledge}

A knowledge extraction technique’s categories include clustering, classification, association, and trend analysis.

\subsection{Sentiment Analysis} 
Artificial Intelligence (AI) and Machine Learning (ML) are actually making a huge difference. It is expected that the 21st century will observer the blast of all their potential in each part of human life. The decision-making, celebrity branding and market research are definitely not special cases since they additionally need AI and ML to improve their businesses’ models and performance\cite{Jimenez-Marquez2019}. One of the sub layers to AI is sentiment analysis which utilizes many ML algorithms.
Humans are social creatures. They are skillful in the use of a set of instruments to communicate. They regularly consult financial talk forums prior to making an investment decision; approach some people for their views of a recently-opened restaurant or a recently-released movie; in addition, lead internet searches read consumer reviews and expert reports prior to making a major purchase such as a home, an appliance, or a vehicle. Individuals rely on the suppositions of others to make better decisions, specifically in a region where they do not have sufficient knowledge or experience. 
In light of the increasing popularity and availability of view-rich internet resources, e.g., social media outlets (such as Twitter and Facebook), personal blogs, and online review websites, it is today simpler than before to find alternatives of others (truly, a huge number of them) on an infinite number of things ranging from the newest devices to the political and public figures. Although not everyone communicates suppositions on the internet—generally due to social correspondence channels’ rapidly-developing abilities and number—the numbers are exponentially increasing. 
It is difficult to define the word “sentiment.” It is typically associated with or mistaken for various terms, such as belief, conviction, view, and opinion. It suggests a settled set that reflects the emotions of an individual  \cite{Mejova2009}.Sentiment has a few remarkable properties which make it distinct from various ideas that should be recognized in the text. Typically, it is required to arrange texts based on the topics. This can include the management of all scientific subject categorizations. Again, sentiment classification serves as a rule and manages two classes (i.e., positive and negative), a polarity range (e.g., star ratings of movies), or even an opinion quality range \cite{Pang2008}. In addition the majority of the semantics-based models use sentiment lexicon to 
perceive emotional keywords. Be that as it may, in light of the fact that the slant of the whole content is made a decision about dependent on a couple of emotional keywords, the logical relations among trending events that fundamentally impact public opinions and feelings are typically dismissed\cite{Wu2020}. So for answering to these kinds of neglections, we used Naïve Bayes algorithm alongside the lexicon based and brought both results to compare.

\subsection{Methods for polarity identification}
As mentioned in the previous section, one can apply polarity identification at different levels, including a word, a term, a sentence, or a document. The word level is the highest polarity identification level. Once the word level of polarity identification is performed, one can collect into the following higher level, and then the following until the desired conglomeration degree of sentiment analysis is achieved. Two dominant word/term-level polarity identification procedures have been employed, each of which offers its interest and hindrances aspects :\cite{RameshShardaDursunDelen2019} \\
1. The use a lexicon as a reference library (which is developed automatically or manually by someone for a particular task or by an institution for the purpose of general use) \cite{Esuli2006}\cite{Strapparava2004}\cite{Kim2004}\cite{Liu2005}.\\
2.The use of a set of training documents as the knowledge source of the polarity of terms in a particular domain (inducing predictive models of viewed textual documents) \cite{Dhaoui2017}.

\subsection{Classification Techniques}
A few methods (or algorithms) are utilized in classification modeling, such as decision tree analysis, neural networks, statistical analysis, genetic algorithms, Bayesian classifiers, and rough sets.\cite{RameshShardaDursunDelen2019}\cite{Feldman2007}.

\subsubsection{Naïve Bayes Method for Classification}
Na{\"{\i}}ve Bayes is a simple classification technique based on probability (a machine-learning method applied to prediction problems of classification type) that is derived from the Bayes theorem. In this method, the output variable should be of nominal values. For this reason, the present study employed this technique as the machine-learning approach. Despite the fact that the input variables may be a combination of numeric and nominal types, it is required to discretize the numeric output variable by using a binning method prior to using it in a Bayes classifier. The term “Na{\"{\i}}ve” arises from the strong and, to some extent, unrealistic independence assumption among input variables. In simple terms, in a Naïve Bayes classifier, the input variables are assumed to not be dependent on each other. It also assumes the presence and absence of a specific variable in the predictor combinations to not be associated with other variables’ presence or absence.\\
One can develop classification models of Naïve Bayes type very effectively (completely accurately) and efficiently (i.e., relatively quickly with small computational effort) in a supervised machine-learning setting. In other words, through a collection of training data (which is not necessarily to be large), one can obtain the parameters of classification models of Naïve Bayes type by utilizing the maximum likelihood method -t that is, due to the assumption of independence, Naïve Bayes models can be developed without strictly complying with the Bayes theorem’s entire rules and requirements. \cite{RameshShardaDursunDelen2019}

\subsection{Data gathering and steps}
Today, as \cite{Caesarius2018} mentioned, "Big Data'' opportunities are a lot and in fact they provide manual ways for illogically dealing with sentiment analysis and add to the need for making robotized devices for the analysis of textually-expressed consumer sentiment. The present study selected Twitter as the case-study social media since it can be seen as an internet-based short message service (SMS) extension. IAs Jack Dorsey, the co-founder and co-creator of Twitter, said:\\
"...We came across the word 'twitter', and it was just perfect. The definition was 'a short burst of inconsequential information,' and 'chirps from birds'. And that's exactly what the product was." \\Twitter functions as a utility via which individuals can send SMSs throughout the world. It allows for continuously becoming heard and receiving answers. As the audience of Twitter includes a huge number of people, responses are commonly very quick. This platform facilitates humans’ fundamental social instincts. Through sharing on Twitter, users can easily express their opinions on everything at any time. Connected friends or followers (on Twitter) immediately obtain the information on the current situations in the lives of people. This, in turn, contributes to another emotion of humans—i.e., the innate need for knowing the current life situations of people. The user interface (UI) of Twitter is not only real-time but also easy to use. It is understood instinctively and naturally -i.e., the Twitter UI has a very intuitive nature.\\
If the tweets of users can properly be mined and analyzed, Twitter can serve as an excellent advertisement and marketing tool for so many things including successful social media marketing strategies \cite{P2020}. However, the information that is provided by Twitter includes a larger domain. As Twitter is naturally non-symmetric in terms of followers and followings, it helps better understand user interests than its influence on the social network. One can consider an interest graph as a way of learning the connections of individuals and their various interests. The degree calculation of the association or correlations of the interests of individuals and the potential advertisements is among the most essential applications of an interest graph. On the basis of such correlations, users can be aimed to obtain a maximum response to advertisement campaigns, along with the recommendations of followers.\\
A vast sample of customer tweets of fifteen twitter brands and influencer pages are mined and utilized for the comparison of lexicon-based and machine learning techniques to the sentiment analysis via the Naïve algorithm (lexicon-based) and the Na{\"{\i}}ve Bayes algorithm (i.e., a machine learning method) and for obtaining the desired outcomes.\\
These brands are chosen since they are all located in the USA, and they have a good, appropriate, and active twitter account that can be dug to obtain information, and had many campaigns over the years. Regarding celebrities, almost the same reason was the case, and their reputation and connection to the companies were a major concern.\\
The present study utilized R studio, R language, and Twitter developer API for data mining as it is an open-source instrument, has many useful packages, and has result visualization ability. Appropriate data were gathered by searching hashtags and keywords relating to companies and celebrities. Then, after they were cleansed and made them meaningful, the analysis part was carried out through the two above-mentioned methods. \\

We have 5 companies and 2 correspond celebrities per each company that has been used by those companies before as the influencer of the ad campaign. They all mentioned in Figure 3
\FloatBarrier
\begin{figure}[H]
\centering
\includegraphics[max size={4in}{4in}, keepaspectratio]{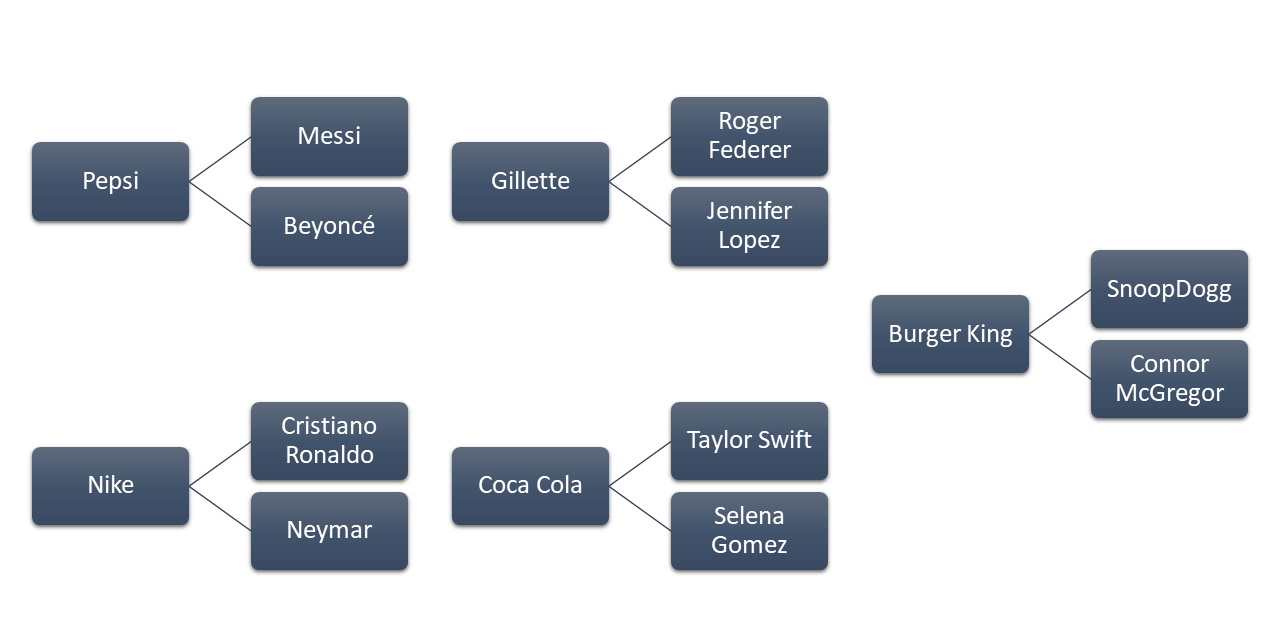}
\caption{{Name of the companies and related celebrities}}
\label{caption.conceptual model}
\end{figure}

The steps can be summarized as:\\
1- Understand the API of Twitter\\
2- Generate a connection with Twitter API \\
3- Generate a new app\\
4- Install the entire required packages and libraries in R\\
5- Supply the secret, key, token, and token secret through the Twitter API for the purpose of authentication and Twitter-R connection\\
6- Gather relating tweets as a textual collection (or corpus)\\
7- Clean the textual collection as follows:\\
\# Omit HTML links that are not needed for the sentiment analysis \\
\# Eliminate retweet entities in the stored tweets (i.e., text)\\
\# Remove every "\#Hashtag"  \\
\# Remove every "@people" \\
\# Remove the entire punctuation  \\
\# Remove numbers (only text is required for the analytics) \\
\# Eventually, remove the unrequired spaces (e.g., white spaces and tabs) \\
\# Anything else that needs to be removed can be removed, e.g., slang words, through the same methods and function applied to the above-mentioned items \\
\# Convert the entire words in lower case as this provides a uniform pattern. \\
\# Eliminate "NA" tweets in this list of tweets \\
\# Remove repeated tweets from the list of tweets.  

\FloatBarrier
\begin{table}[H]
\caption{{Data cleaning process} }
\label{tw-9dff510fd0a5}
\def\arraystretch{1}
\ignorespaces 
\centering 
\begin{tabulary}{\linewidth}{p{\dimexpr.25\linewidth-2\tabcolsep}p{\dimexpr.25\linewidth-2\tabcolsep}p{\dimexpr.2431\linewidth-2\tabcolsep}p{\dimexpr.2569\linewidth-2\tabcolsep}}
\tbltoprule Brand Names & Number of Tweets before cleansing & Number of Tweets after cleansing \\
\tblmidrule 
Nike &10000&4824\\
Coca Cola&10000&5498\\
Burger King&10000&8252\\
Pepsi	&10000&4724\\
Gillette&10000&4640\\
\tblbottomrule 
\end{tabulary}\par 
\end{table}

\FloatBarrier
\begin{table}[H]
\caption{{Data cleaning process} }
\label{tw-9dff510fd0a5}
\def\arraystretch{1}
\ignorespaces 
\centering 
\begin{tabulary}{\linewidth}{p{\dimexpr.25\linewidth-2\tabcolsep}p{\dimexpr.25\linewidth-2\tabcolsep}p{\dimexpr.2431\linewidth-2\tabcolsep}p{\dimexpr.2569\linewidth-2\tabcolsep}}
\tbltoprule Celebrity Names & Number of Tweets before cleansing & Number of Tweets after cleansing \\
\tblmidrule 
Cristiano Ronaldo&10000&2828\\
Neymar&10000&6036\\
Beyoncé&10000&3263\\
Messi&10000&4054\\
Jennifer Lopez&10000&4668\\
Roger Federer&10000&4743\\
Connor McGregor&10000&6138\\
SnoopDogg&10000&5623\\
Taylor Swift&10000&3121\\
Selena Gomez&10000&5774\\
\tblbottomrule 
\end{tabulary}\par 
\end{table}

\textbf{}\\

\textbf{Specific sentiment estimation steps in the first method}\\
8- Download both negative and positive English opinion/sentiment words (nearly 16000 words)\\
9- Add some industry-specific and/or particularly emphatic terms, depending on the requirements\\
10- Generate a function in R that calculates the raw sentiment by the simple matching algorithm by considering the following:\\
\# Remove digits, punctuations, and control characters\\
\# Convert them all into the lower sentence case\\
\# Divide each of the sentences by using space-delimiter words \\
\# Obtain the Boolean matches of the words with the negative and positive opinion-lexicon\\
\# Obtain the score in the form of the total positive sentiment dedicated from the total negative sentiment\\
\# Return a frame of data containing the respective sentence and score\\

\textbf{Specific sentiment estimation steps in the second method}\\
11- Install  Rstem and sentiment packages\\
12- Make use of the classify\_emotion function that returns a class data frame object, consisting of seven columns (including disgust anger, joy, fear, surprise, sadness, and best\_fit) and a row for each of the documents\\
13- Replace such NA values  \\
14- Employ another function, i.e., classify\_polarity(), that is provided by the sentiment package, for the classification of  tweets into two groups, namely neg (negative sentiment) and pos (i.e., positive sentiment)  (It should be noted that the overall tweet sentiment is considered as neutral when this ratio is found to be 1)\\
15- Create consolidated results in a data frame from these two functions \\
16- Sort and rearrange the data within the frame\\
17- Generate a single function so that it is used by the tweets of each business and the sentiment can be plotted for each of the businesses \\
18- Likewise, plot the polarity distribution of the tweets\\
19- Try to obtain a sense of the tweets’ overall content through the word clouds.

\section{Results and Conclusions}

This study adopted a lexicon-based algorithm that used the Naïve algorithm for the polarity separation of tweets based on the collection of words that were fed to the system as negative and positive lexicon resources. However, this method has no deep analysis and no efficient sentiment analysis procedure. Thus, the next phase was included, in which the machine learning method was applied via the Naïve Bayes algorithm as it suits nominal analysis. Then, through the addition of r sentiment libraries and obtaining the benefit of the lexicon references, the tweets were classified into six groups of essential feelings, including joy, anger, fear, sadness, disgust, and surprise. \cite{RameshShardaDursunDelen2019}

\subsection{Results of the first method}

\FloatBarrier
\begin{table}[H]
\caption{{Burger King and related celebrities Mean Sentiment Score} }
\label{tw-9dff510fd0a5}
\def\arraystretch{1}
\ignorespaces 
\centering 
\begin{tabulary}{\linewidth}{p{\dimexpr.25\linewidth-2\tabcolsep}p{\dimexpr.25\linewidth-2\tabcolsep}p{\dimexpr.2431\linewidth-2\tabcolsep}p{\dimexpr.2569\linewidth-2\tabcolsep}}
\tbltoprule Organization(brand) & MeanSentimentScore \\
\tblmidrule 
Burger King	&0.0307804\\
SnoopDogg	&0.2443535\\
Connor McGregor&	0.0301401\\
\tblbottomrule 
\end{tabulary}\par 
\end{table}
As can be seen in Table 3, Snoopdogg was found to have a more positive sentiment within the minds of consumers, while Connor and Burger king were found to have a closer sentiment score. 


\FloatBarrier
\begin{figure}[H]
\begin{tabular}{cccc}
\subfloat[Burger King]{\includegraphics[width = 2in]{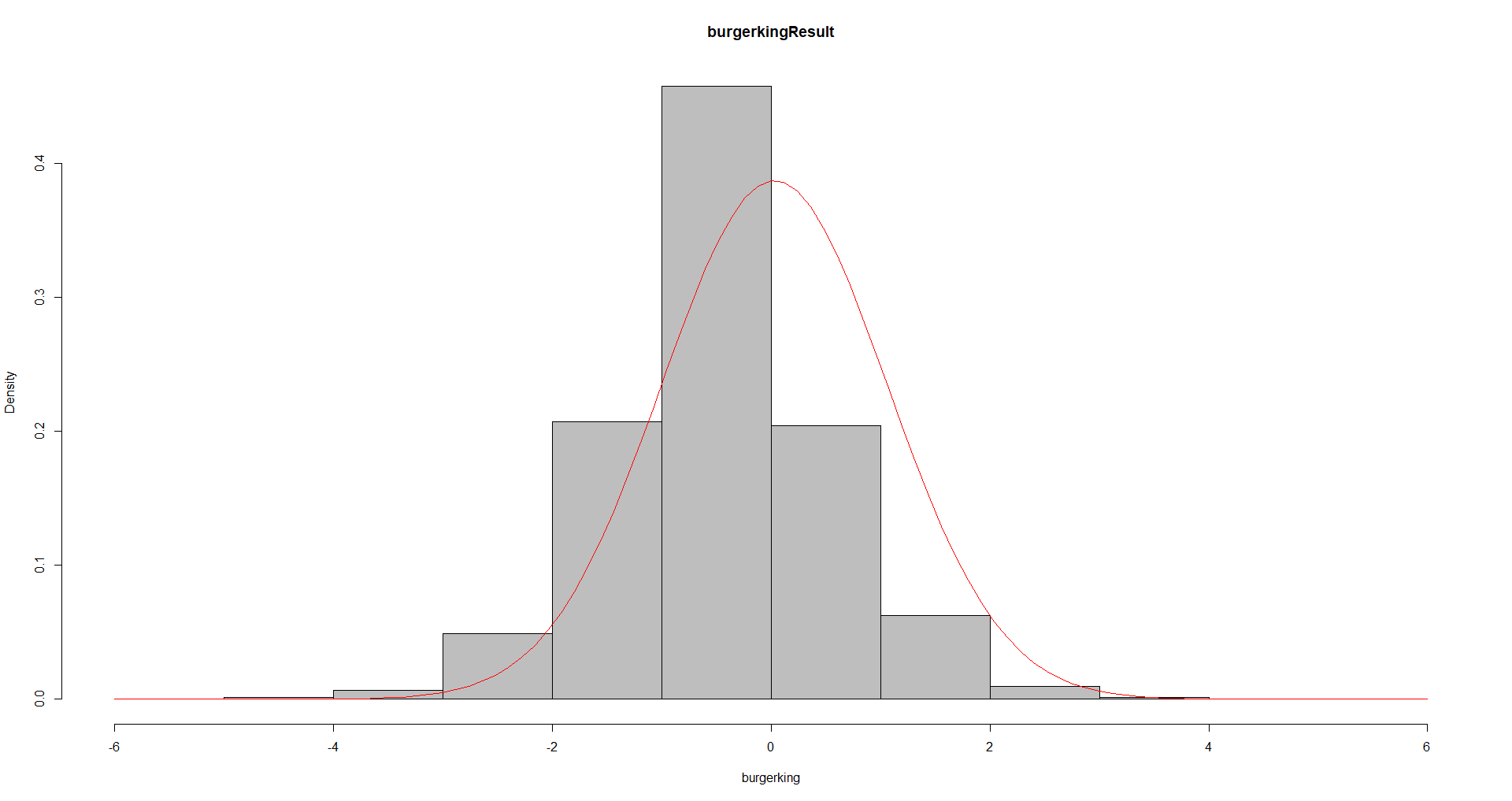}} &
\subfloat[Snoopdogg]{\includegraphics[width = 2in]{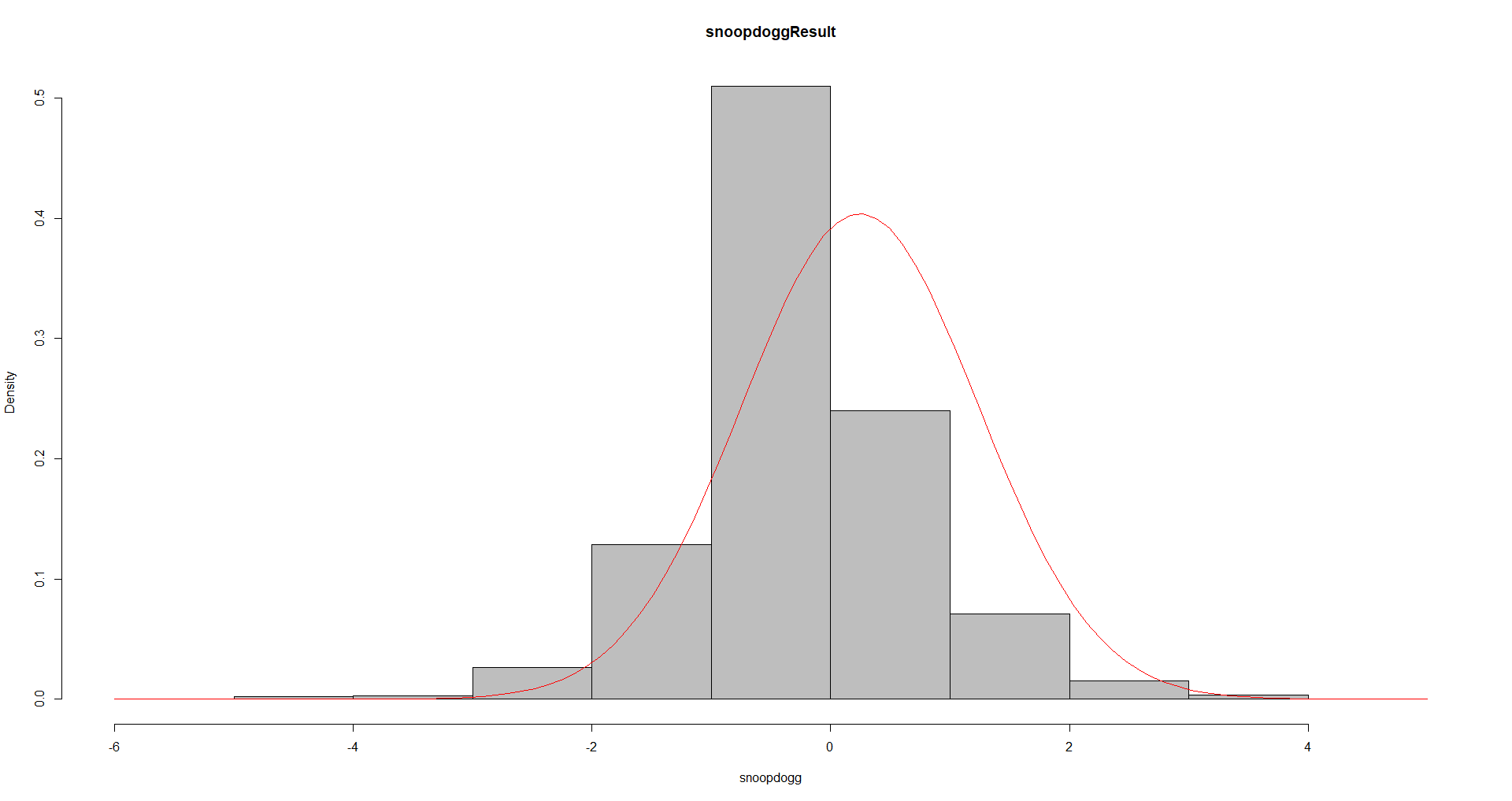}}&
\subfloat[Connor McGregor]{\includegraphics[width = 2in]{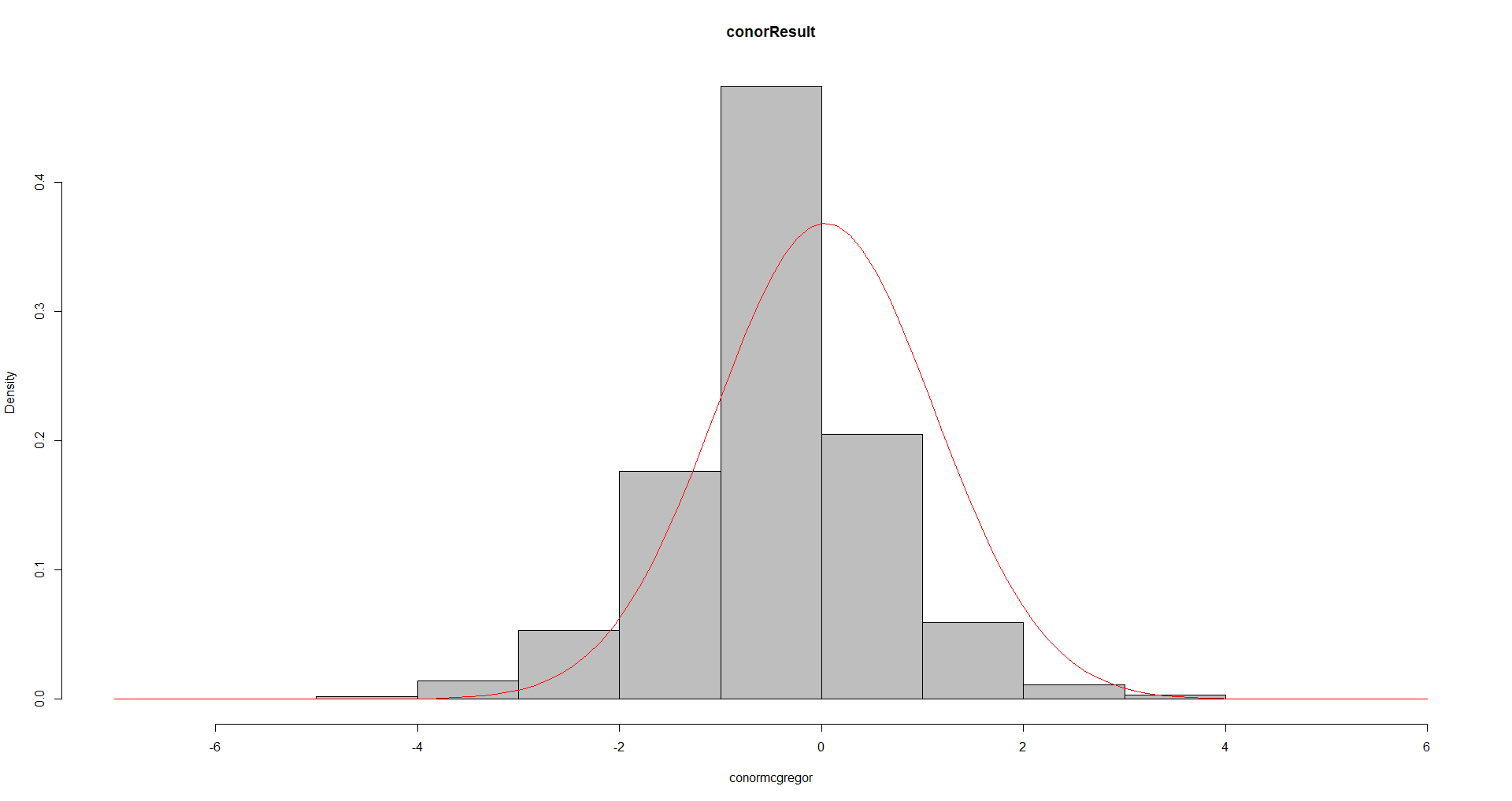}}\\
\subfloat[Coca Cola]{\includegraphics[width = 2in]{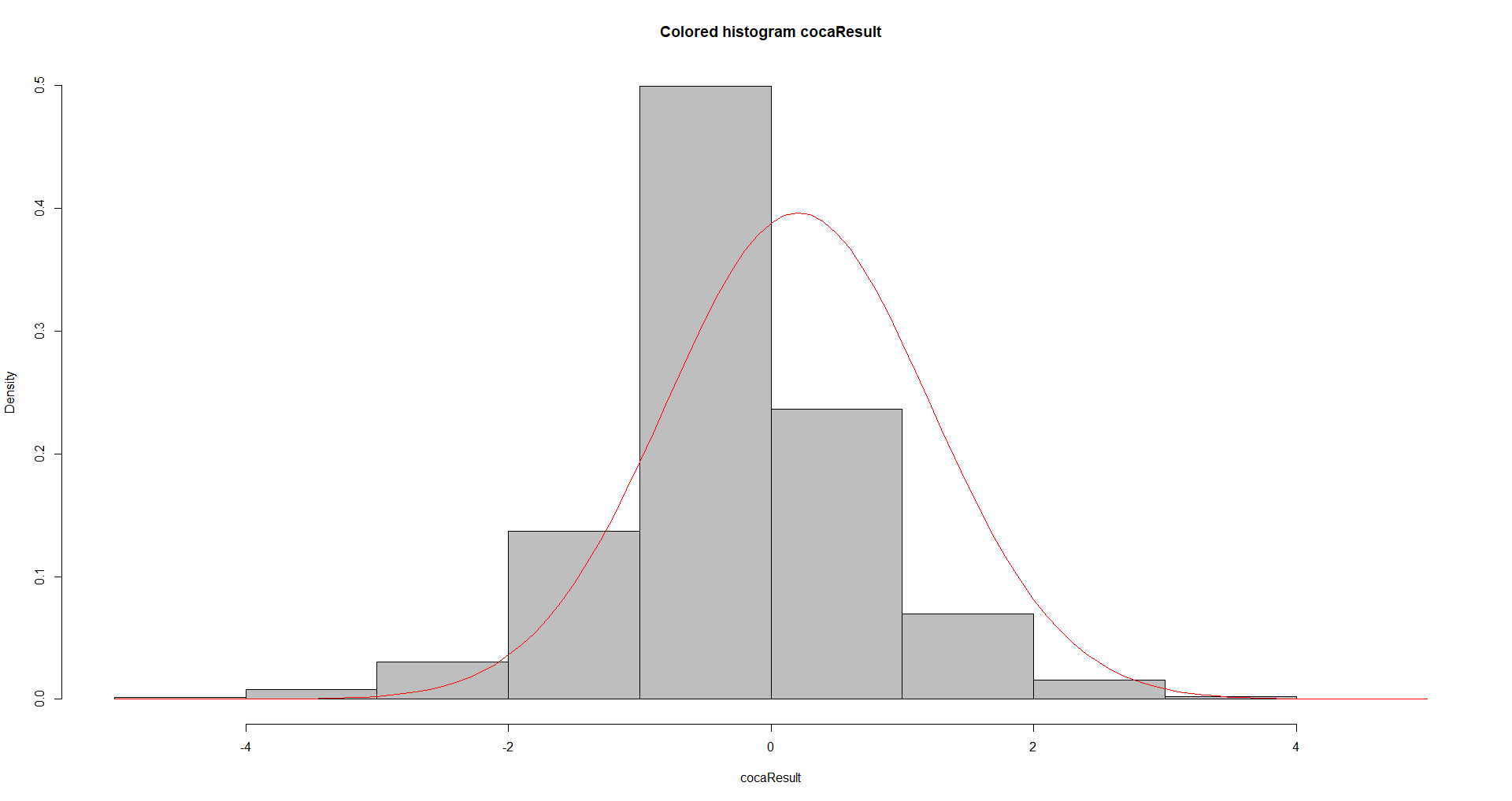}}&
\subfloat[Taylor Swift]{\includegraphics[width = 2in]{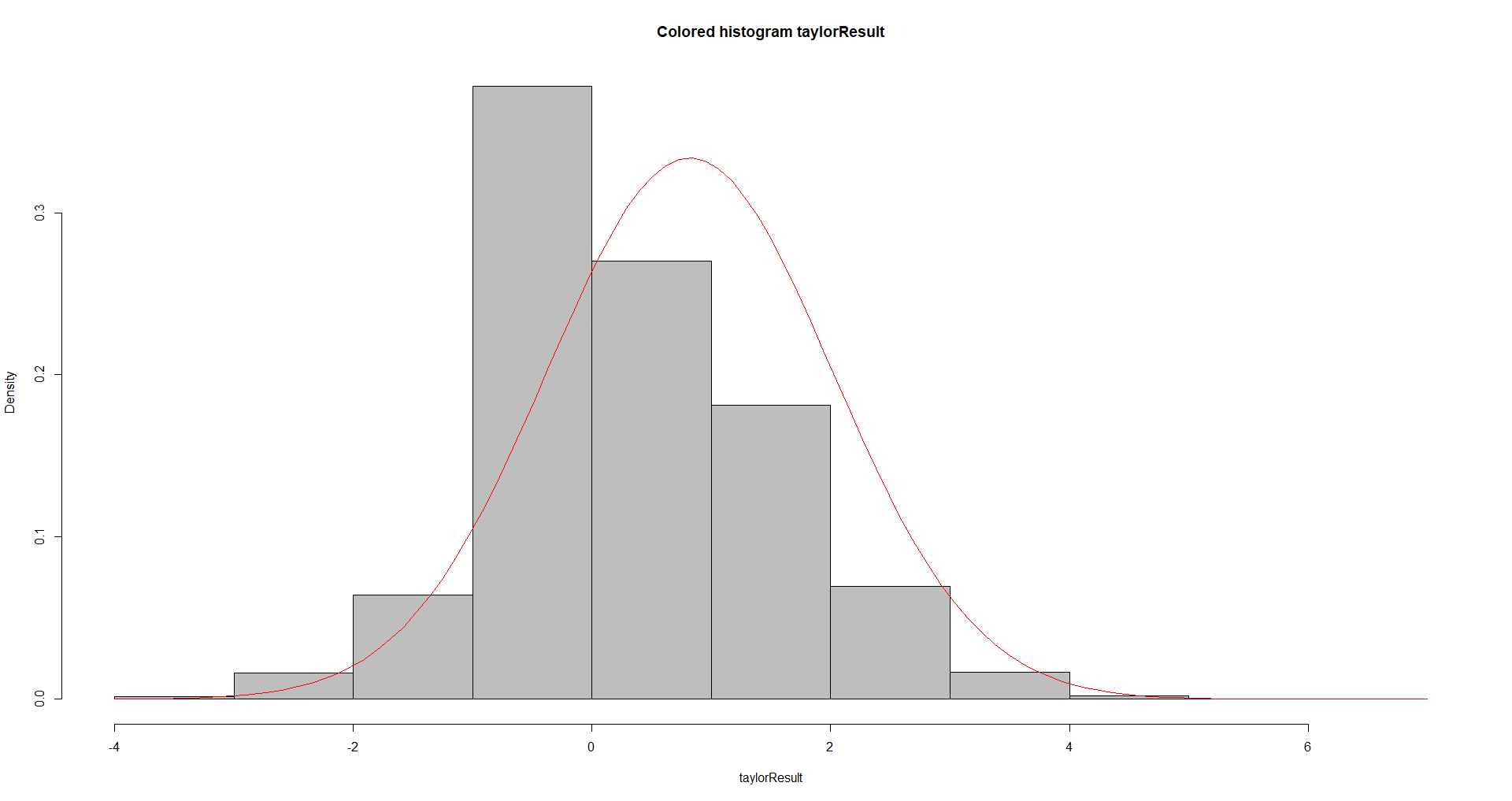}} &
\subfloat[Selena Gomez]{\includegraphics[width = 2in]{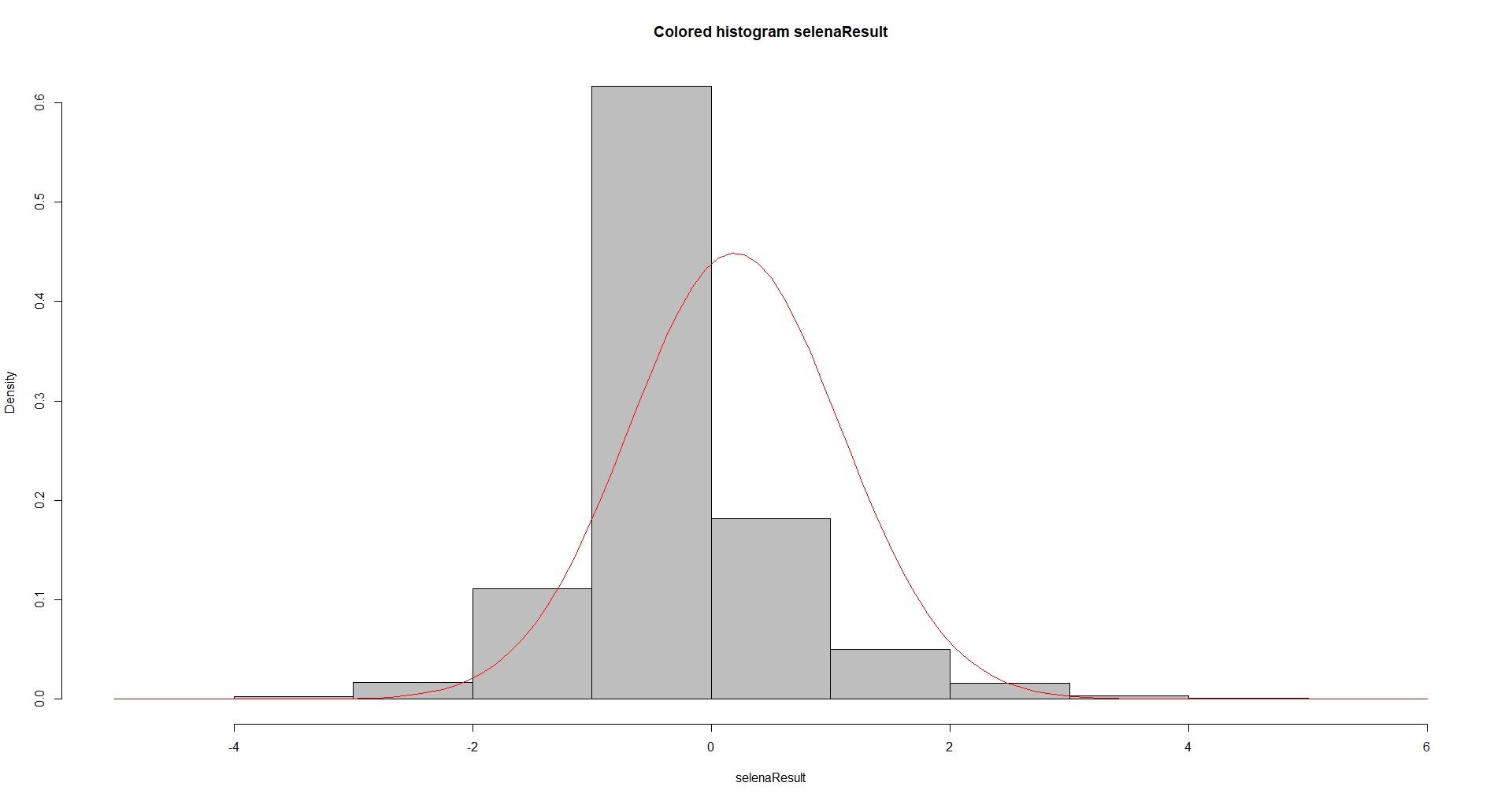}}\\
\subfloat[Pepsi]{\includegraphics[width = 2in]{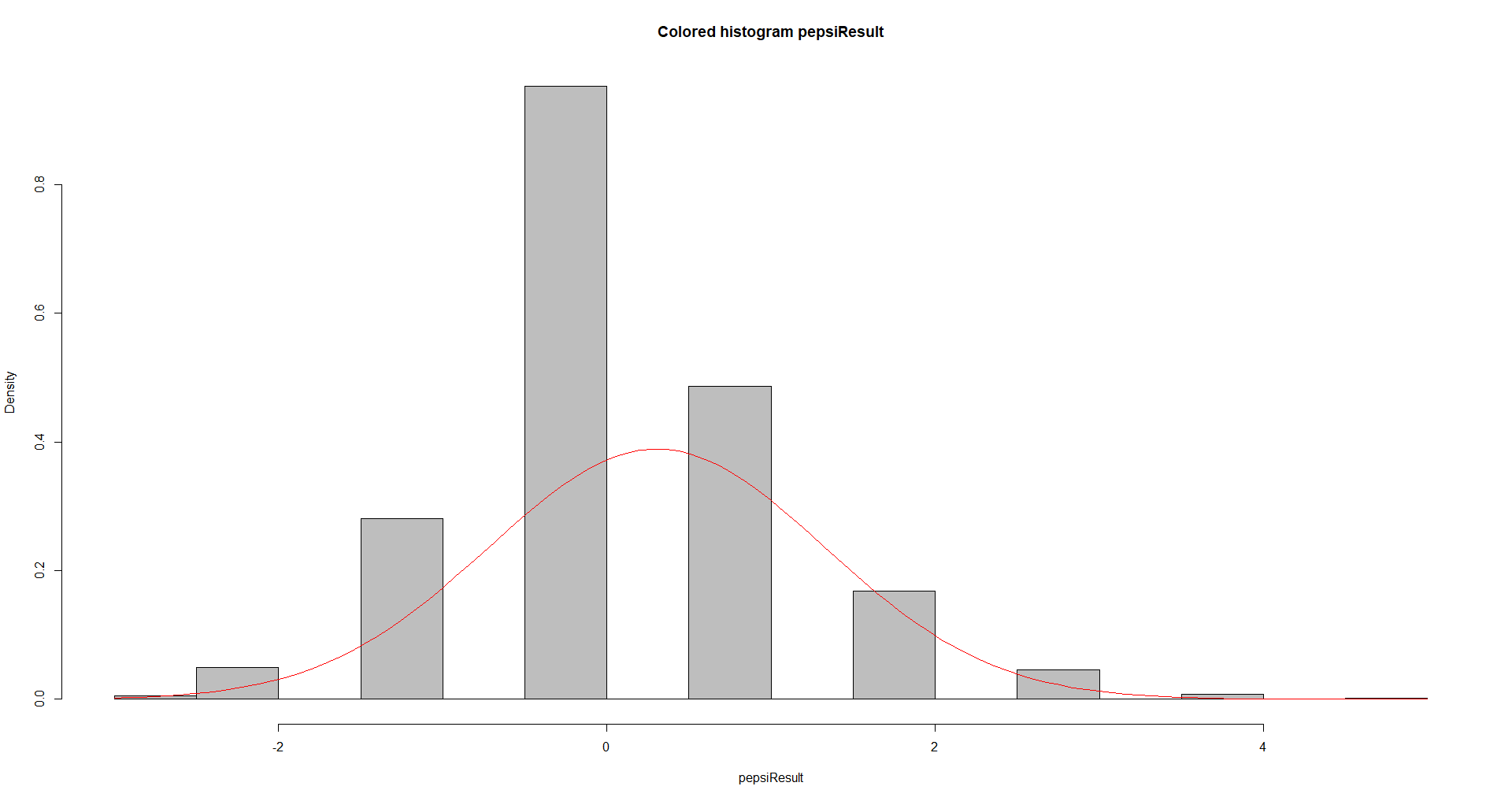}}&
\subfloat[Messi]{\includegraphics[width = 2in]{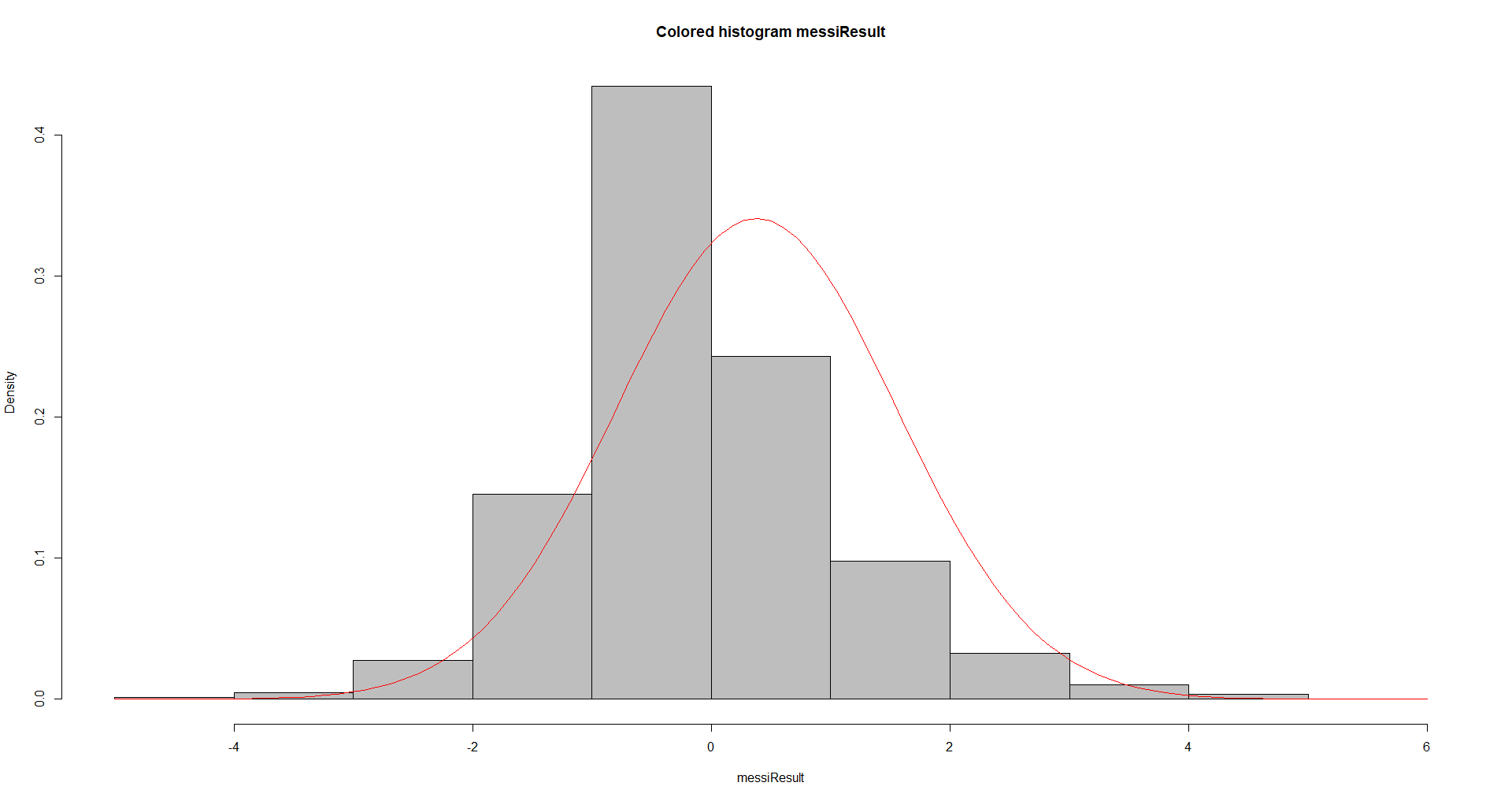}}&
\subfloat[Beyoncé]{\includegraphics[width = 2in]{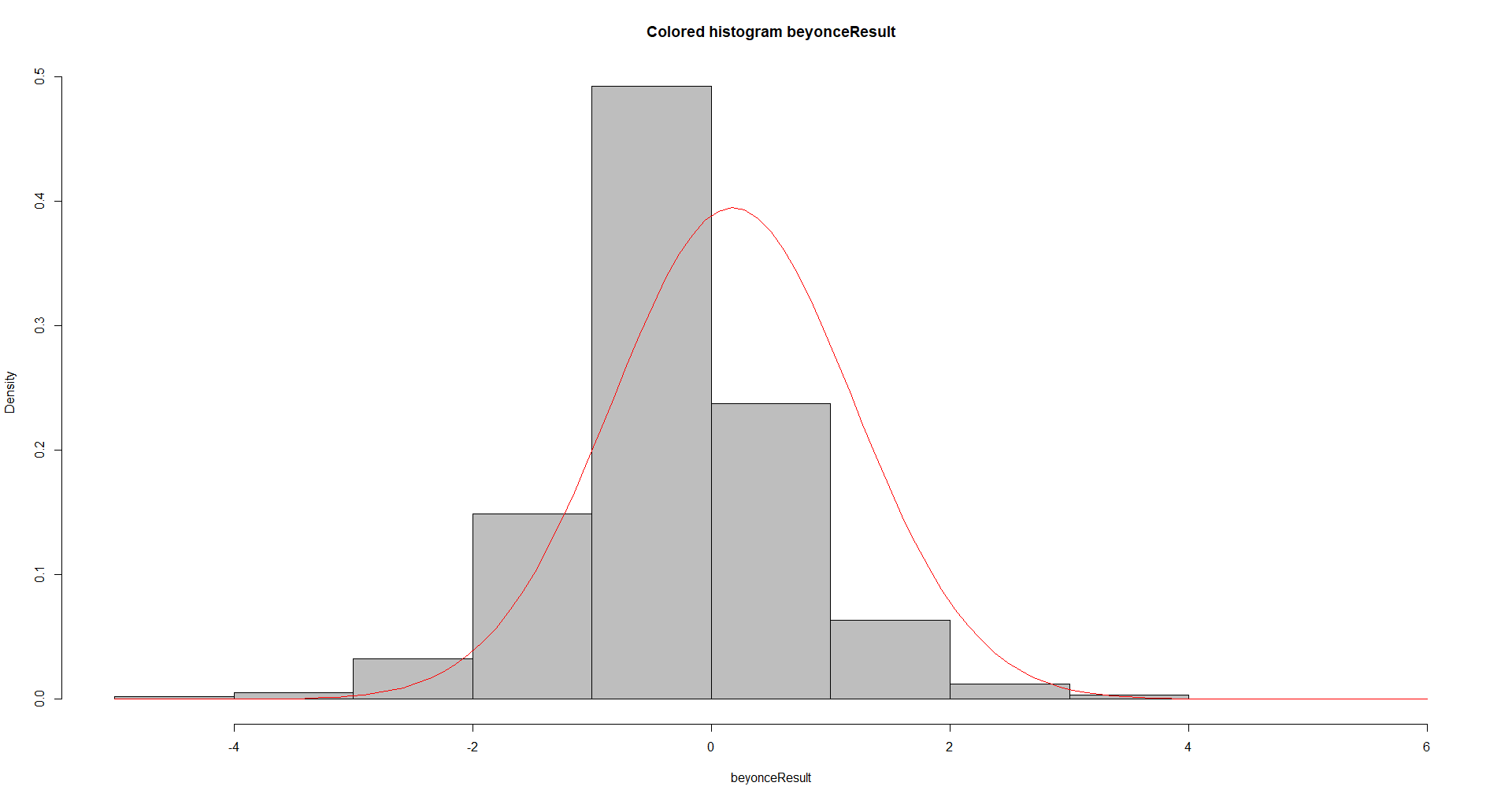}}\\
\subfloat[Gillette]{\includegraphics[width = 2in]{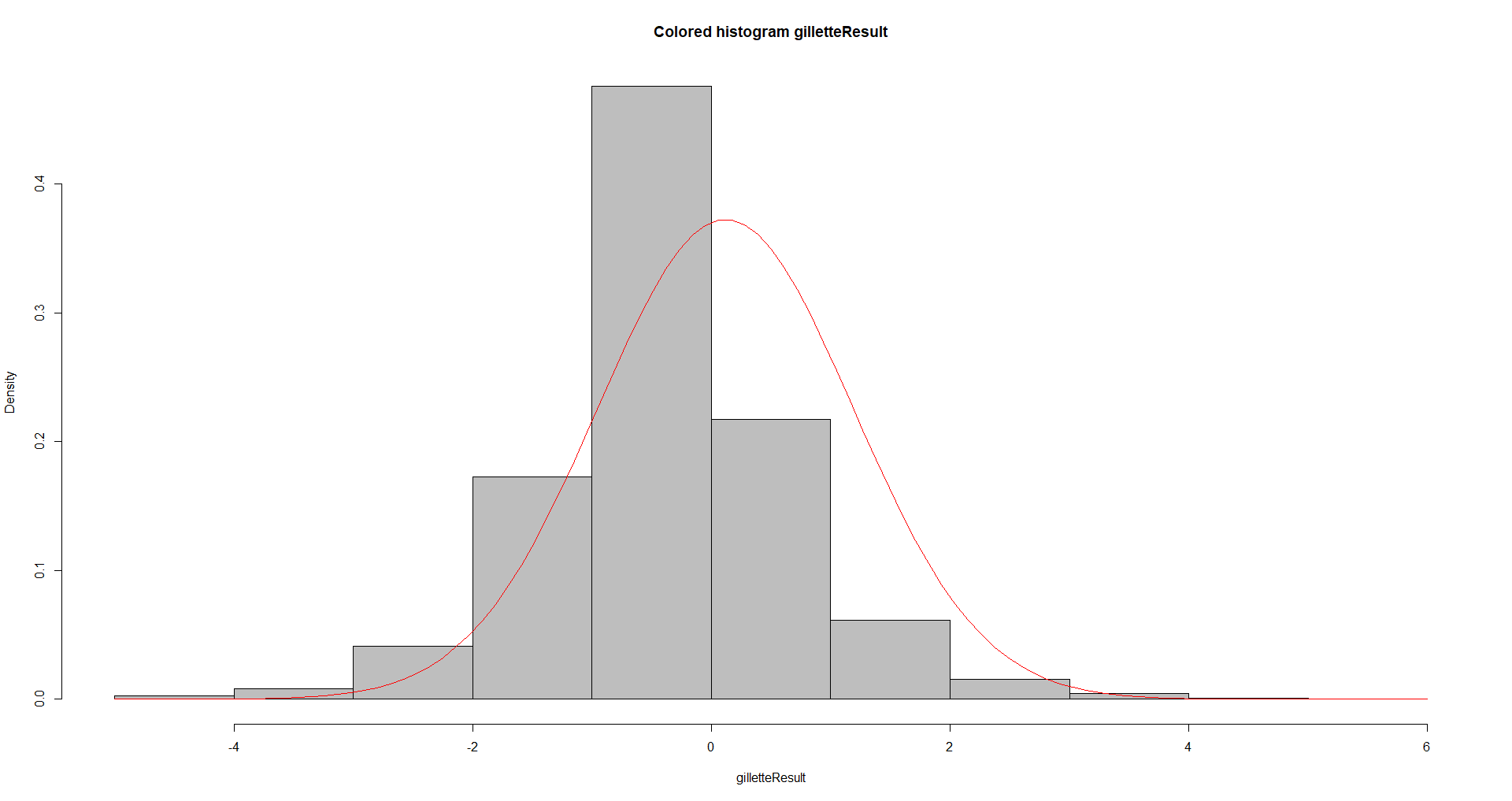}}&
\subfloat[Roger Federer]{\includegraphics[width = 2in]{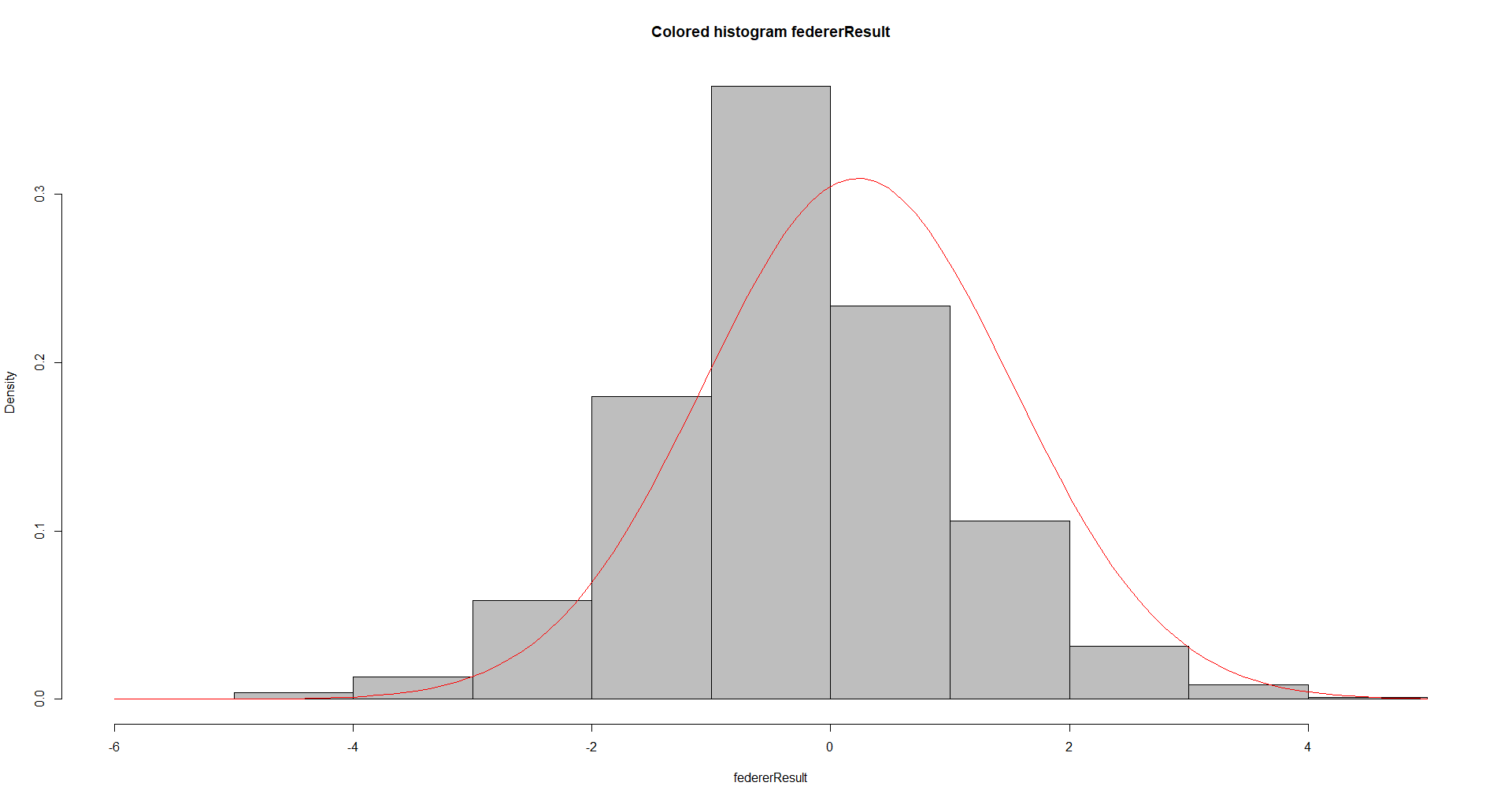}}&
\subfloat[Jennifer Lopez]{\includegraphics[width = 2in]{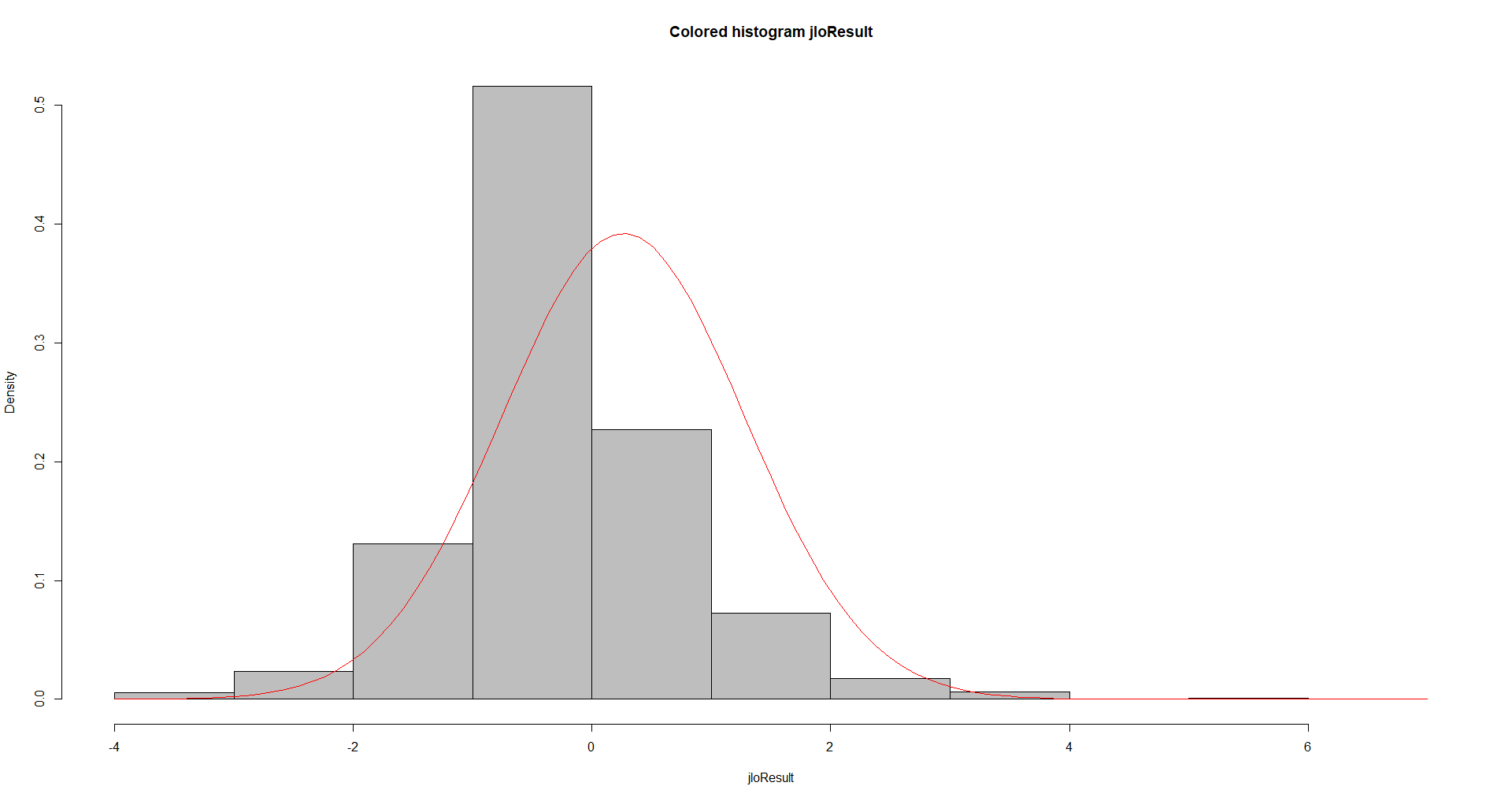}}\\
\subfloat[Nike]{\includegraphics[width = 2in]{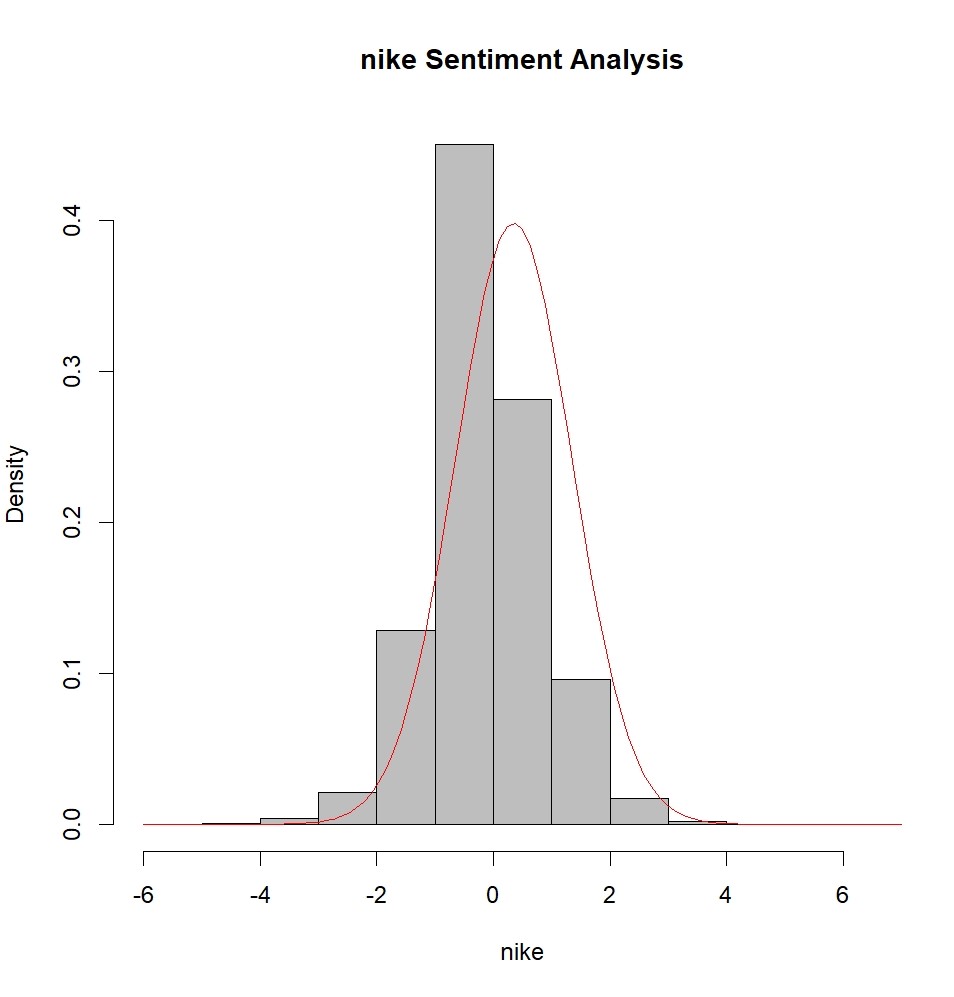}}&
\subfloat[Cristiano Ronaldo]{\includegraphics[width = 2in]{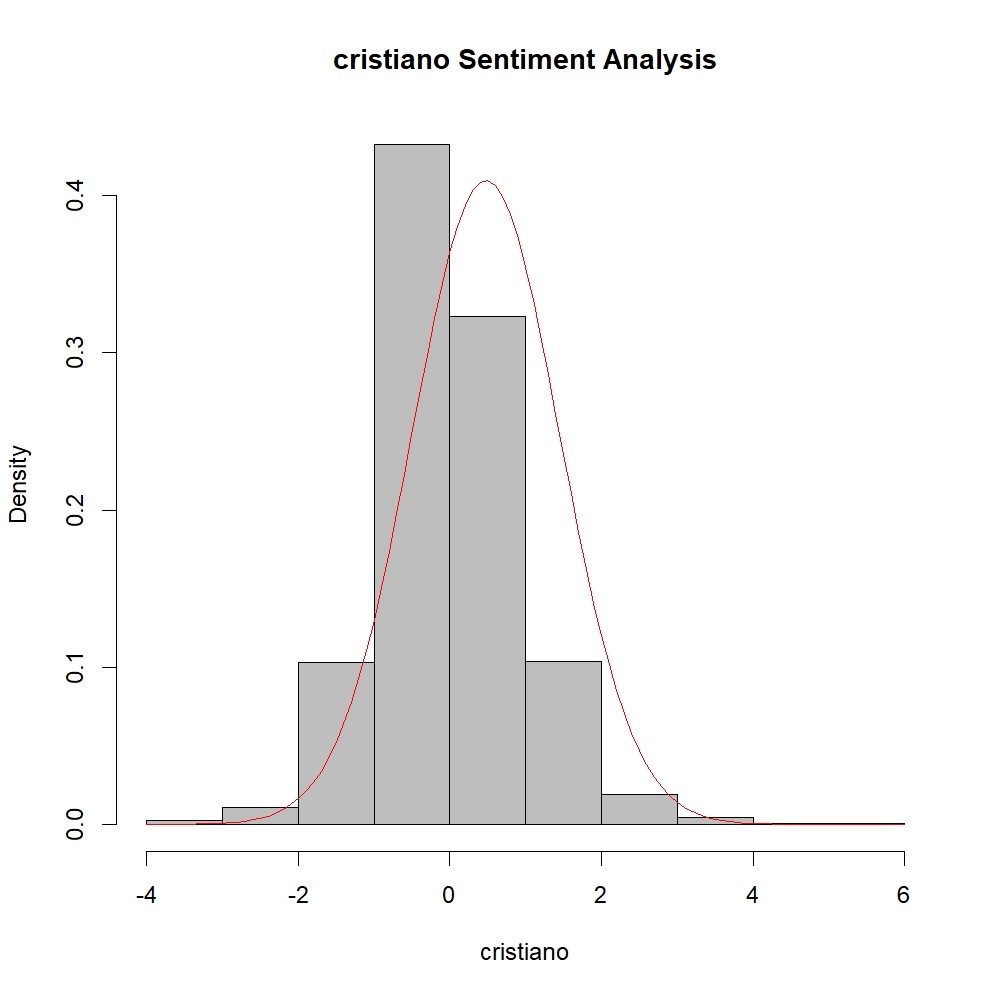}}&
\subfloat[Neymar]{\includegraphics[width = 2in]{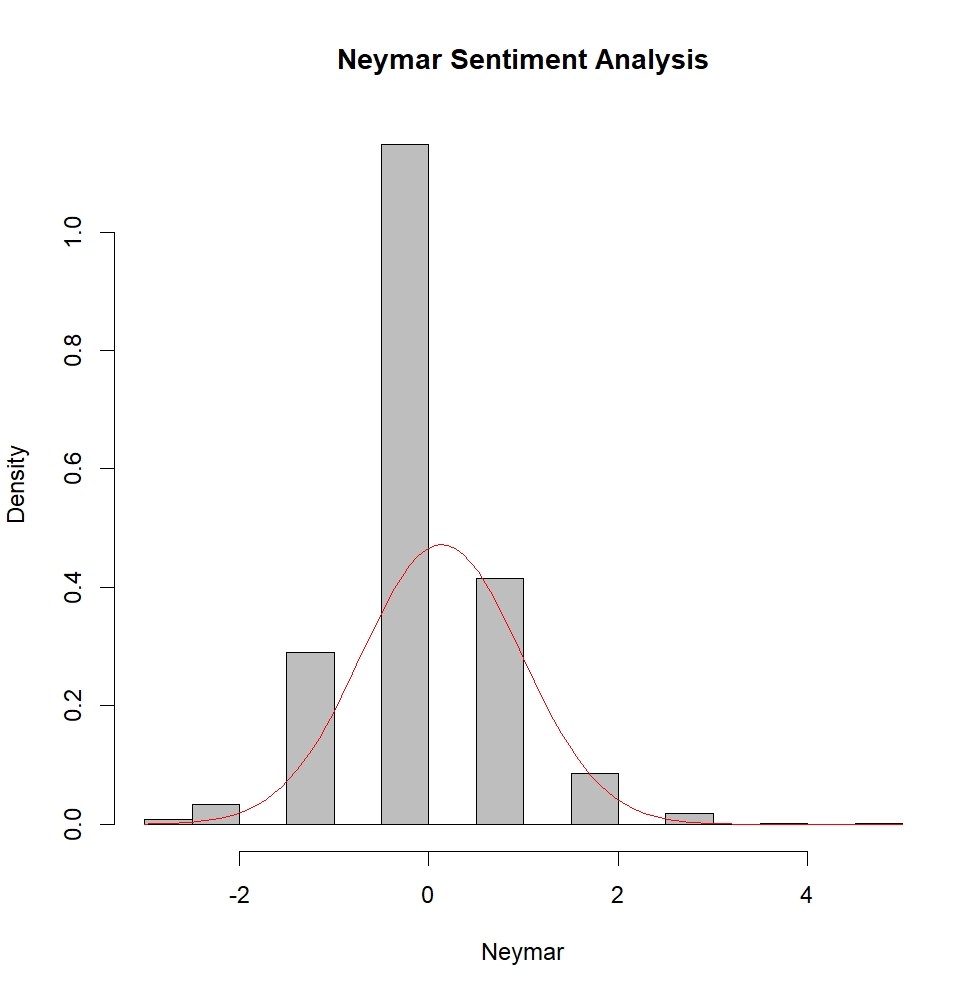}}

\end{tabular}
\caption{Sentiment Score bar plots with normal distribution curves for all brands of interest}
\end{figure}

\FloatBarrier
\begin{table}[H]
\caption{{Pepsi and related celebrities Mean Sentiment Score} }
\label{tw-9dff510fd0a5}
\def\arraystretch{1}
\ignorespaces 
\centering 
\begin{tabulary}{\linewidth}{p{\dimexpr.25\linewidth-2\tabcolsep}p{\dimexpr.25\linewidth-2\tabcolsep}p{\dimexpr.2431\linewidth-2\tabcolsep}p{\dimexpr.2569\linewidth-2\tabcolsep}}
\tbltoprule Organization(brand) & MeanSentimentScore \\
\tblmidrule 
Pepsi	&0.3033446\\
Messi	&0.3793784\\
Beyoncé	&0.1802023\\
\tblbottomrule 
\end{tabulary}\par 
\end{table}
According to Table 4, Pepsi and Messi have induced a substantially more positive feeling than Beyoncé in the minds of customers.

\FloatBarrier
\begin{table}[H]
\caption{{Gillette and related celebrities Mean Sentiment Score} }
\label{tw-9dff510fd0a5}
\def\arraystretch{1}
\ignorespaces 
\centering 
\begin{tabulary}{\linewidth}{p{\dimexpr.25\linewidth-2\tabcolsep}p{\dimexpr.25\linewidth-2\tabcolsep}p{\dimexpr.2431\linewidth-2\tabcolsep}p{\dimexpr.2569\linewidth-2\tabcolsep}}
\tbltoprule Organization(brand) & MeanSentimentScore \\
\tblmidrule 
Gillette	&0.1181034\\
Roger Federer&	0.2283365\\
Jennifer Lopez&	0.2611397\\
\tblbottomrule 
\end{tabulary}\par 
\end{table}
As can be observed in Table 5, as the brand, Gillette had a less positive sentiment than Roger Federer and Jennifer Lopez.

\FloatBarrier
\begin{table}[H]
\caption{{Coca Cola and related celebrities Mean Sentiment Score} }
\label{tw-9dff510fd0a5}
\def\arraystretch{1}
\ignorespaces 
\centering 
\begin{tabulary}{\linewidth}{p{\dimexpr.25\linewidth-2\tabcolsep}p{\dimexpr.25\linewidth-2\tabcolsep}p{\dimexpr.2431\linewidth-2\tabcolsep}p{\dimexpr.2569\linewidth-2\tabcolsep}}
\tbltoprule Organization(brand) & MeanSentimentScore \\
\tblmidrule 
Coca Cola	&0.2064387\\
Taylor Swift	&0.8189683\\
Selena Gomez	&0.1970904\\
\tblbottomrule 
\end{tabulary}\par 
\end{table}
According to Table 6, Taylor Swift had a more positive influence in the minds of people as compared to Selena Gomez and Coca Cola, which had closer sentiment scores.
\FloatBarrier
\begin{table}[H]
\caption{{Nike and related celebrities Mean Sentiment Score} }
\label{tw-9dff510fd0a5}
\def\arraystretch{1}
\ignorespaces 
\centering 
\begin{tabulary}{\linewidth}{p{\dimexpr.25\linewidth-2\tabcolsep}p{\dimexpr.25\linewidth-2\tabcolsep}p{\dimexpr.2431\linewidth-2\tabcolsep}p{\dimexpr.2569\linewidth-2\tabcolsep}}
\tbltoprule Organization(brand) & MeanSentimentScore \\
\tblmidrule 
Nike&	0.3472222\\
Neymar	&0.1358516\\
Cristiano Ronaldo	&0.4756011\\
\tblbottomrule 
\end{tabulary}\par 
\end{table}
As can be seen in Table 7, Cristiano Ronaldo had the largest positive sentiment in the minds of people, followed by Nike and Neymar.

\subsection{Results of the second method}

\FloatBarrier
\begin{figure}[H]
\begin{tabular}{cccc}
\subfloat[Burger King]{\includegraphics[width = 2in]{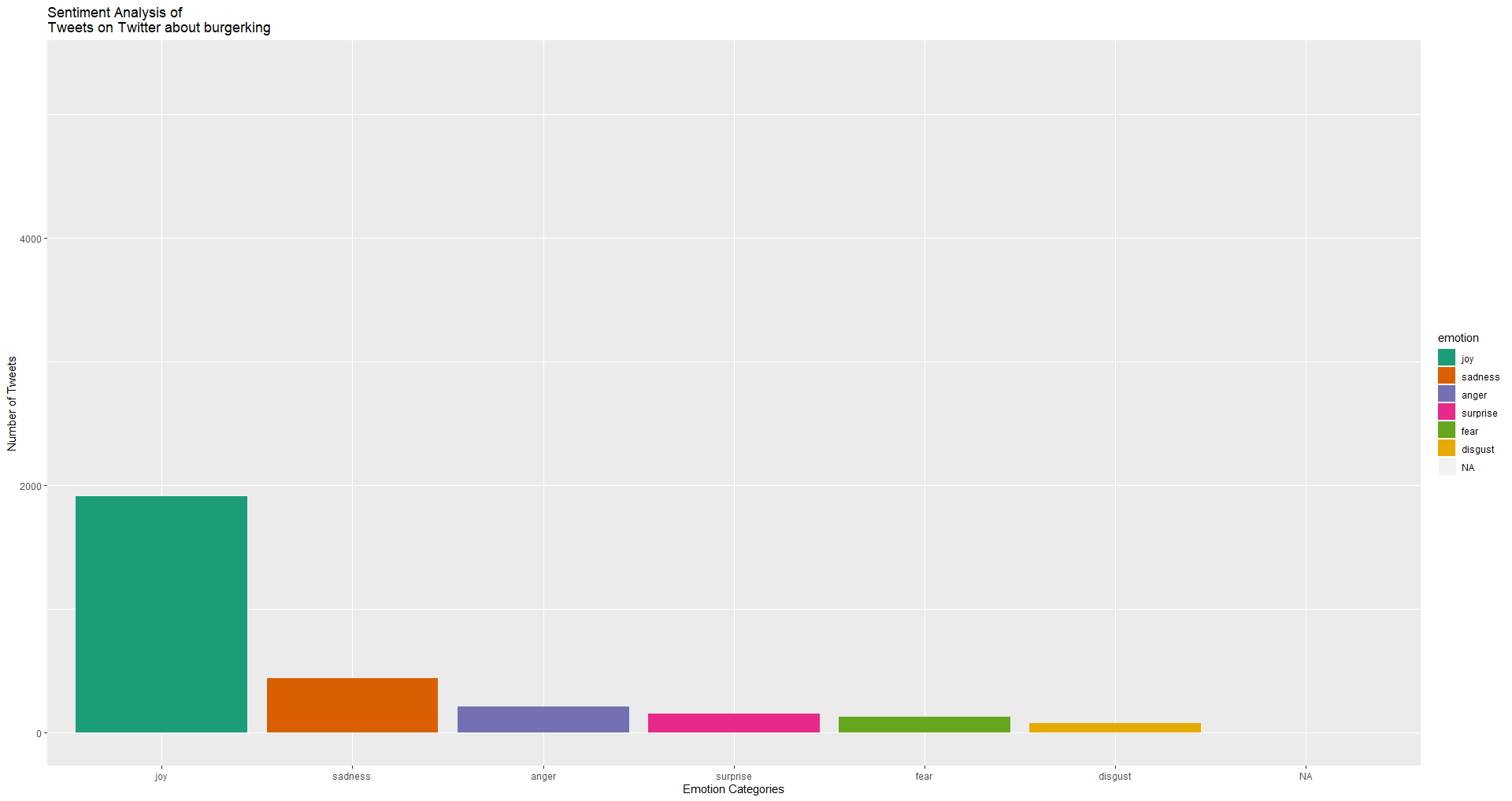}} &
\subfloat[Snoopdogg]{\includegraphics[width = 2in]{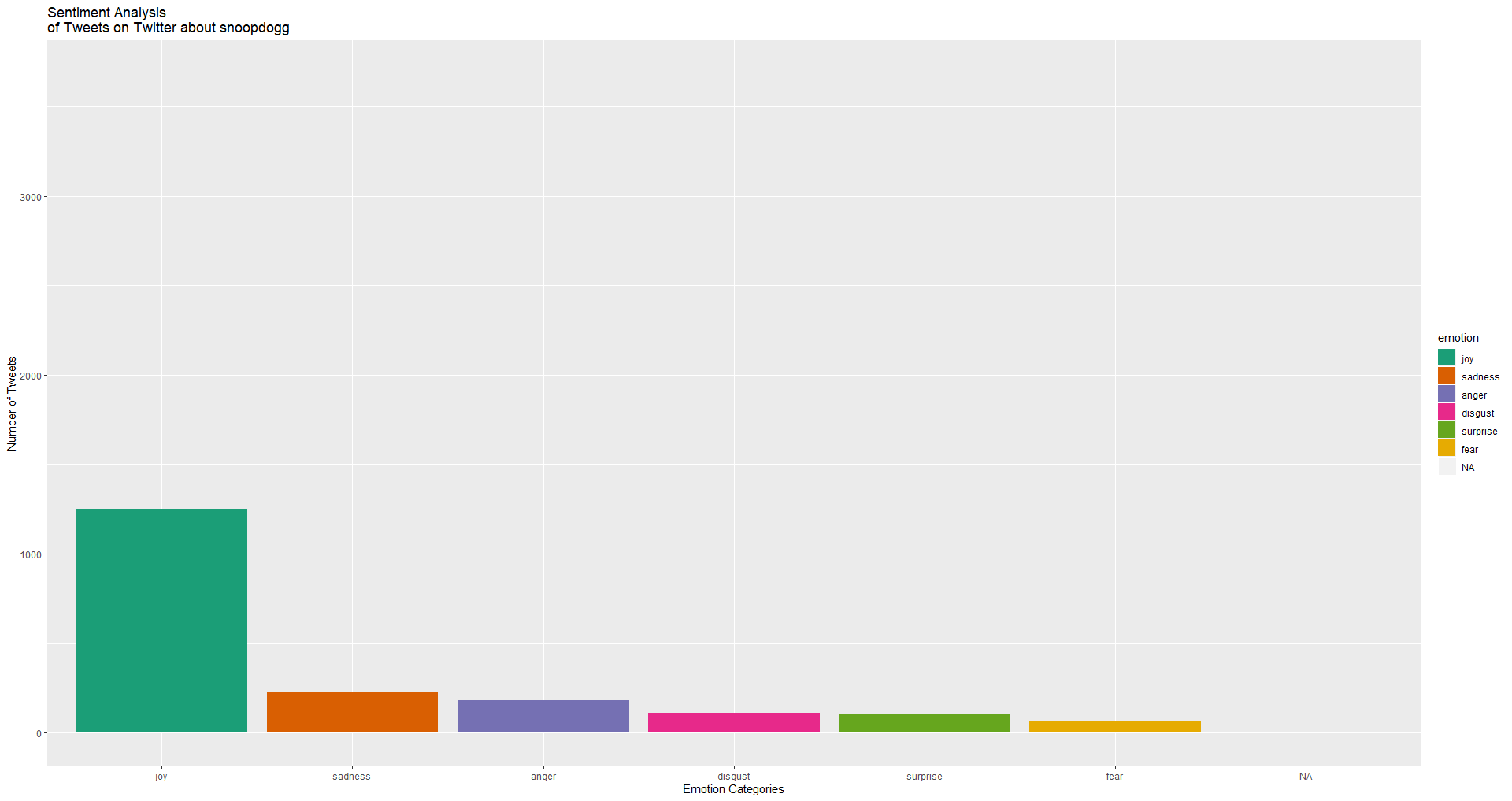}}&
\subfloat[Connor McGregor]{\includegraphics[width = 2in]{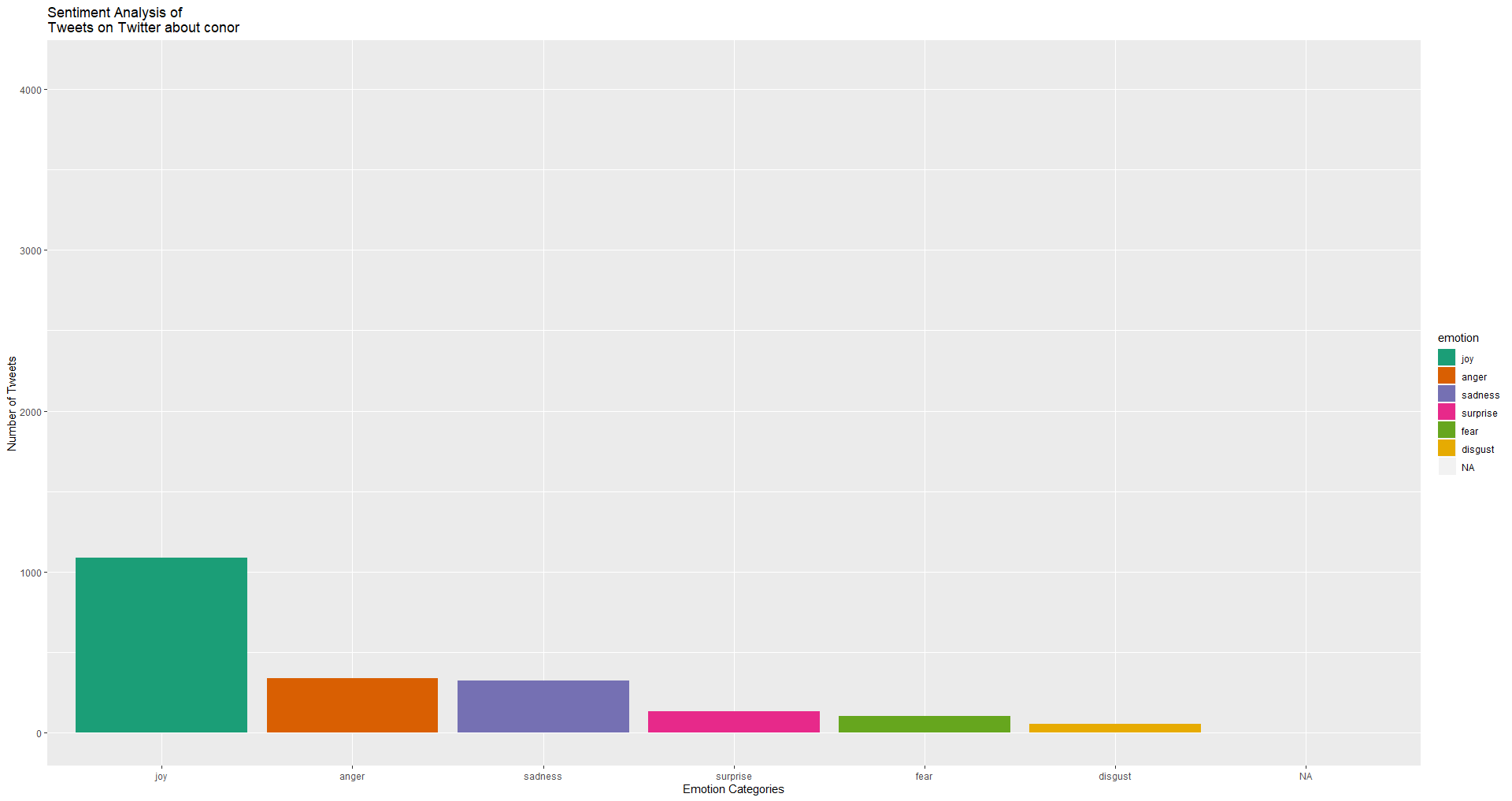}}\\
\subfloat[Coca Cola]{\includegraphics[width = 2in]{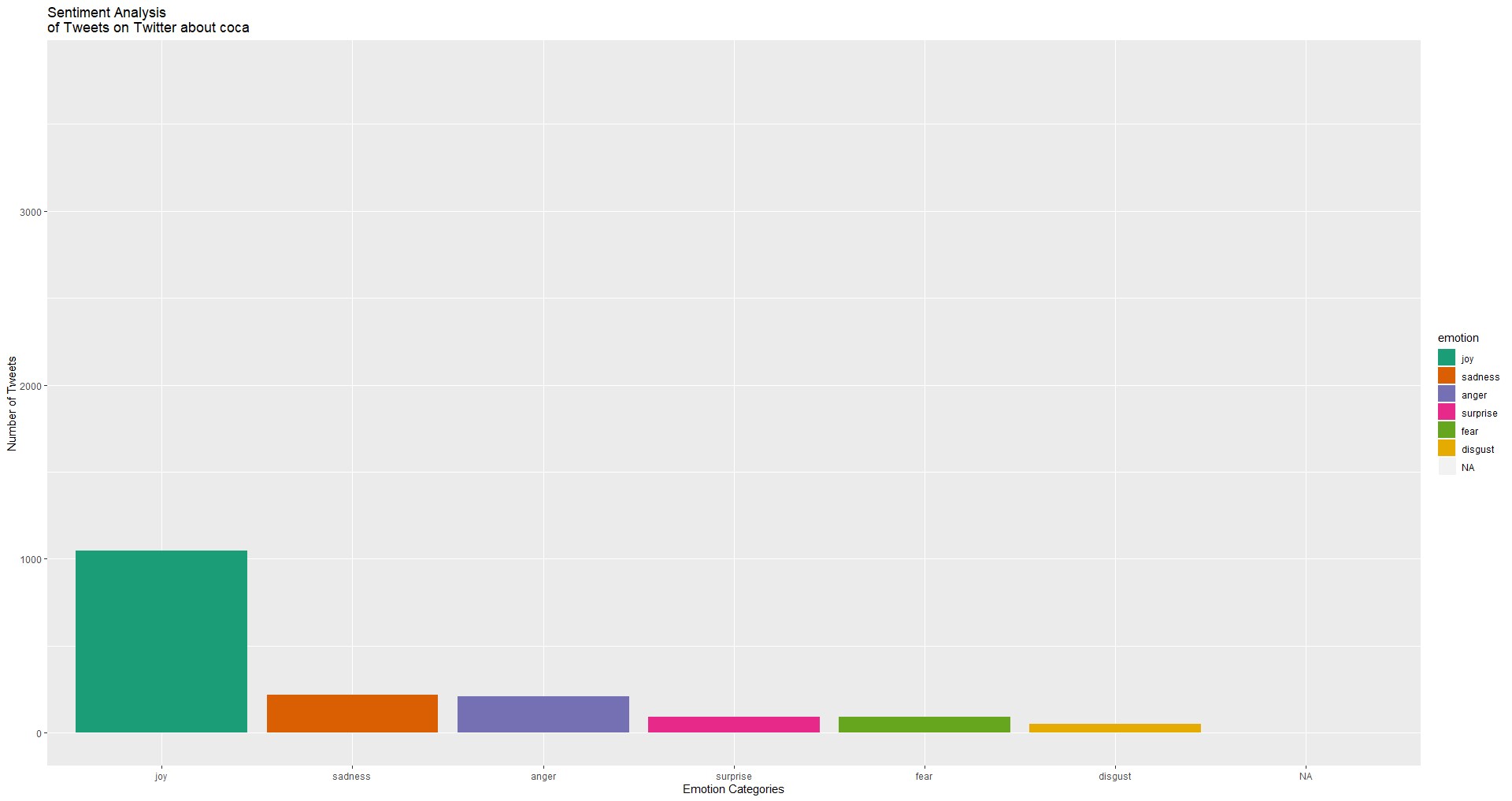}}&
\subfloat[Taylor Swift]{\includegraphics[width = 2in]{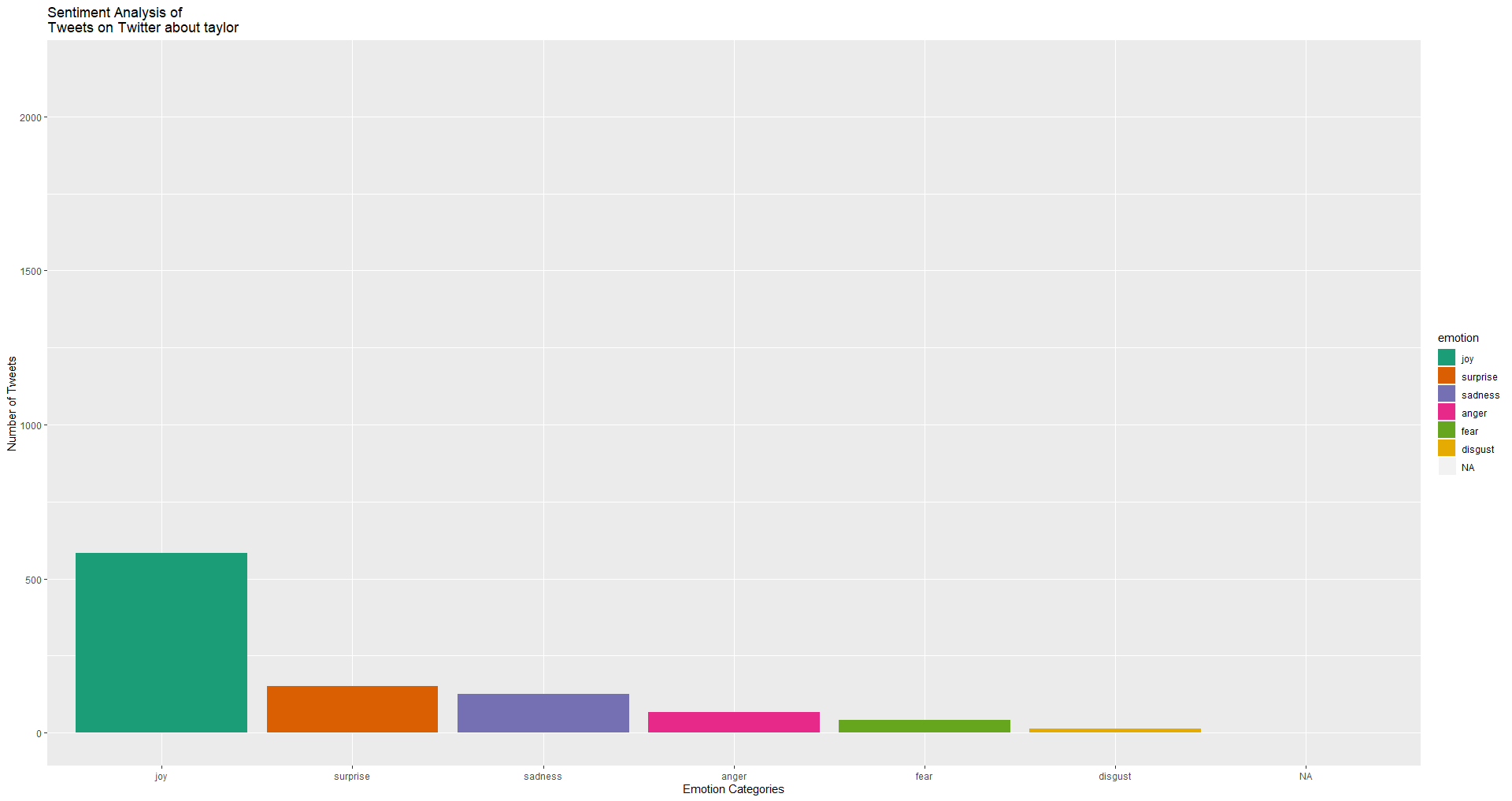}} &
\subfloat[Selena Gomez]{\includegraphics[width = 2in]{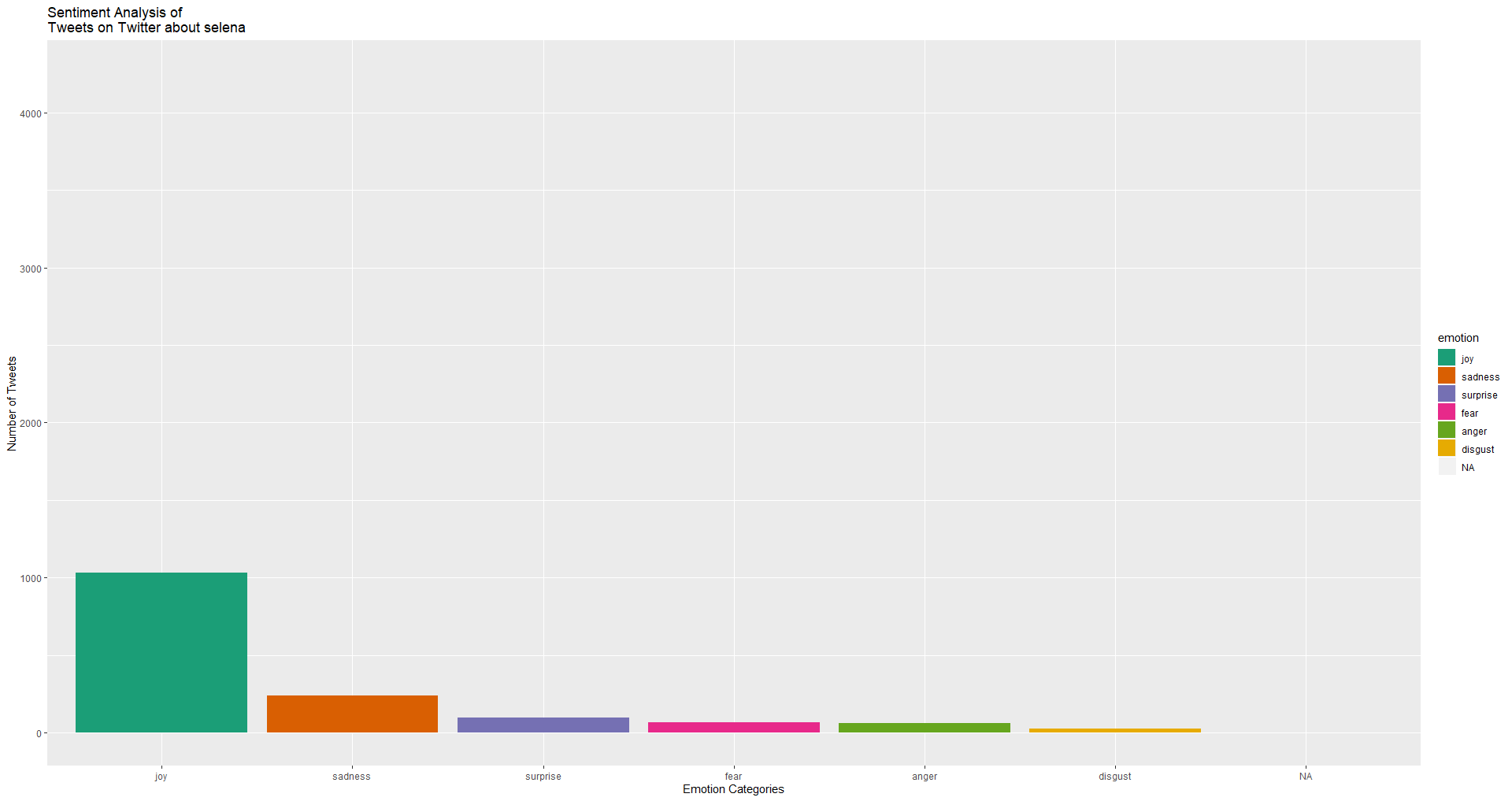}}\\
\subfloat[Pepsi]{\includegraphics[width = 2in]{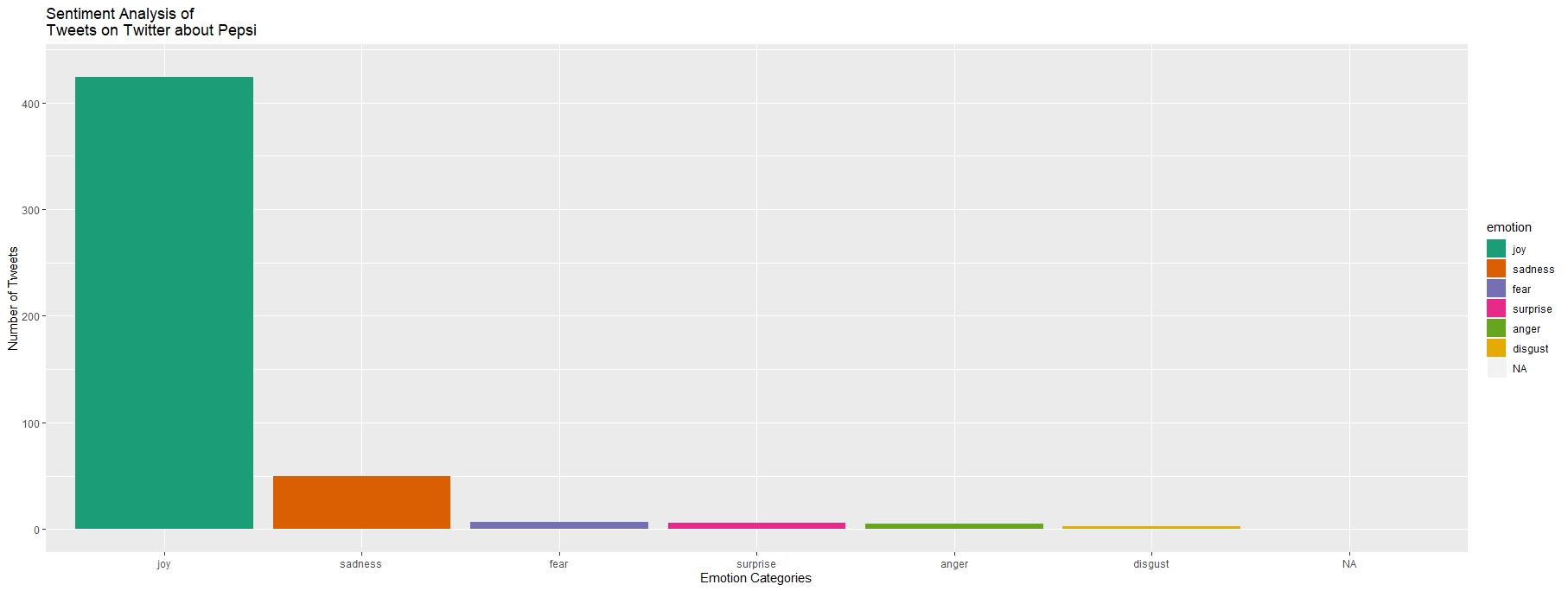}}&
\subfloat[Messi]{\includegraphics[width = 2in]{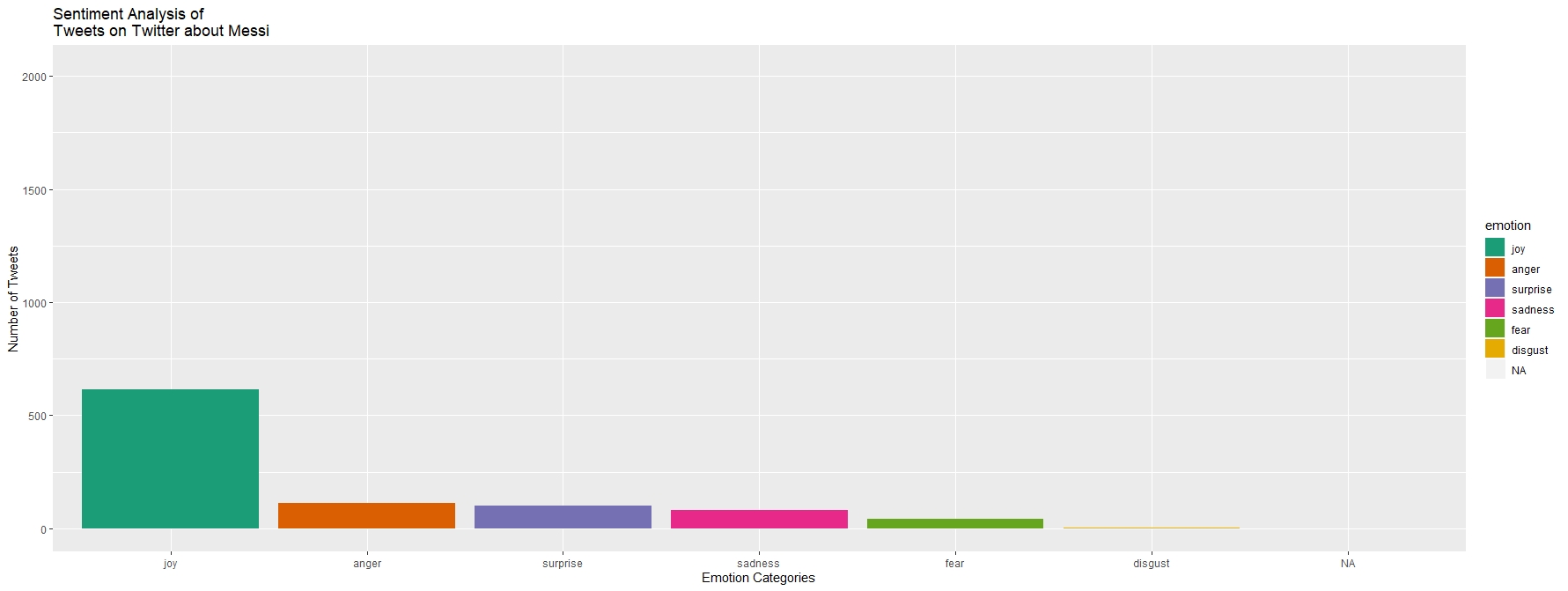}}&
\subfloat[Beyoncé]{\includegraphics[width = 2in]{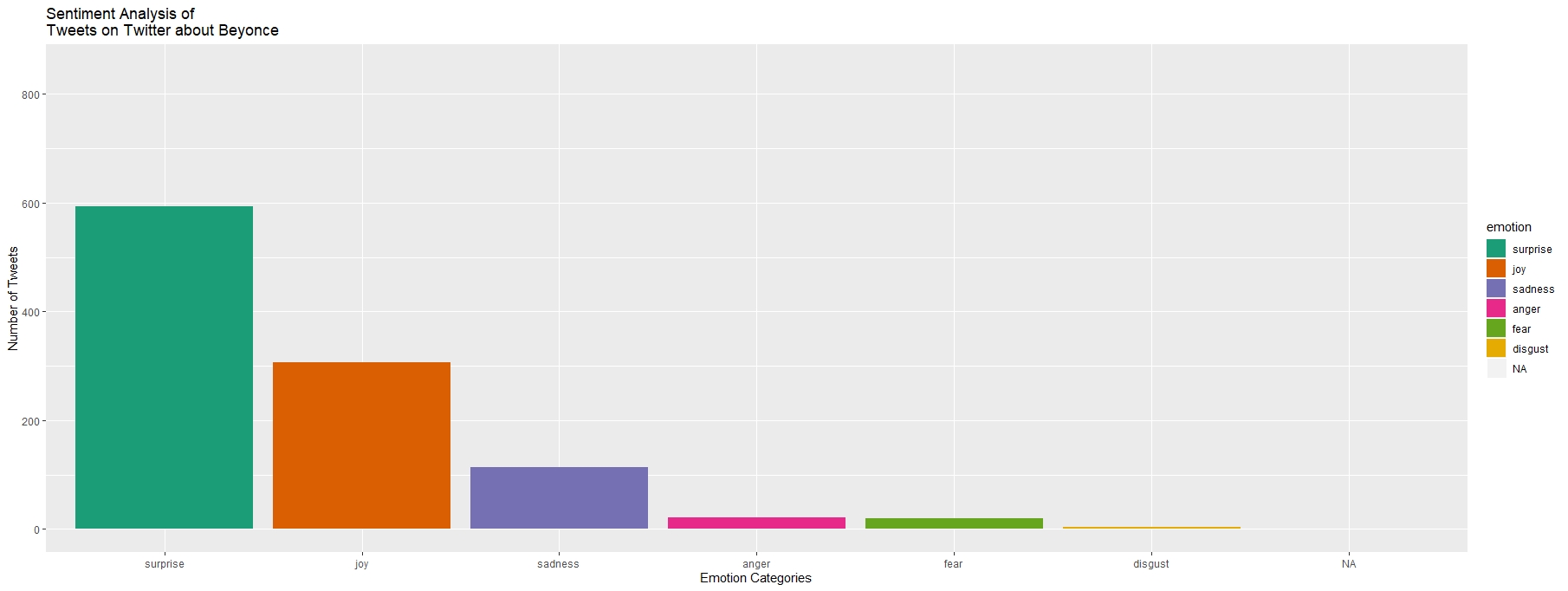}}\\
\subfloat[Gillette]{\includegraphics[width = 2in]{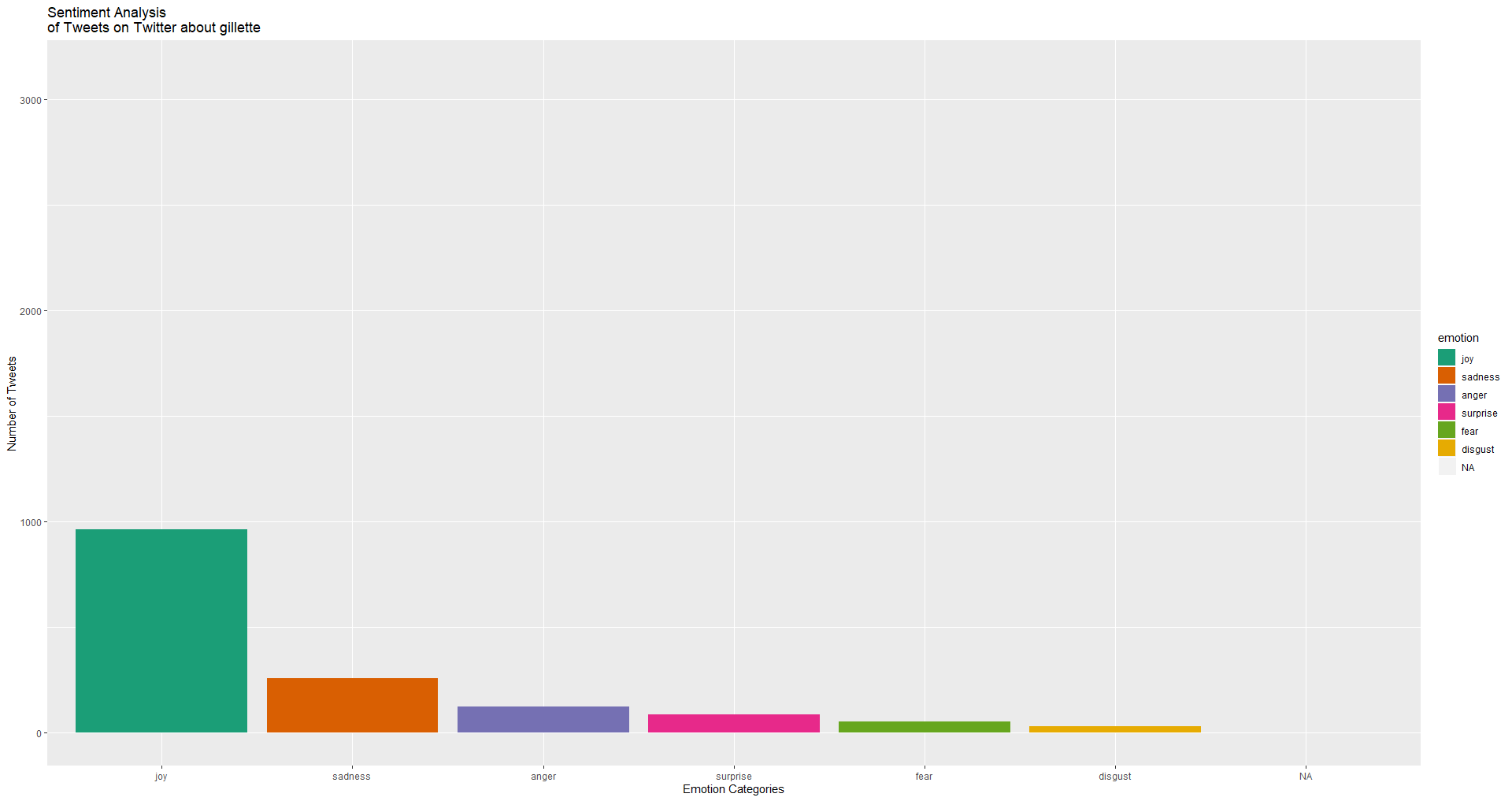}}&
\subfloat[Roger Federer]{\includegraphics[width = 2in]{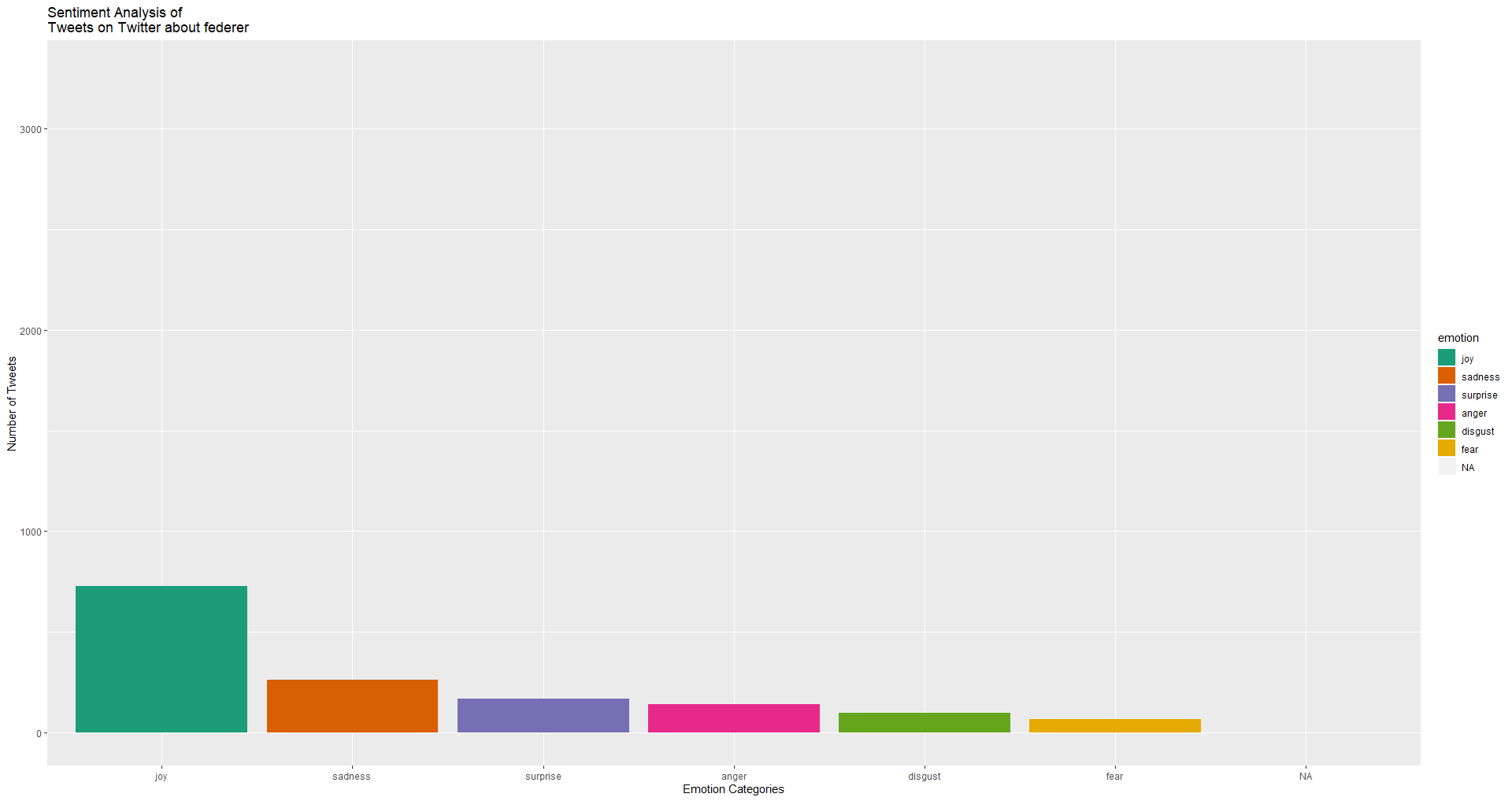}}&
\subfloat[Jennifer Lopez]{\includegraphics[width = 2in]{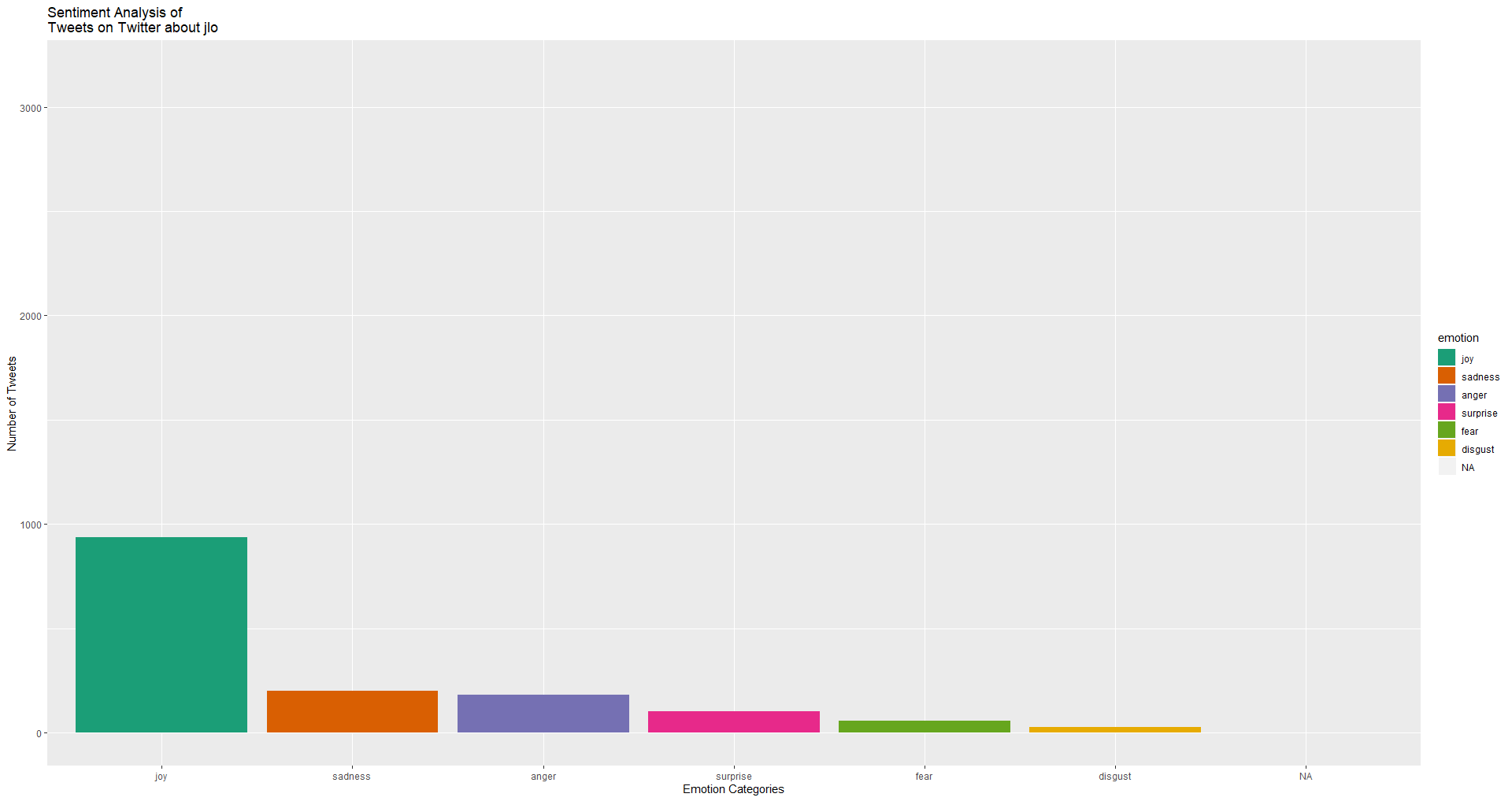}}\\
\subfloat[Nike]{\includegraphics[width = 2in]{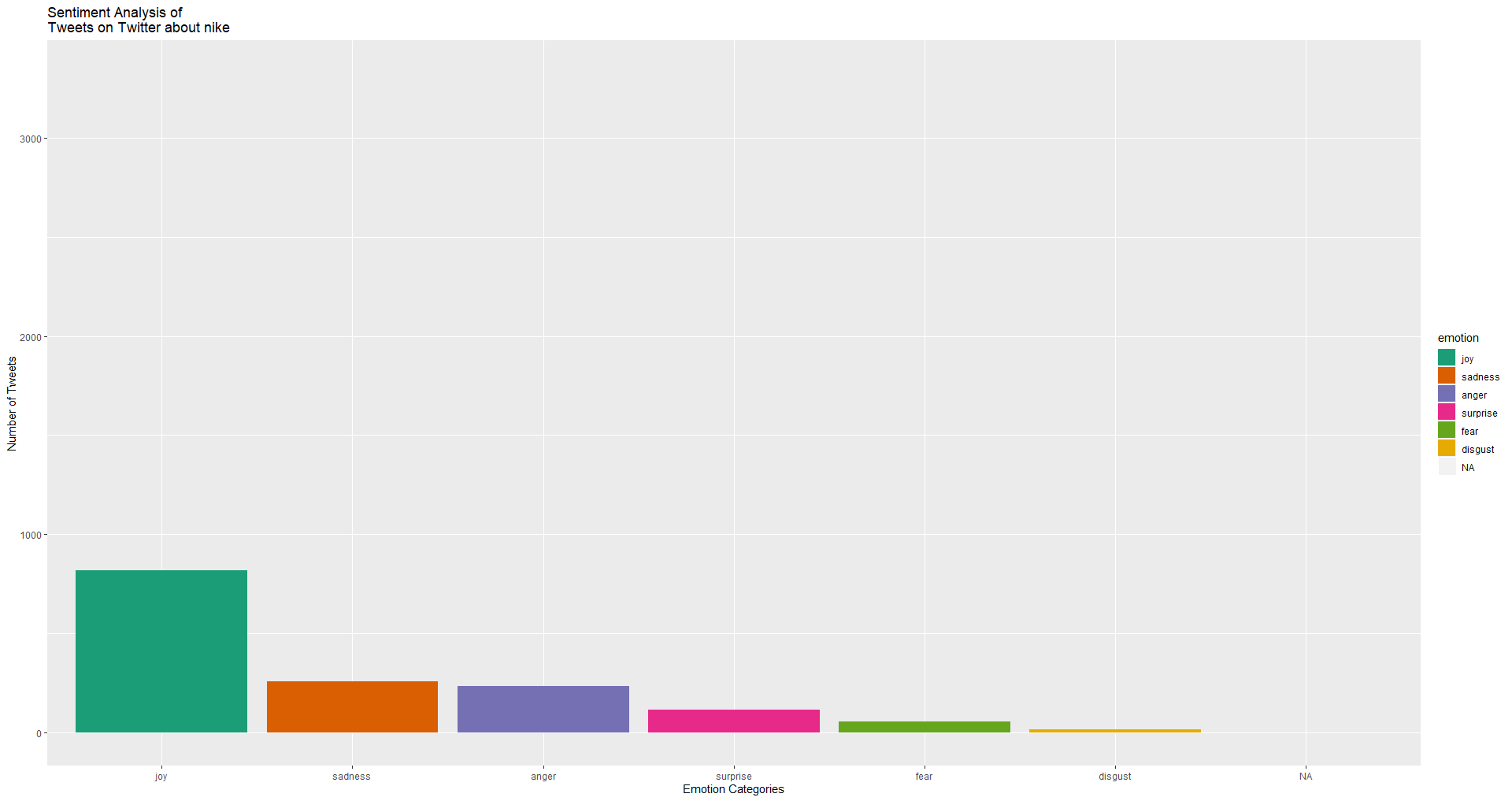}}&
\subfloat[Cristiano Ronaldo]{\includegraphics[width = 2in]{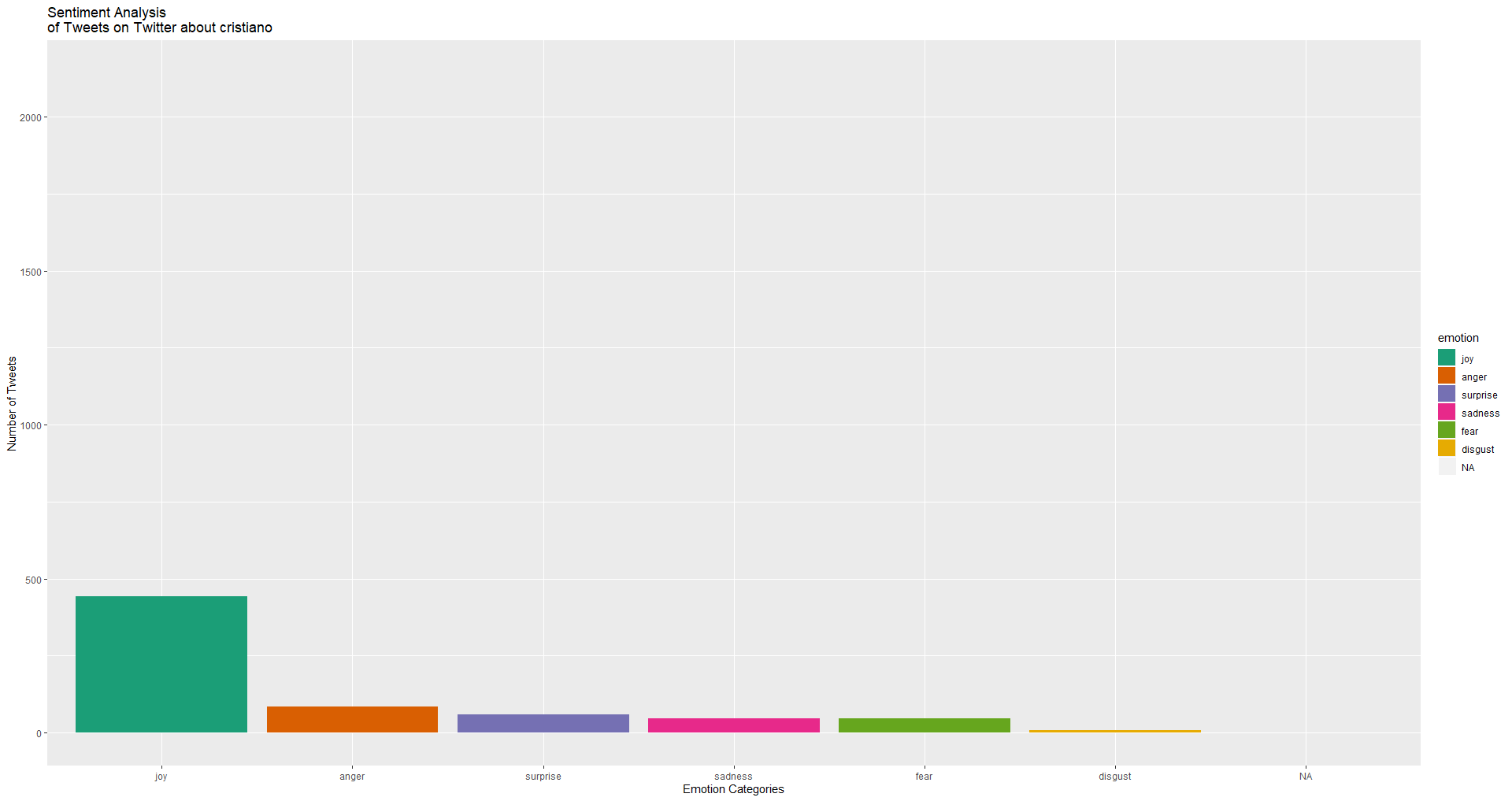}}&
\subfloat[Neymar]{\includegraphics[width = 2in]{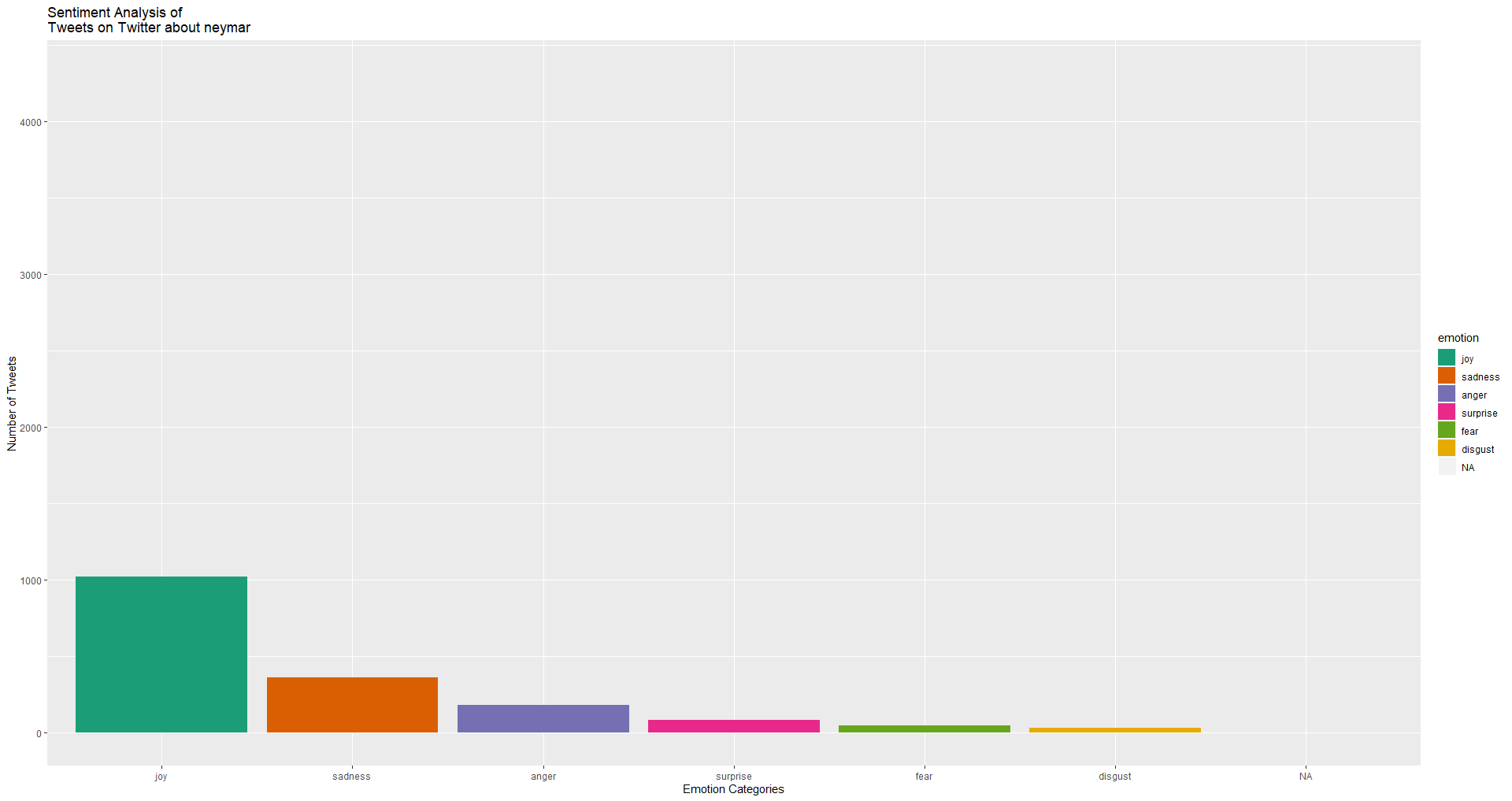}}

\end{tabular}
\caption{Sentiment Analysis plots according to emotions for all brands of interest}
\end{figure}

%
As mentioned in figure 5, for example in sentiment analysis of Pepsi most significant emotion is joy and then sadness and the rest are not that much noticeable.

\FloatBarrier
\begin{figure}[H]
\begin{tabular}{cccc}
\subfloat[Burger King]{\includegraphics[width = 2in]{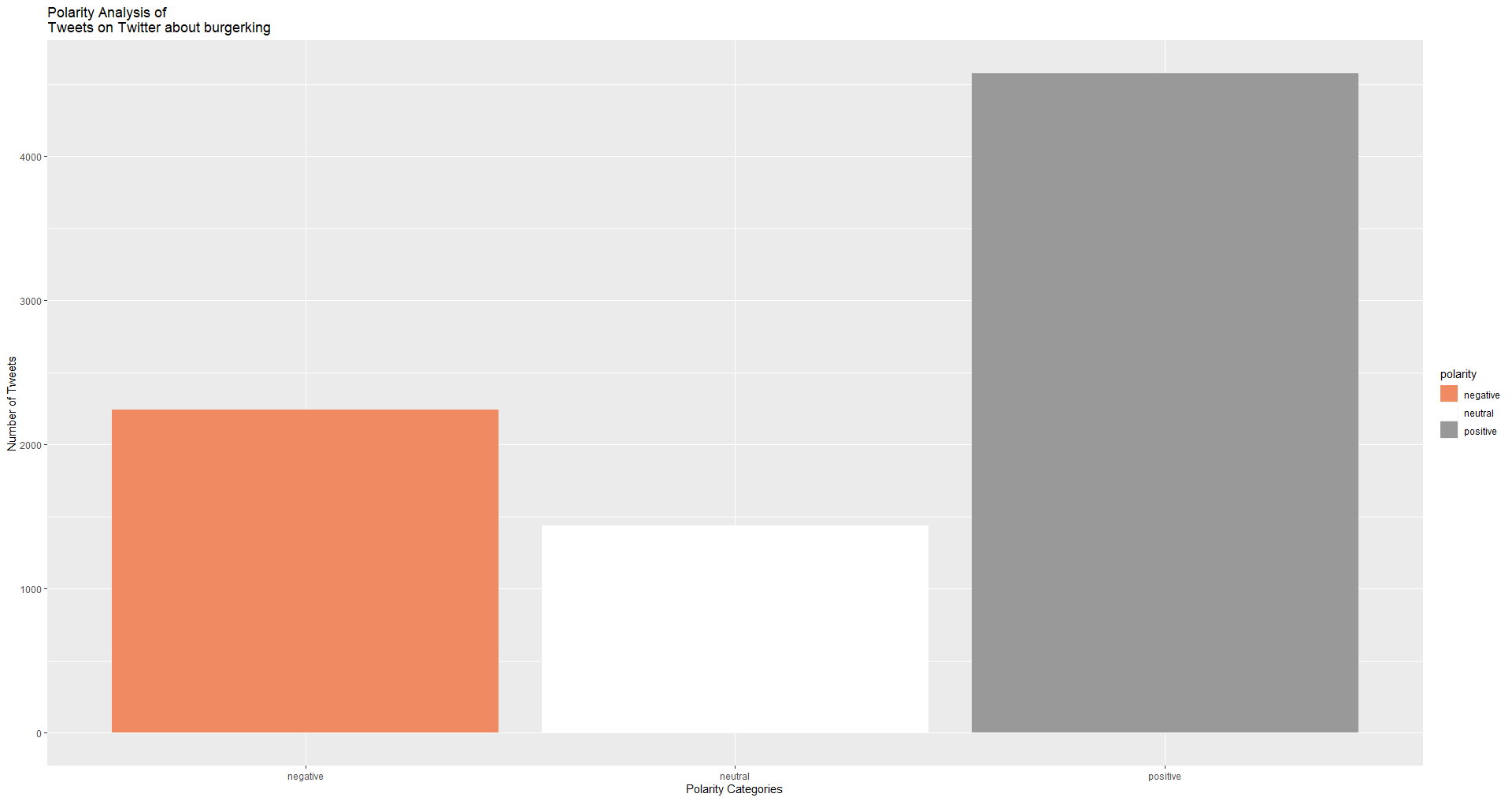}} &
\subfloat[Snoopdogg]{\includegraphics[width = 2in]{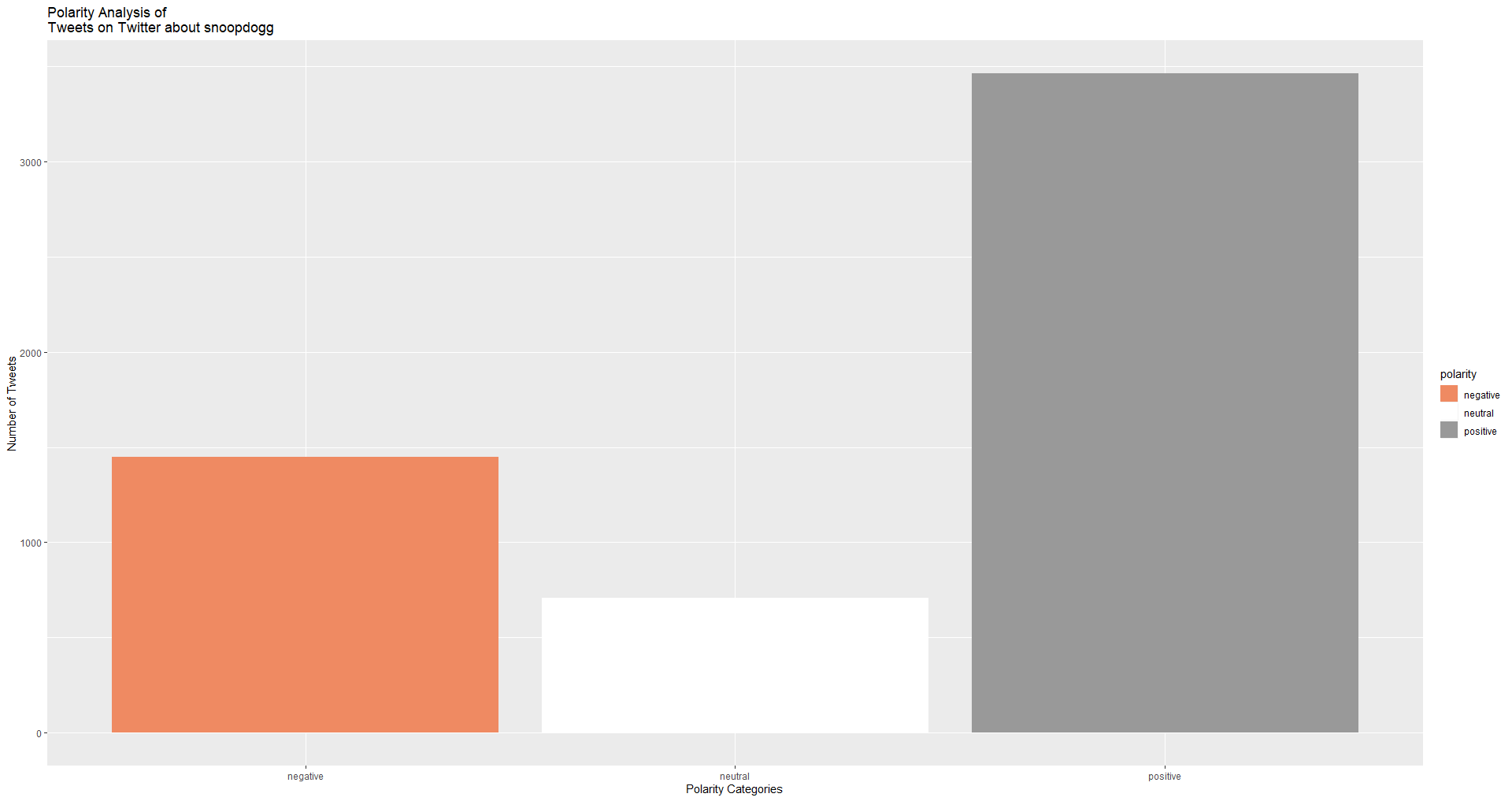}}&
\subfloat[Connor McGregor]{\includegraphics[width = 2in]{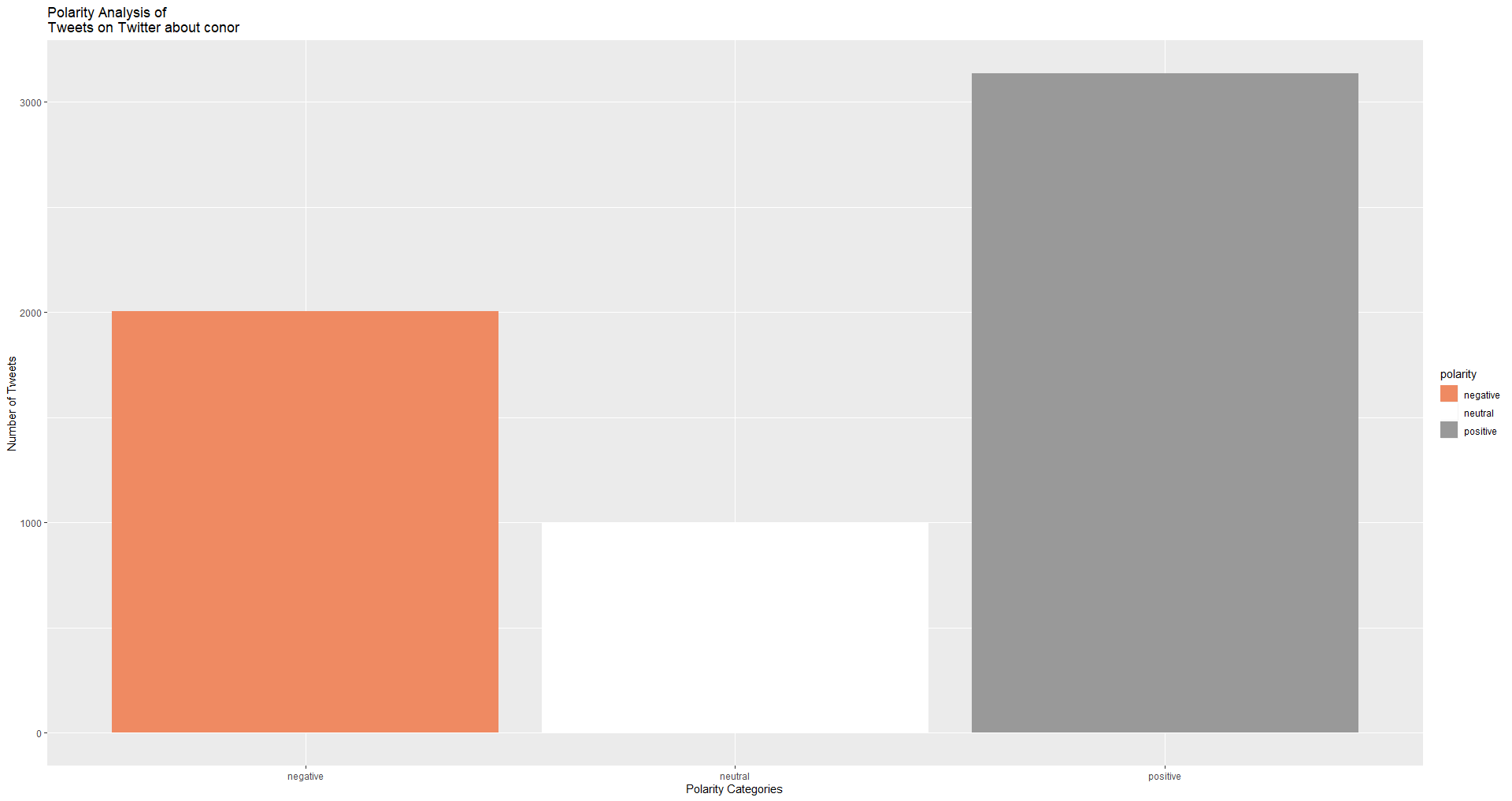}}\\
\subfloat[Coca Cola]{\includegraphics[width = 2in]{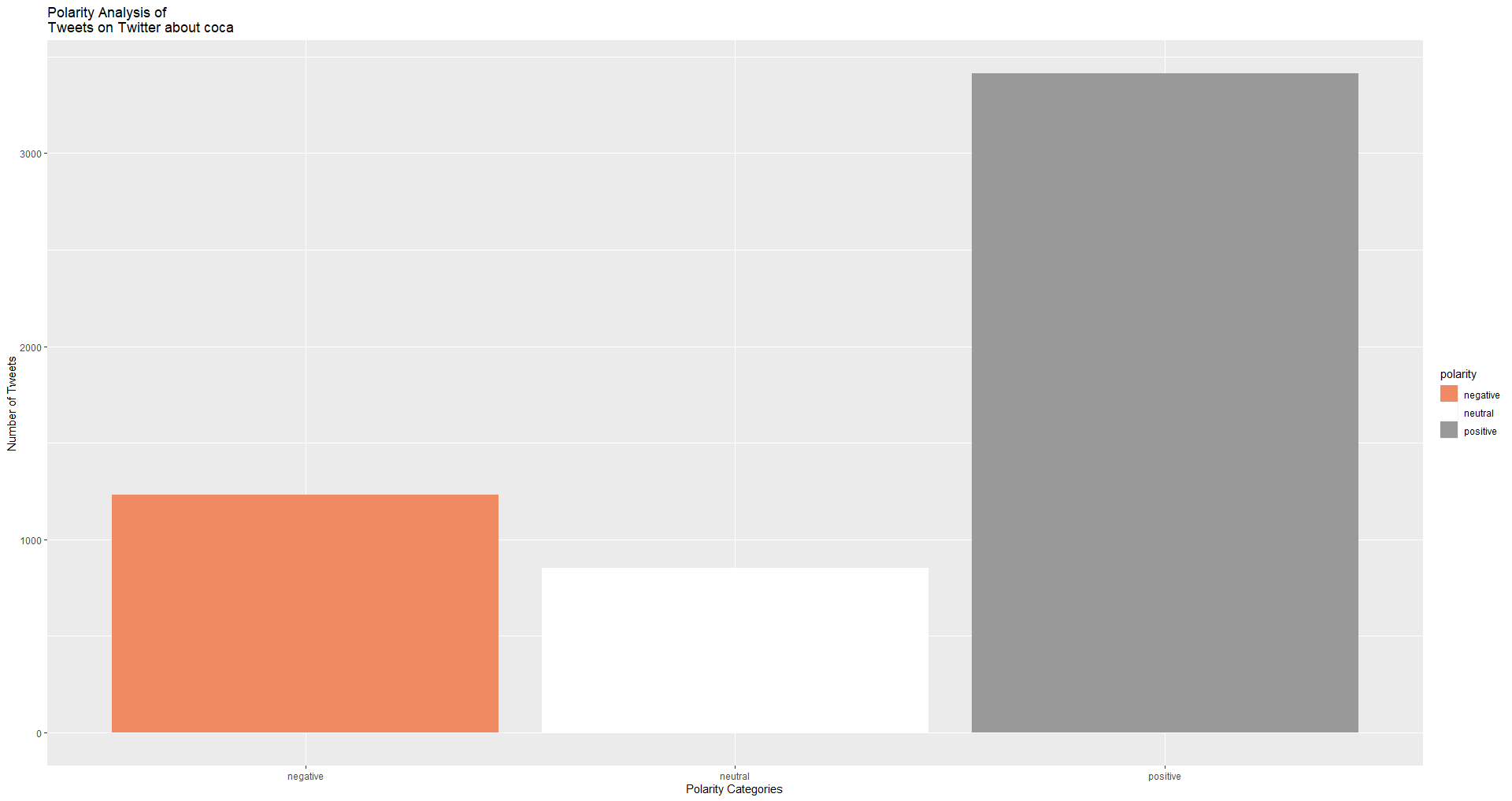}}&
\subfloat[Taylor Swift]{\includegraphics[width = 2in]{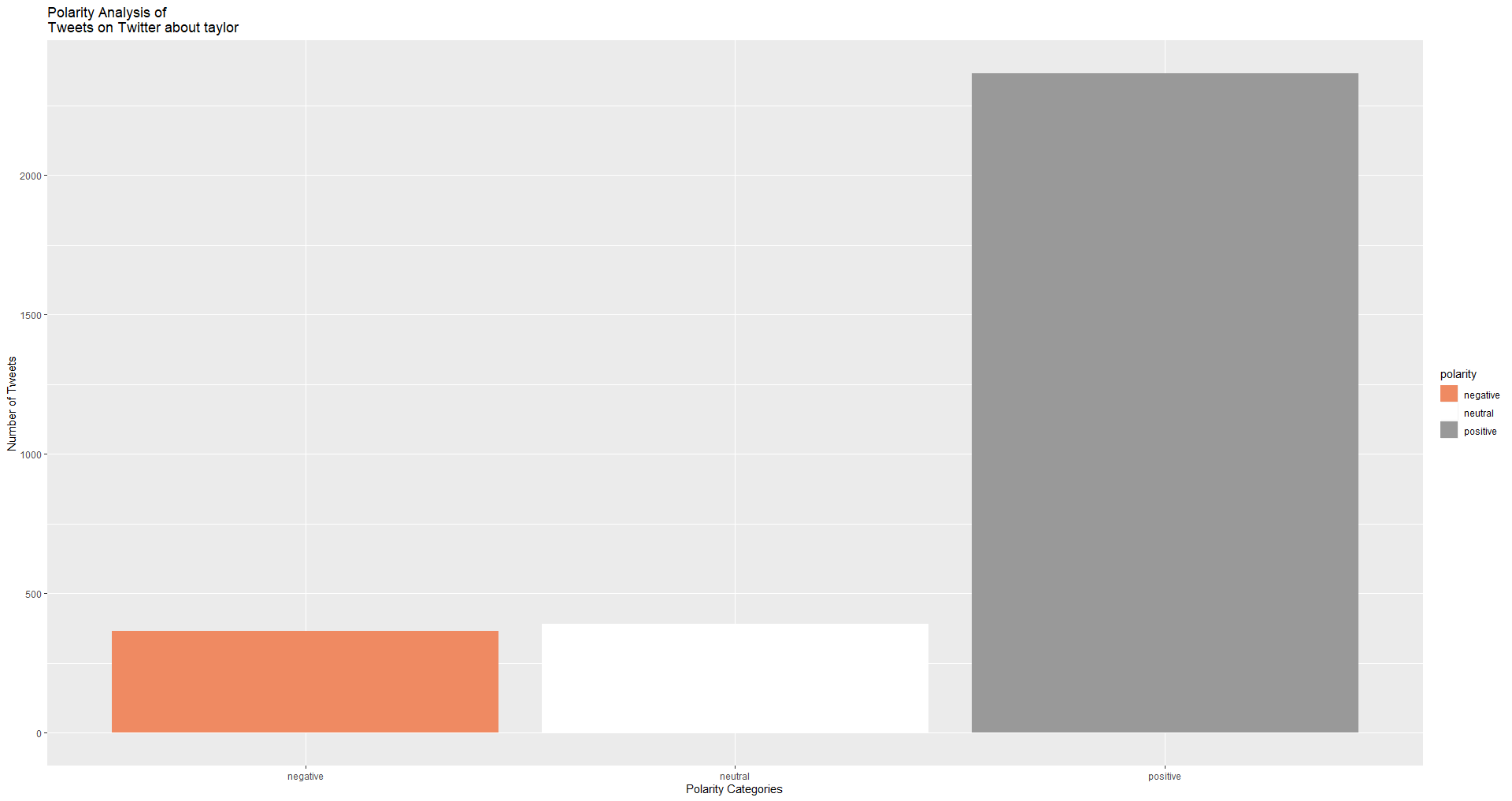}} &
\subfloat[Selena Gomez]{\includegraphics[width = 2in]{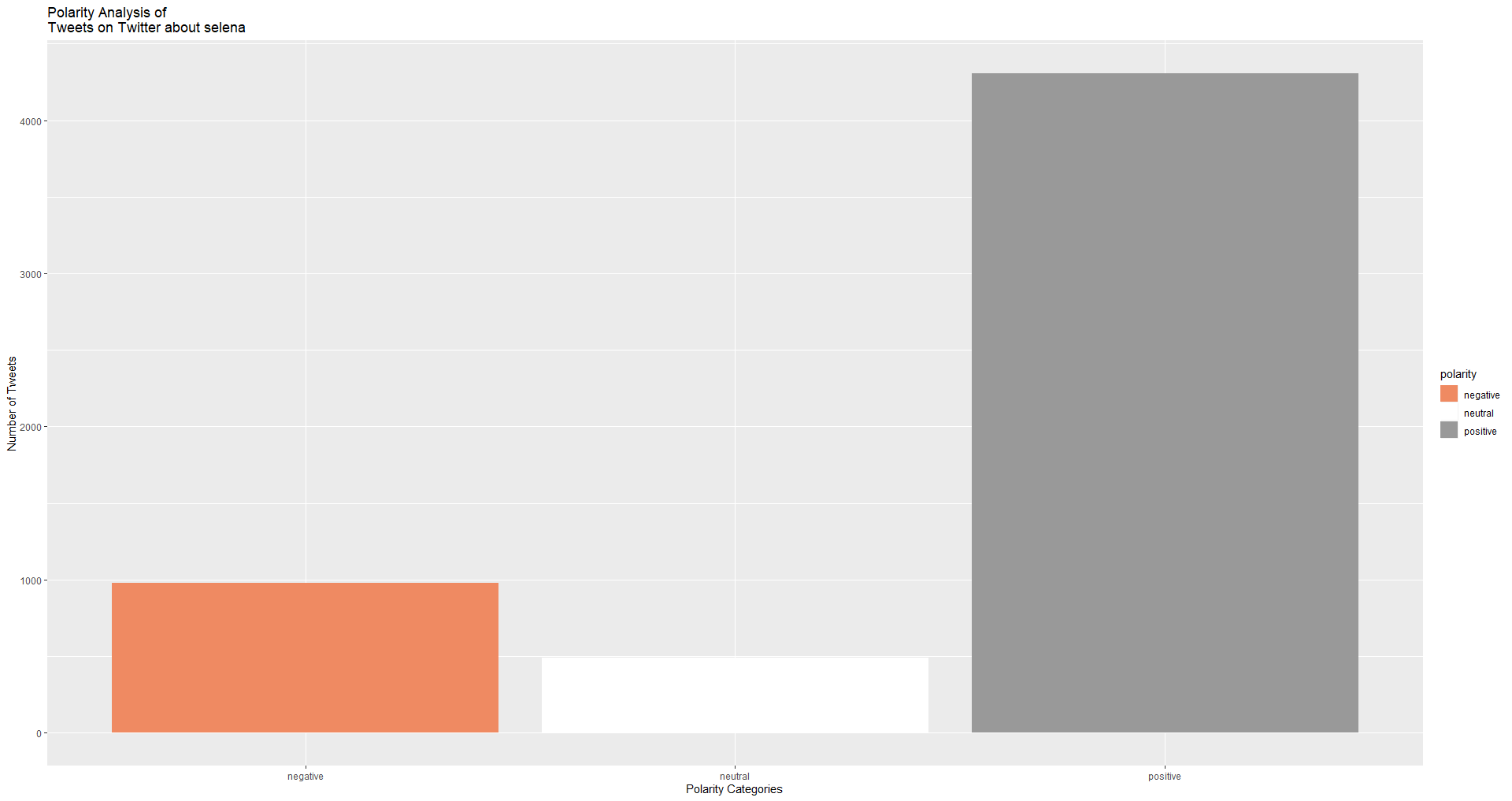}}\\
\subfloat[Pepsi]{\includegraphics[width = 2in]{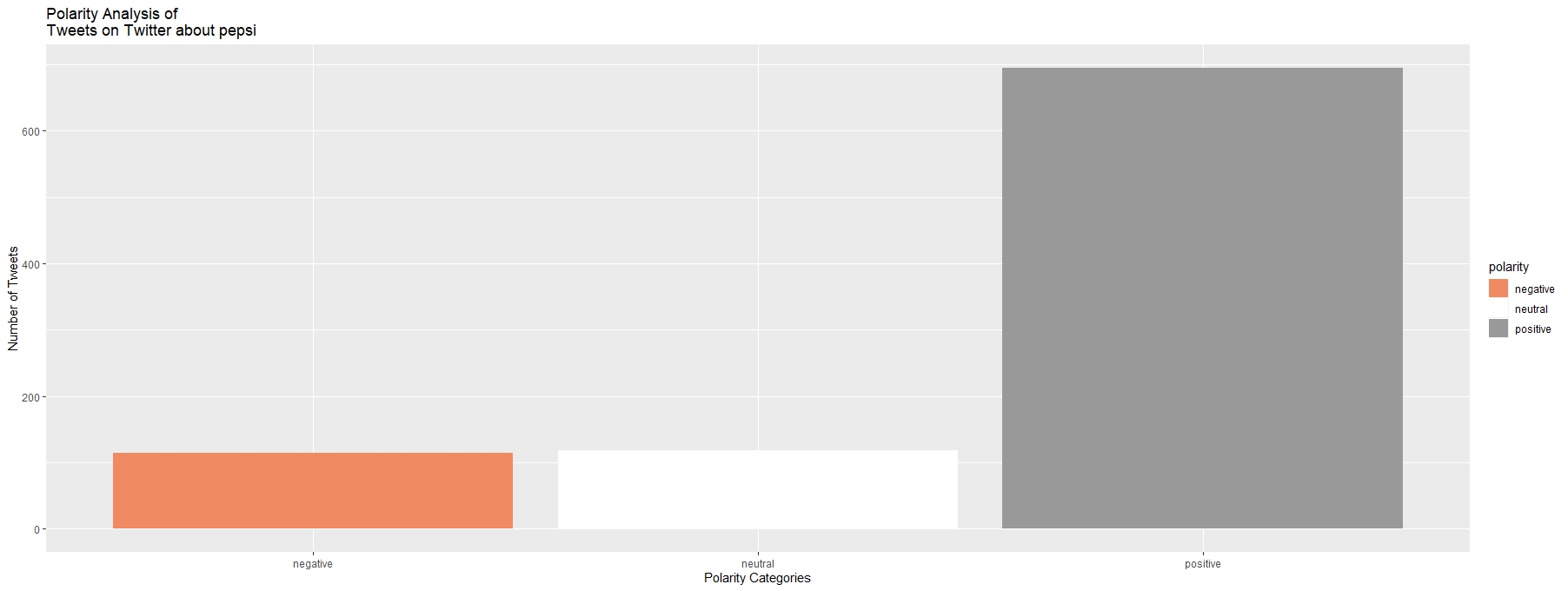}}&
\subfloat[Messi]{\includegraphics[width = 2in]{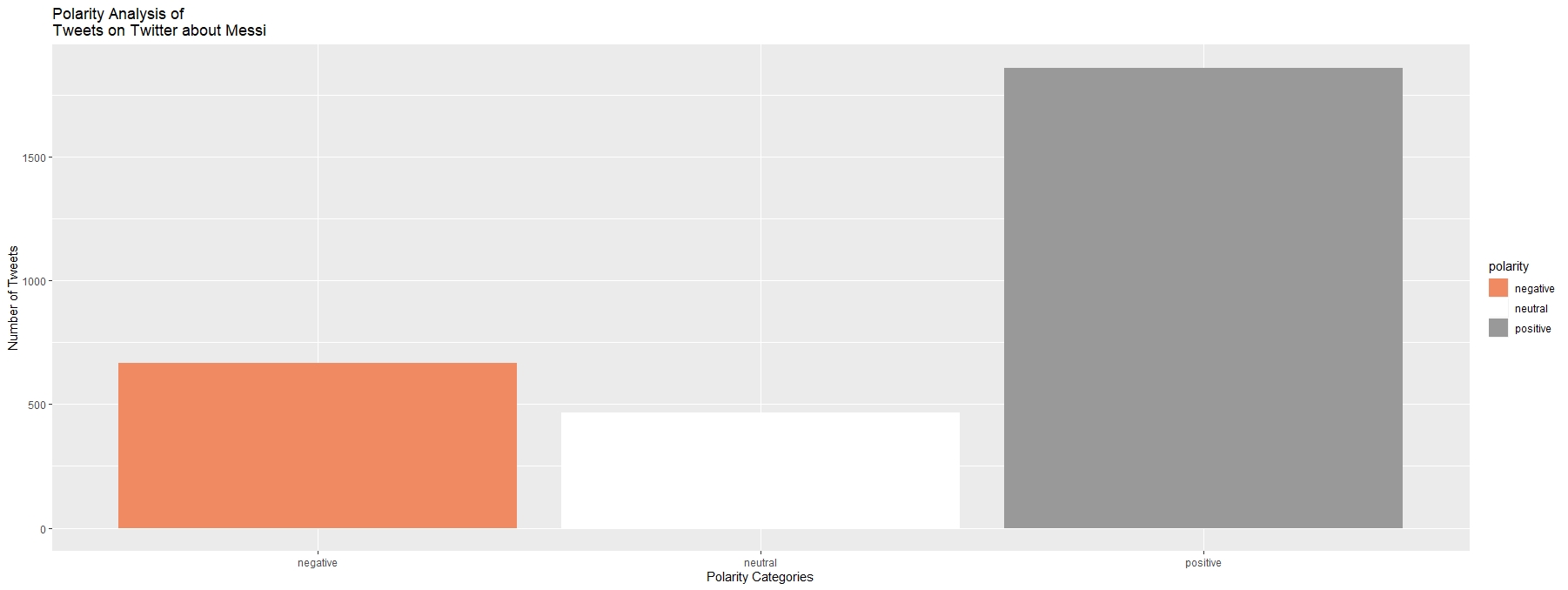}}&
\subfloat[Beyoncé]{\includegraphics[width = 2in]{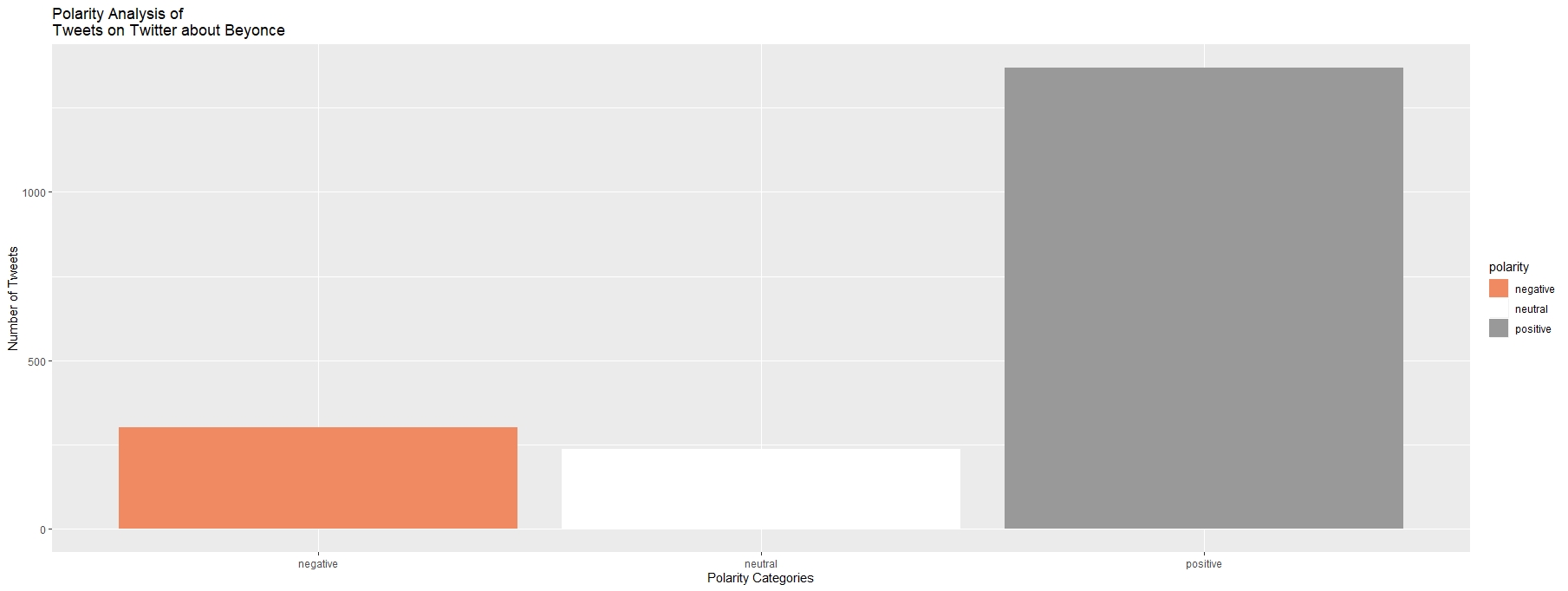}}\\
\subfloat[Gillette]{\includegraphics[width = 2in]{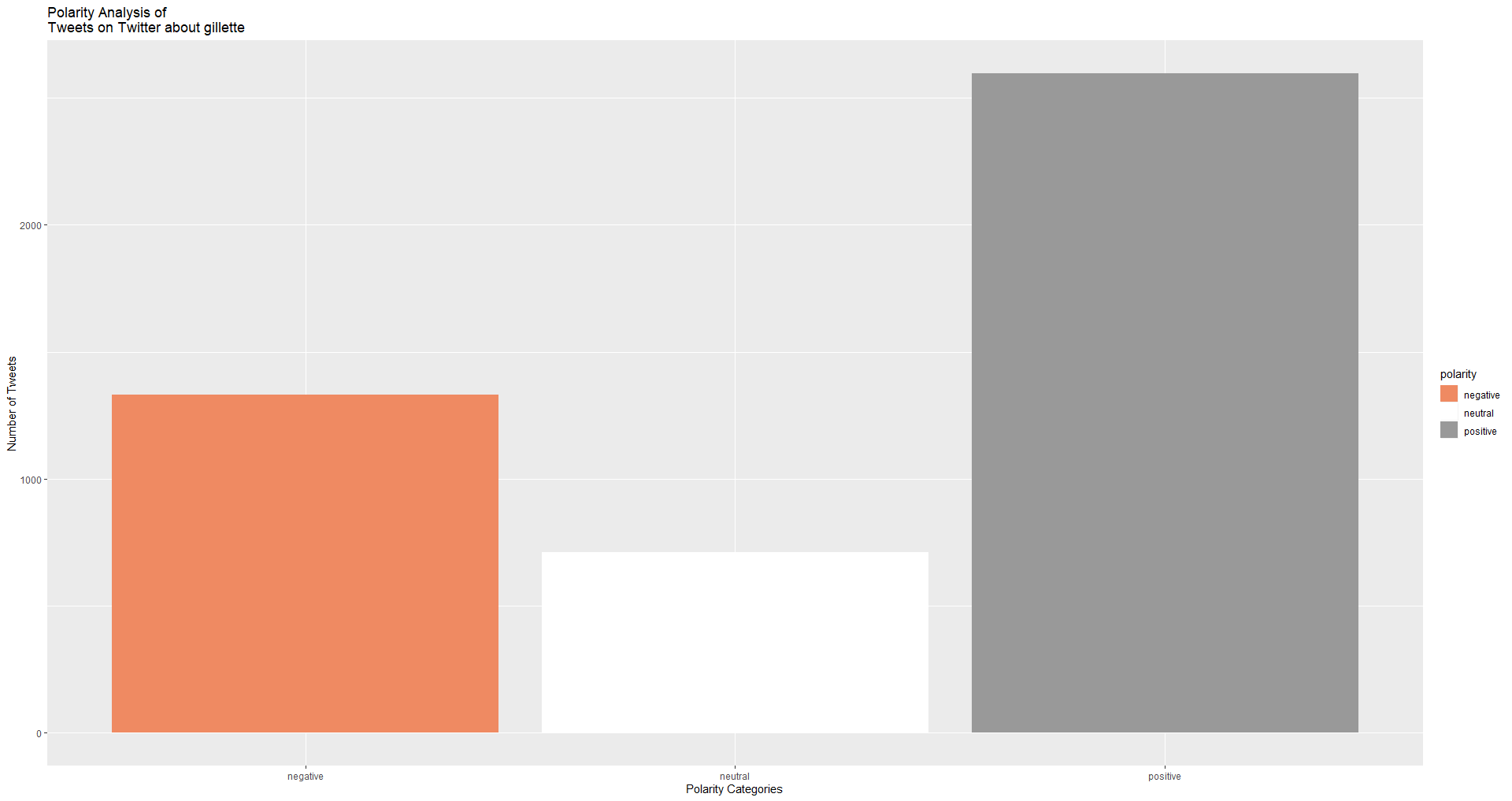}}&
\subfloat[Roger Federer]{\includegraphics[width = 2in]{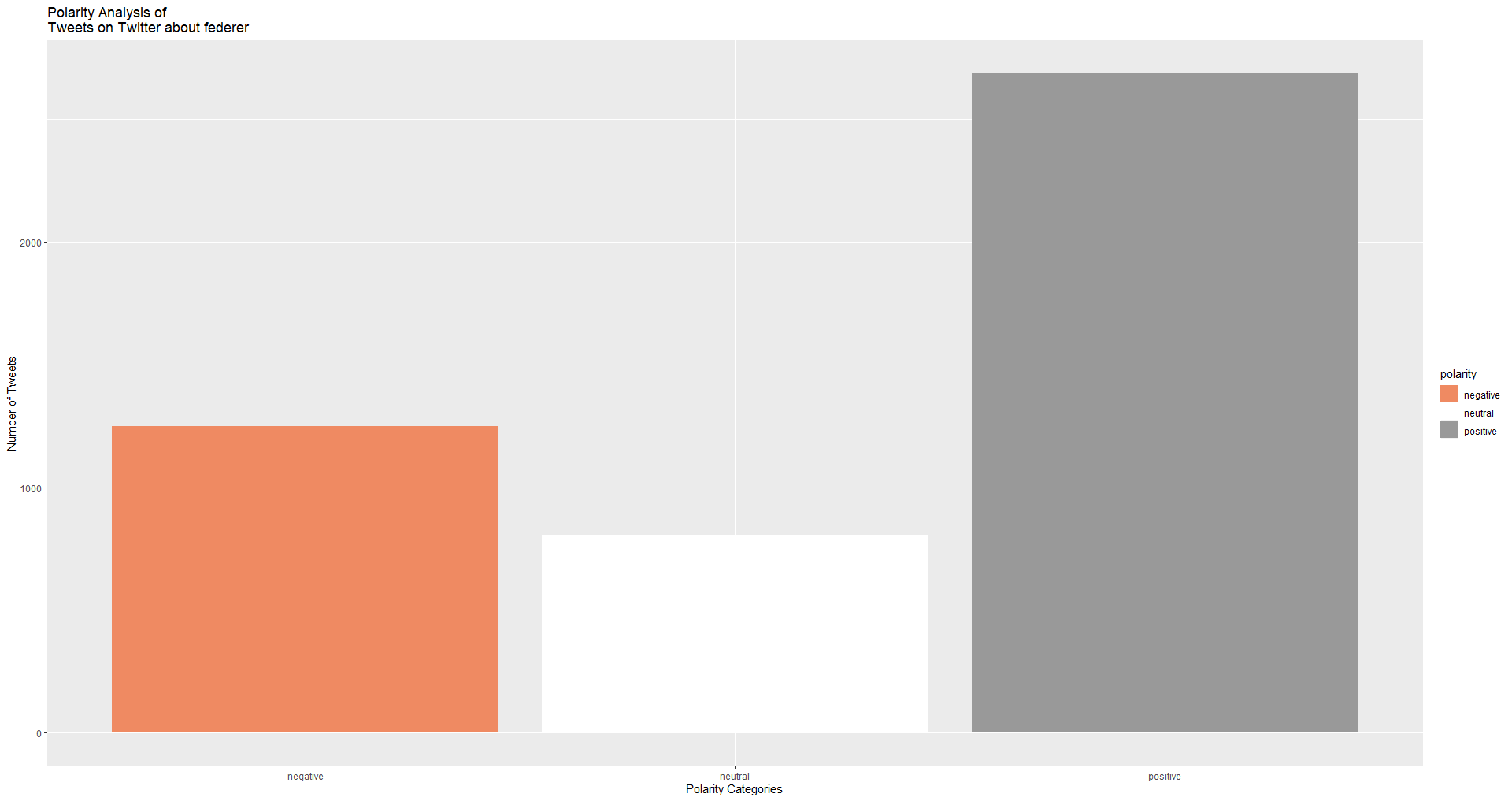}}&
\subfloat[Jennifer Lopez]{\includegraphics[width = 2in]{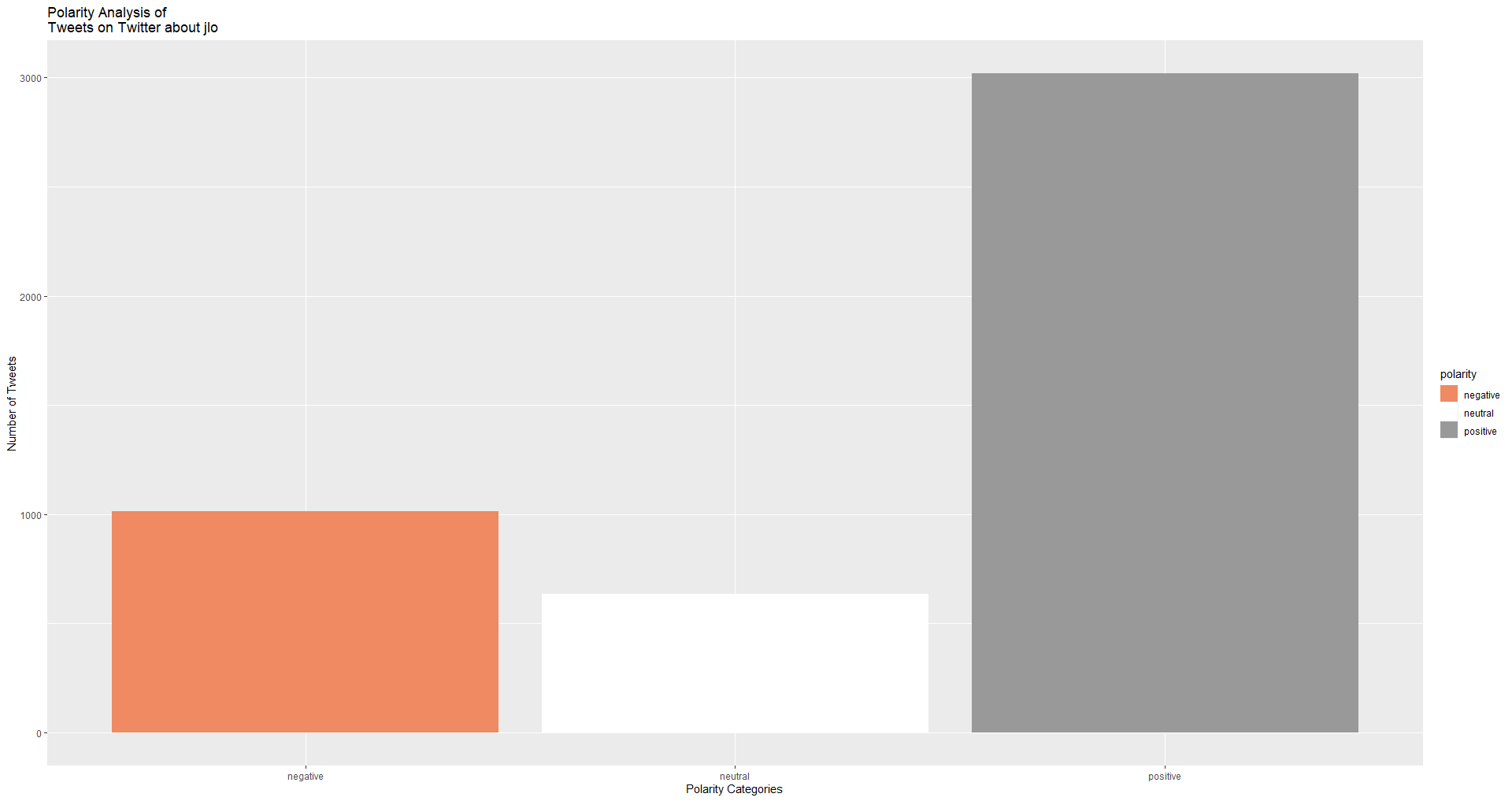}}\\
\subfloat[Nike]{\includegraphics[width = 2in]{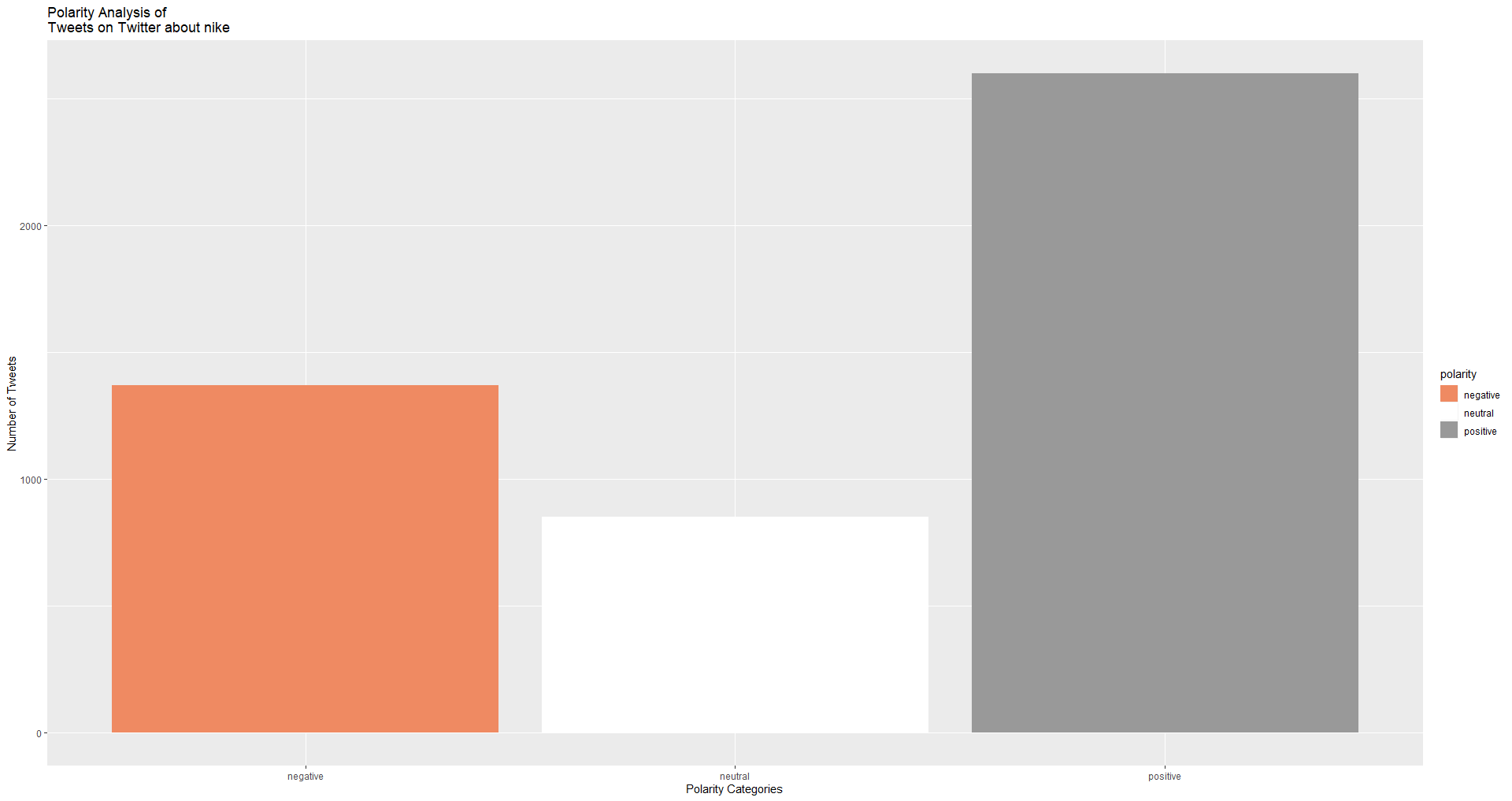}}&
\subfloat[Cristiano Ronaldo]{\includegraphics[width = 2in]{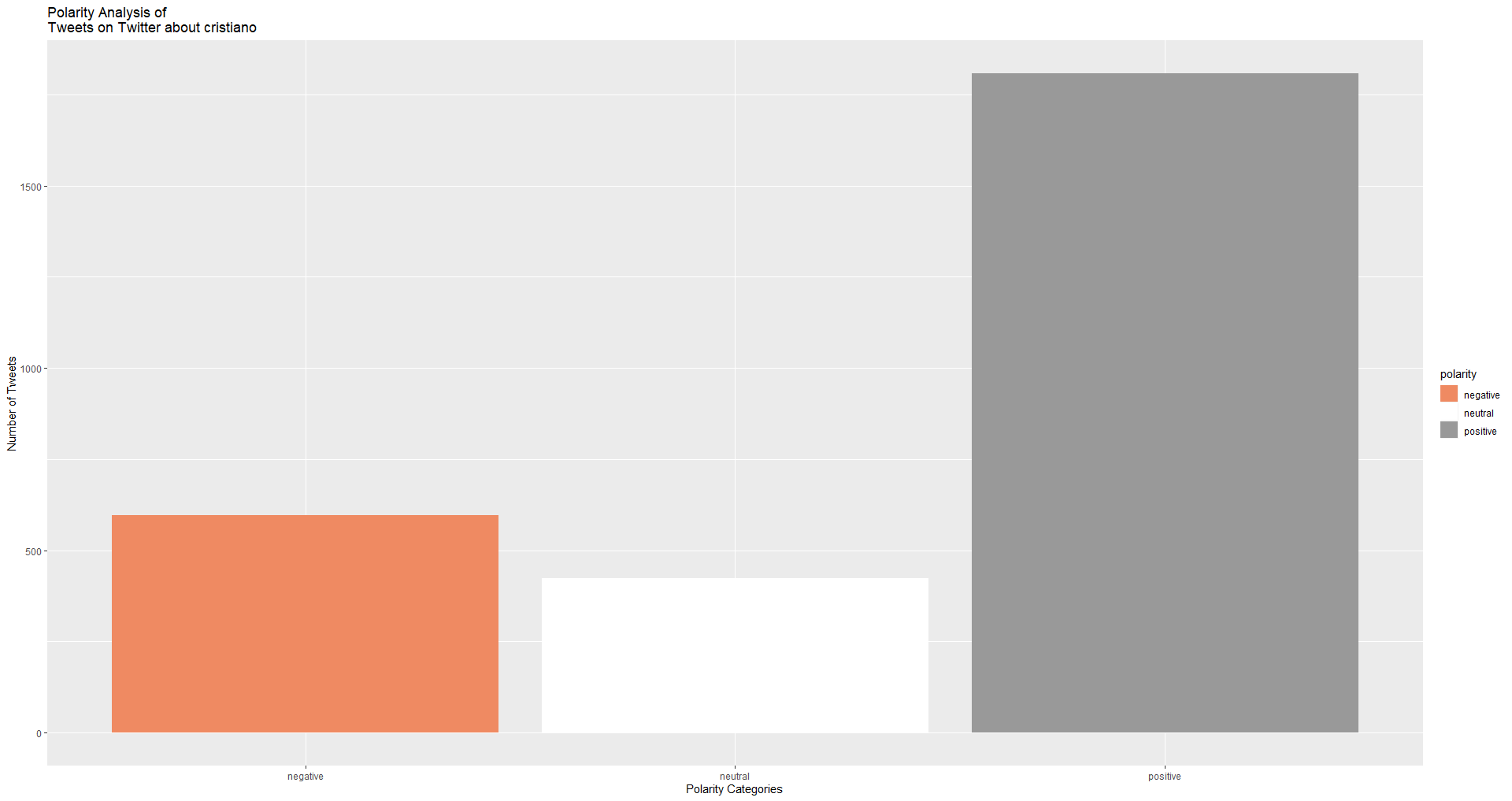}}&
\subfloat[Neymar]{\includegraphics[width = 2in]{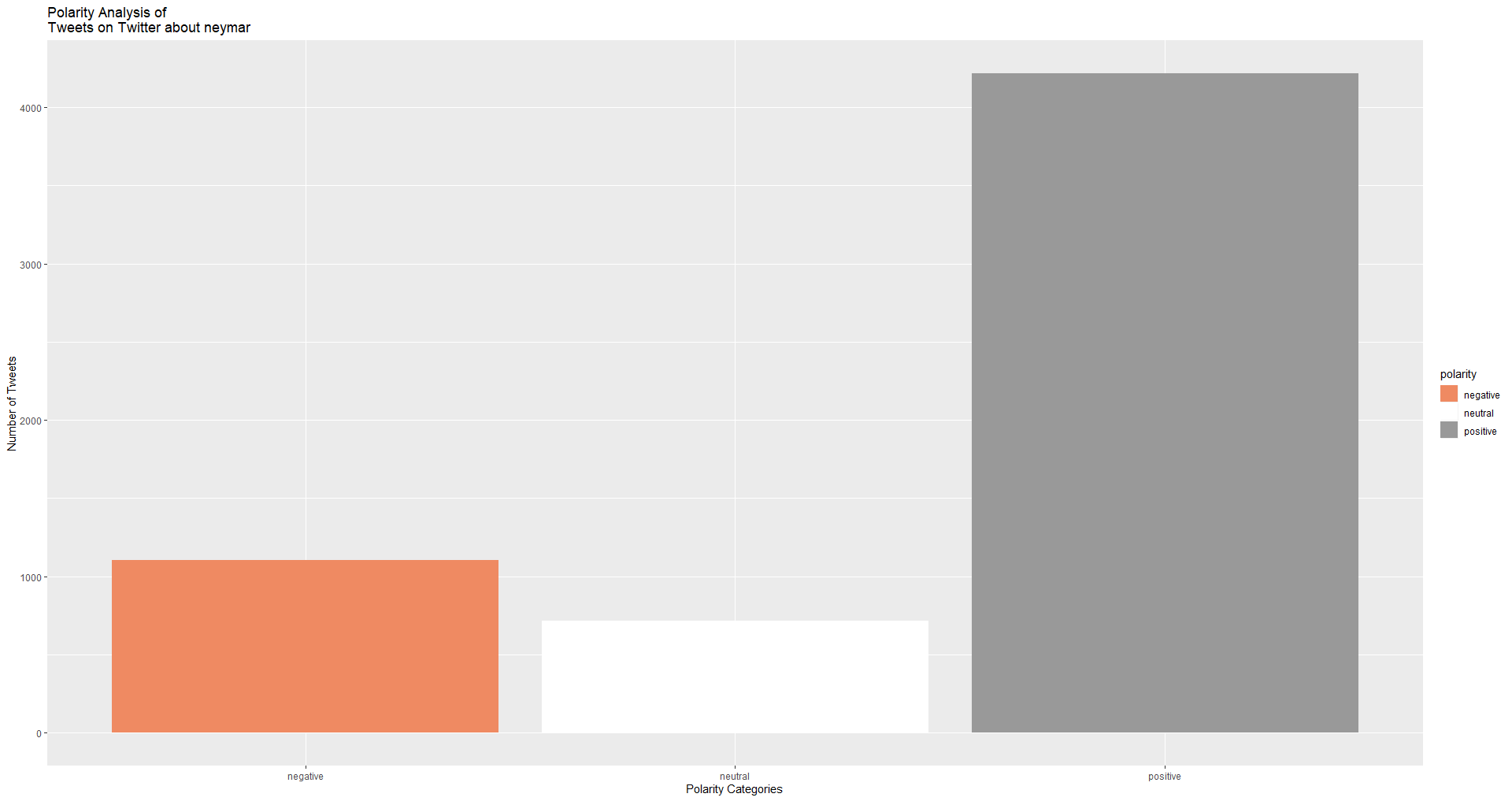}}

\end{tabular}
\caption{Polarity analysis plots for all brands of interest}
\end{figure}

%
%

According to figure 6, for example in polarity analysis of Pepsi, the positive impact of the brand is much more than the negative impact.

\FloatBarrier
\begin{figure}[H]
\begin{tabular}{cccc}
\subfloat[Burger King]{\includegraphics[width = 2in]{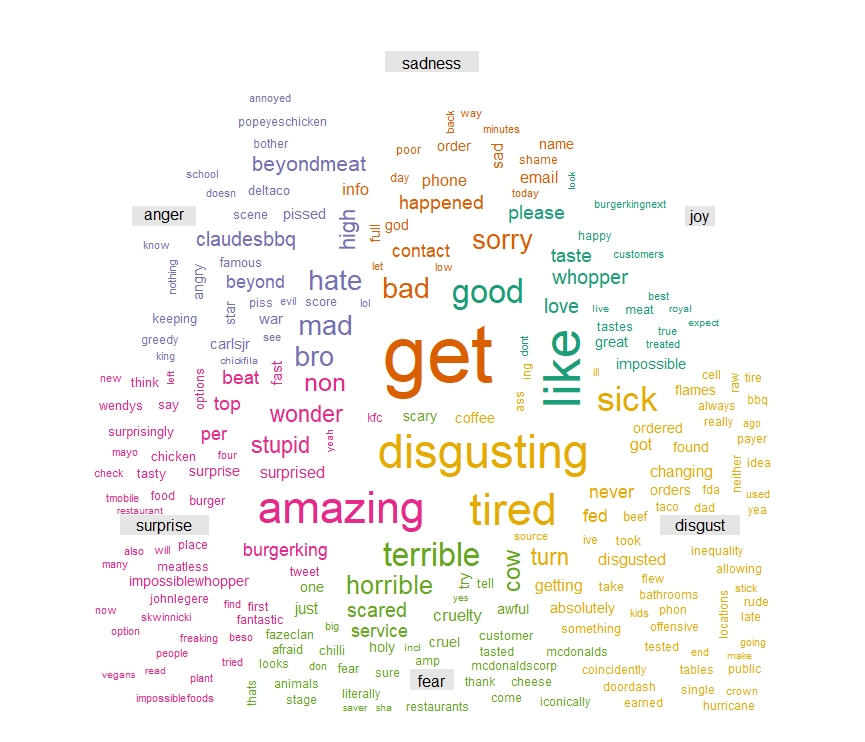}} &
\subfloat[Snoopdogg]{\includegraphics[width = 2in]{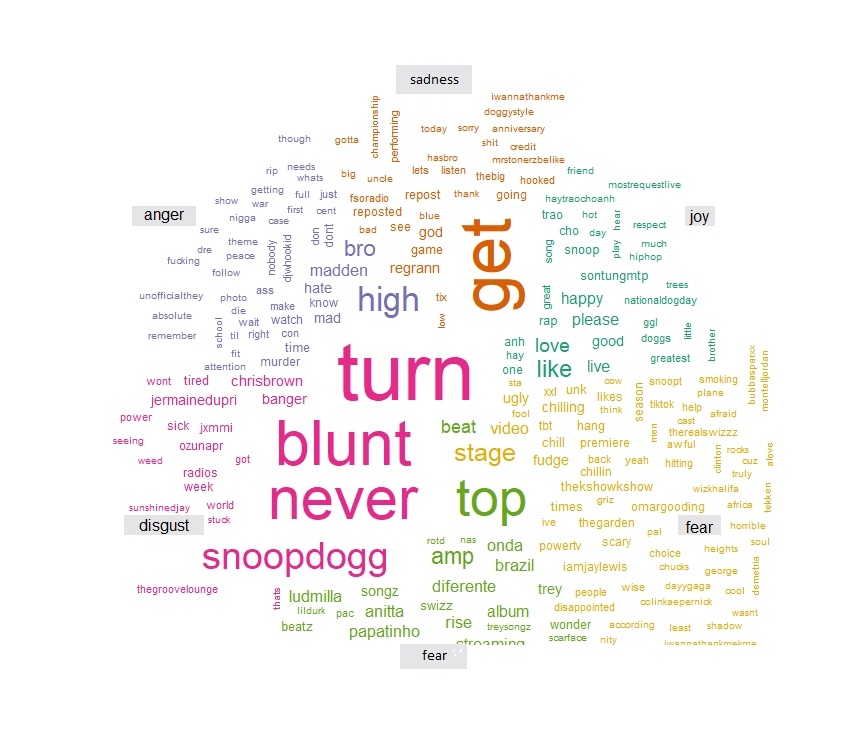}}&
\subfloat[Connor McGregor]{\includegraphics[width = 2in]{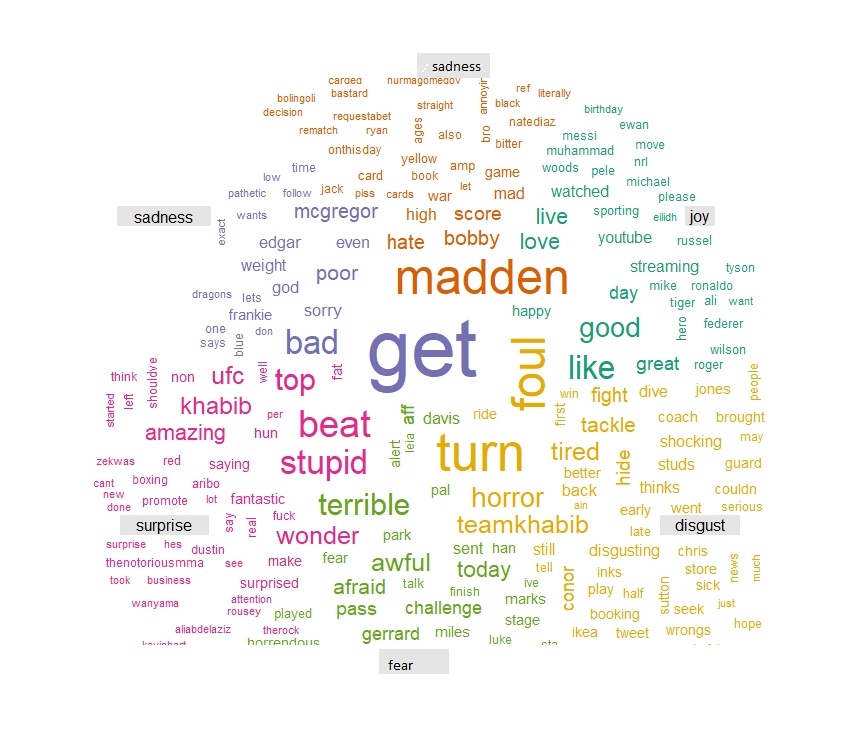}}\\
\subfloat[Coca Cola]{\includegraphics[width = 2in]{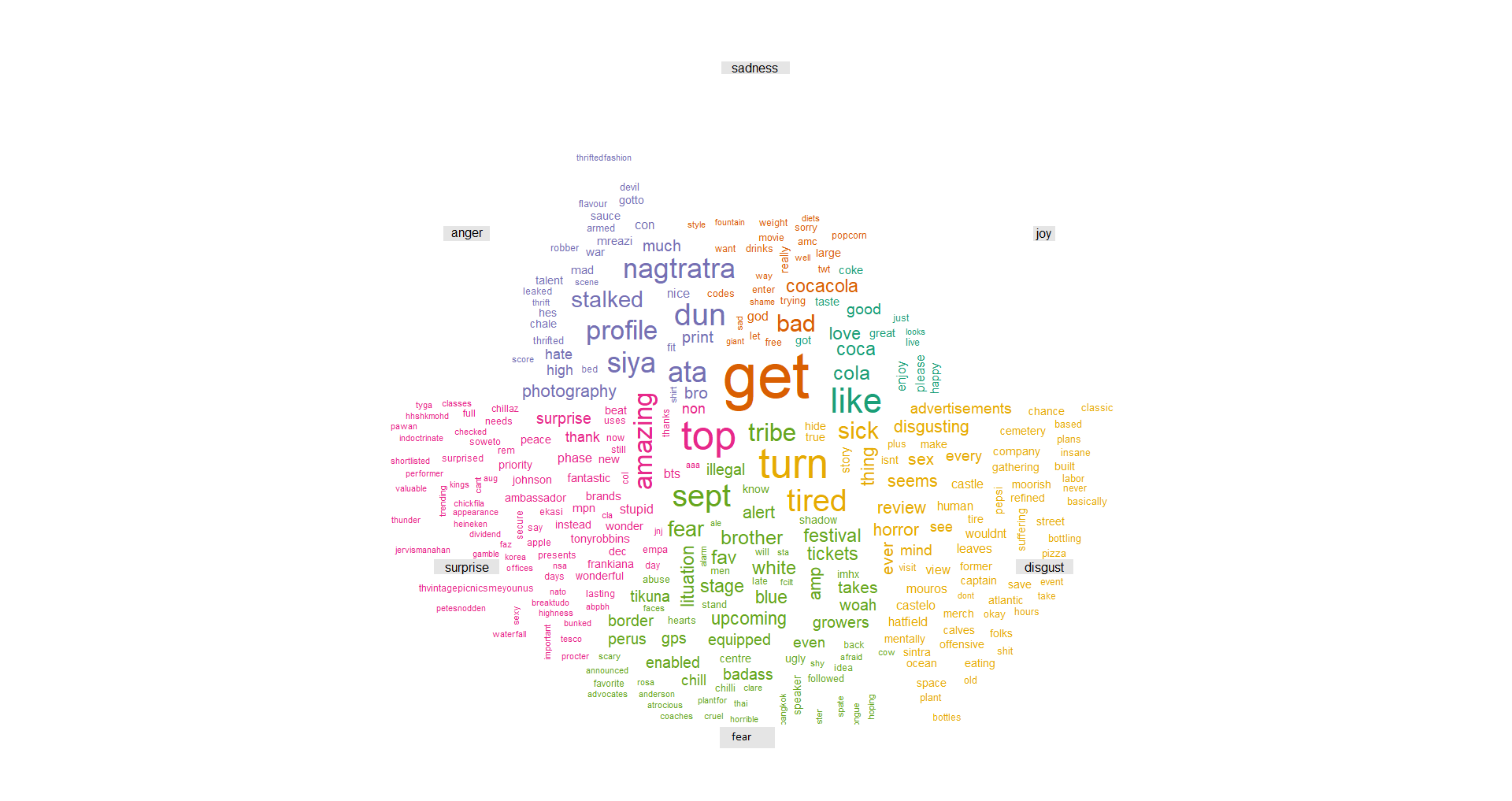}}&
\subfloat[Taylor Swift]{\includegraphics[width = 2in]{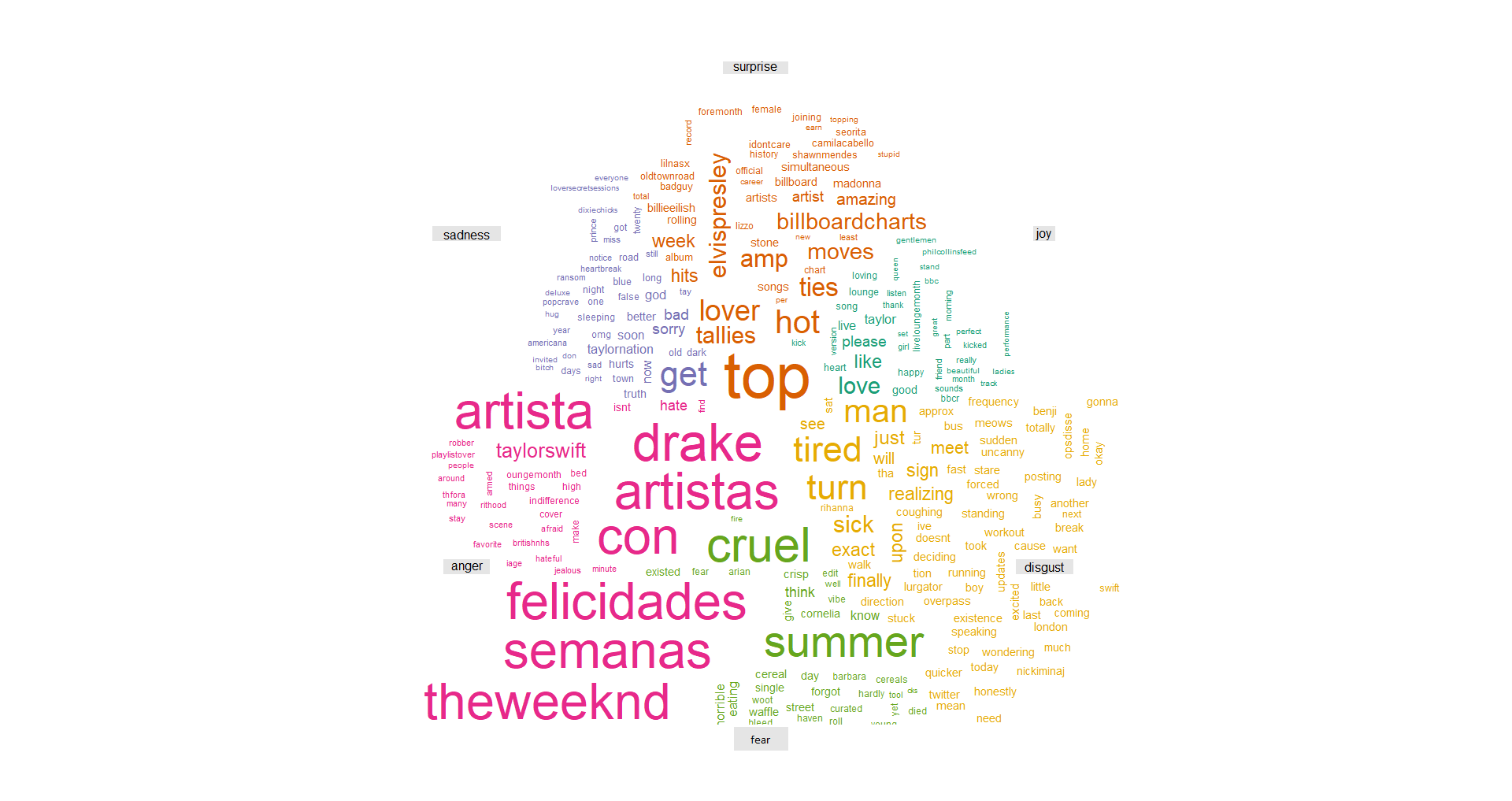}} &
\subfloat[Selena Gomez]{\includegraphics[width = 2in]{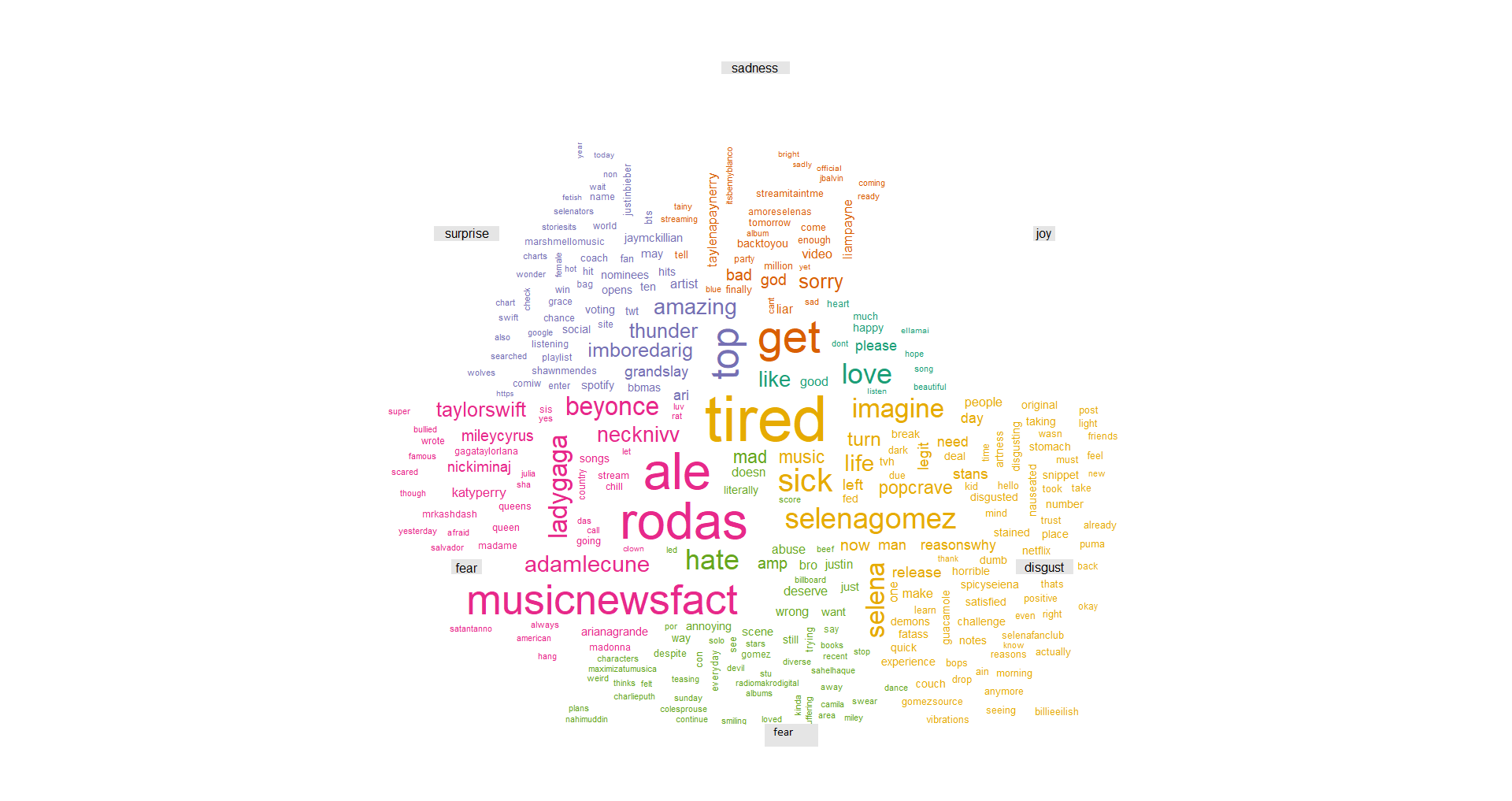}}\\
\subfloat[Pepsi]{\includegraphics[width = 2in]{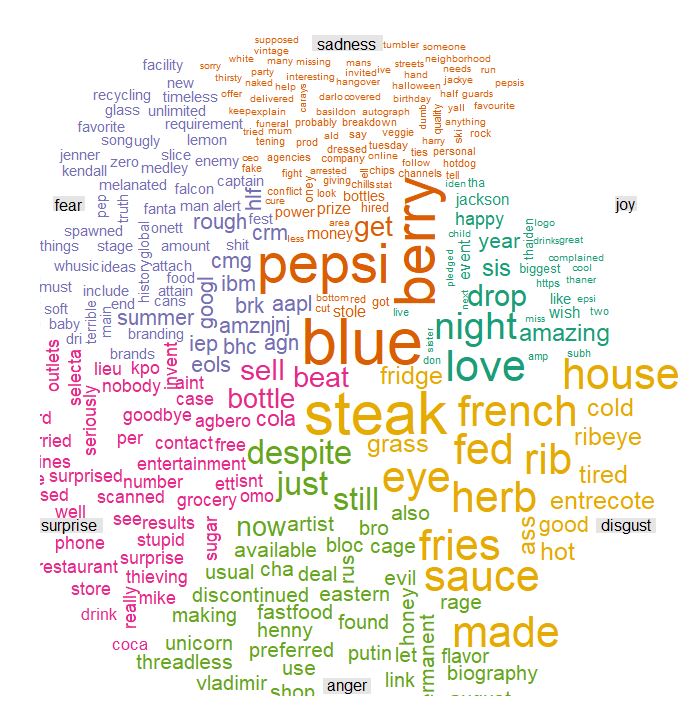}}&
\subfloat[Messi]{\includegraphics[width = 2in]{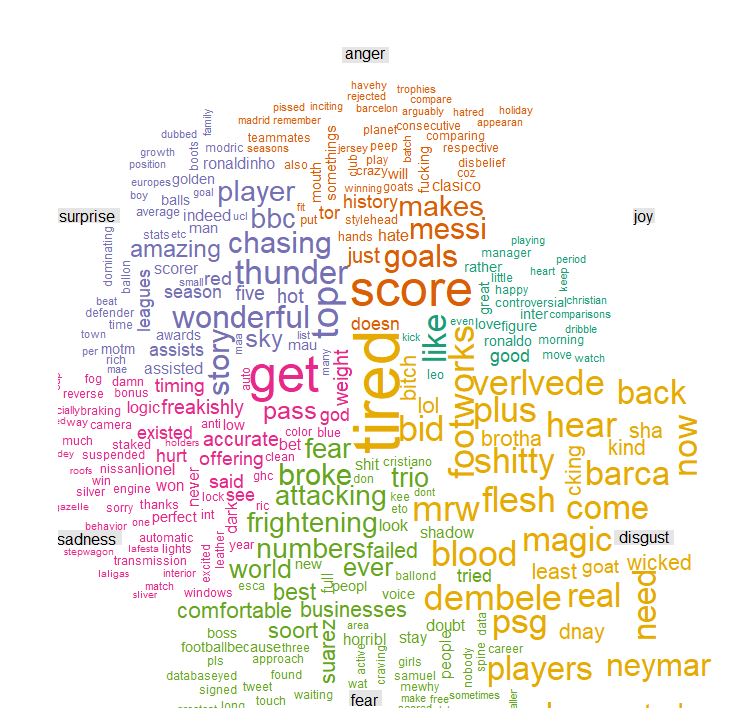}}&
\subfloat[Beyoncé]{\includegraphics[width = 2in]{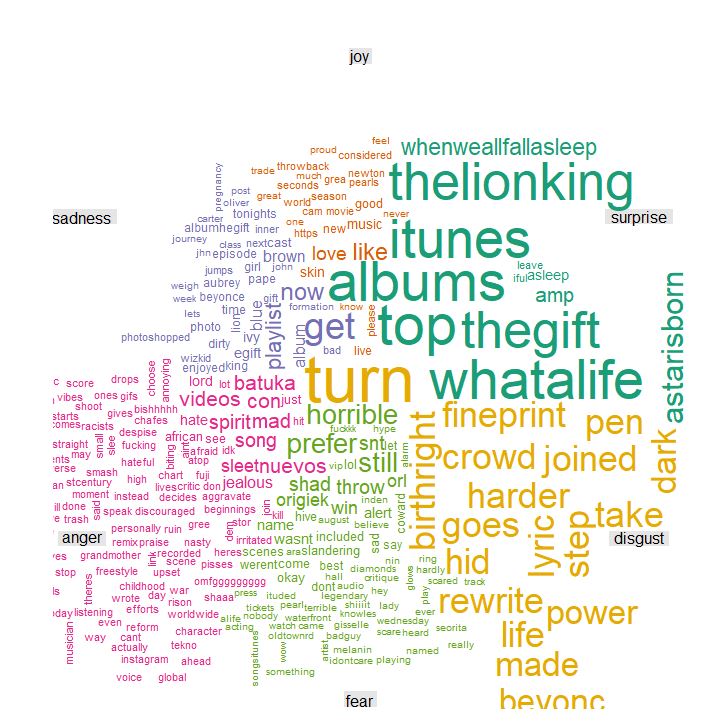}}\\
\subfloat[Gillette]{\includegraphics[width = 2in]{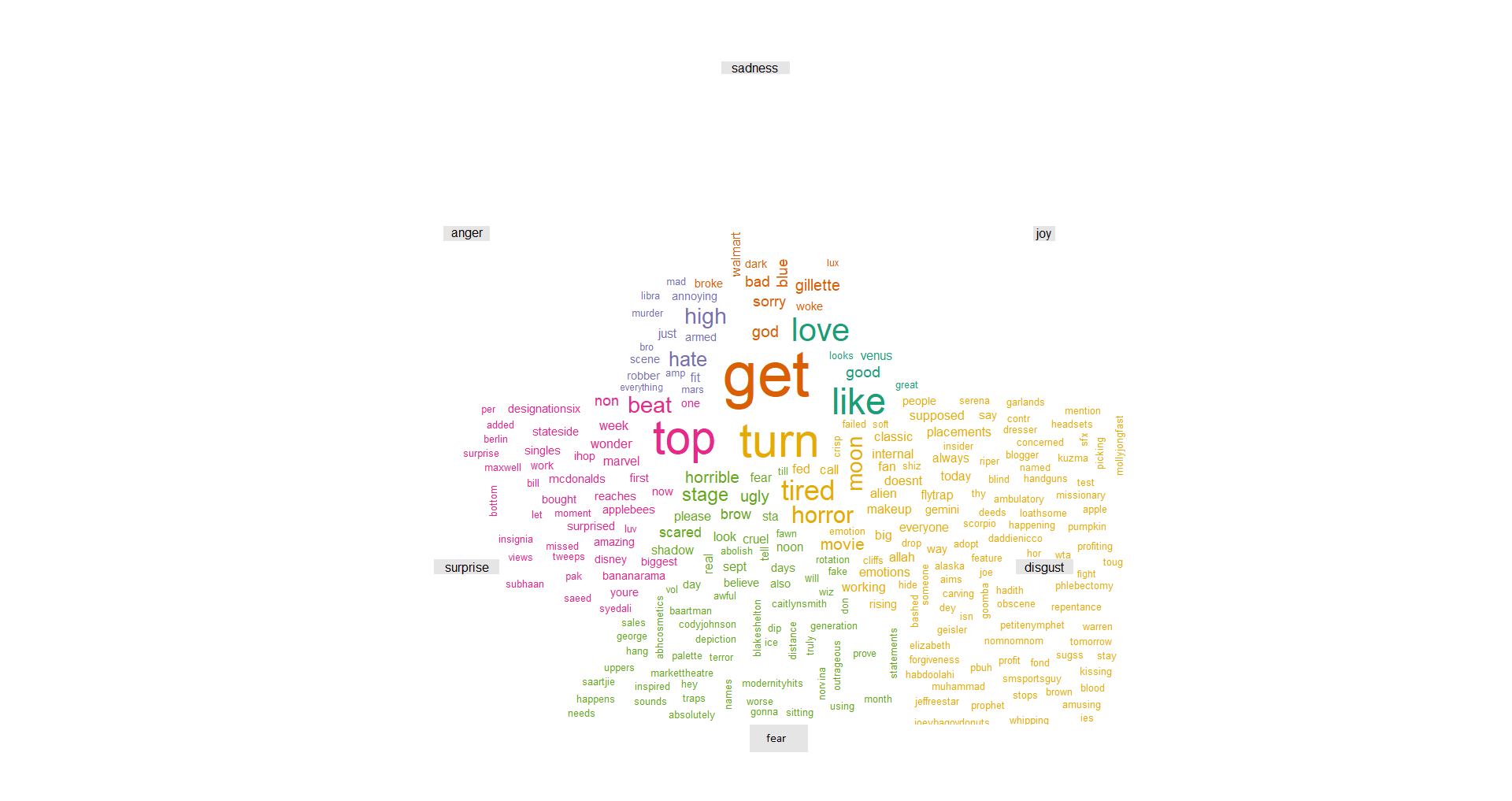}}&
\subfloat[Roger Federer]{\includegraphics[width = 2in]{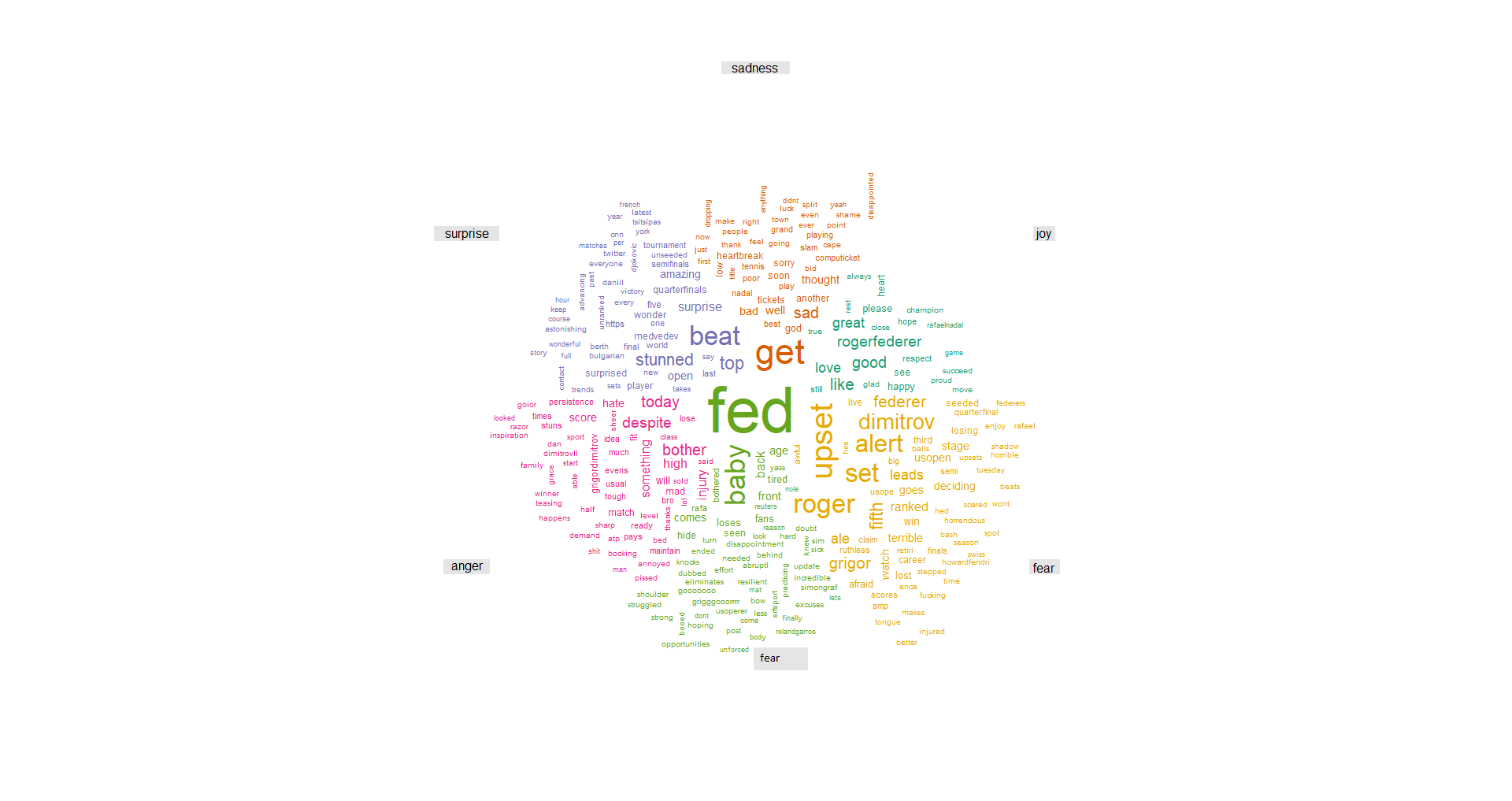}}&
\subfloat[Jennifer Lopez]{\includegraphics[width = 2in]{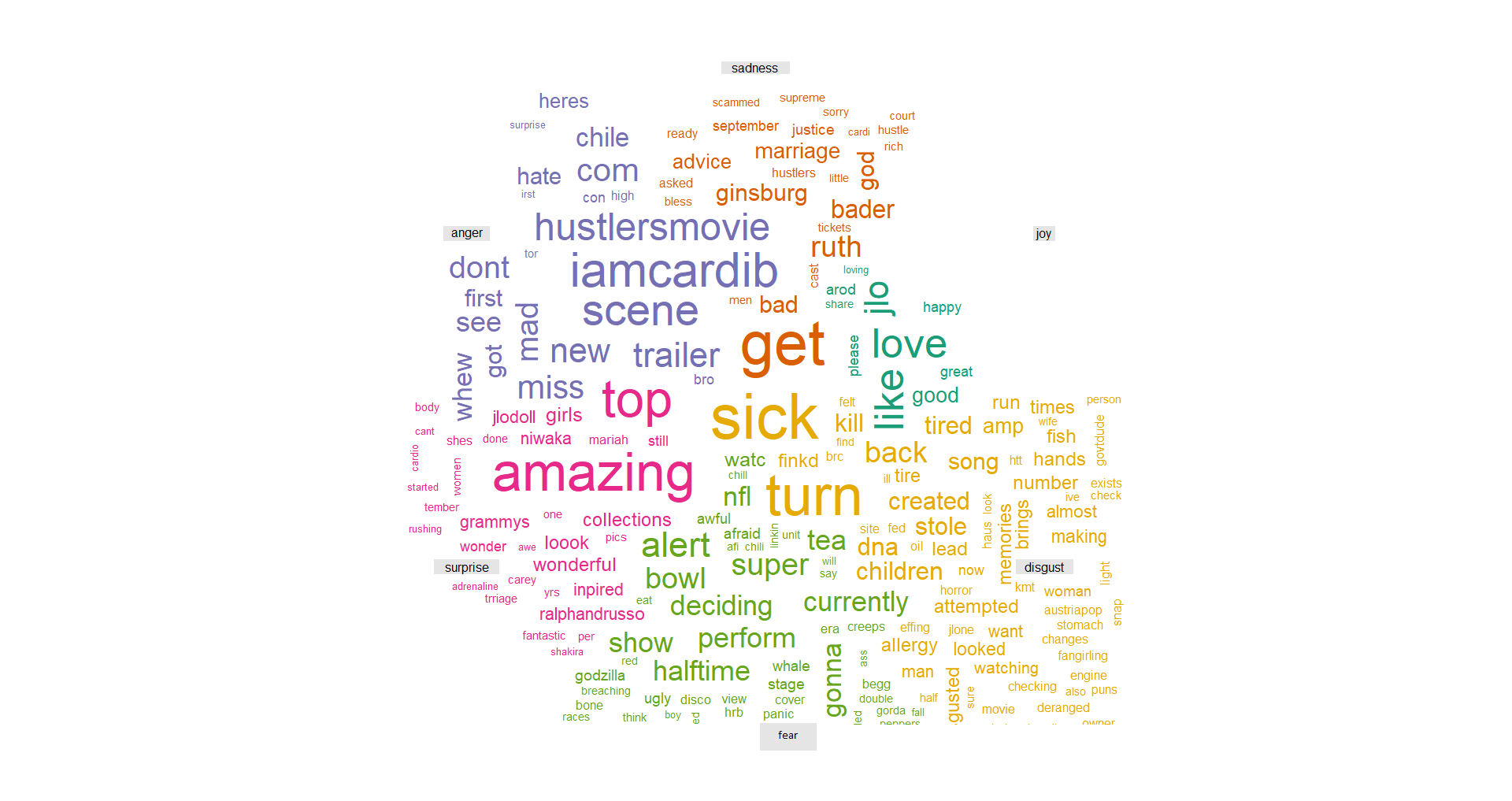}}\\
\subfloat[Nike]{\includegraphics[width = 2in]{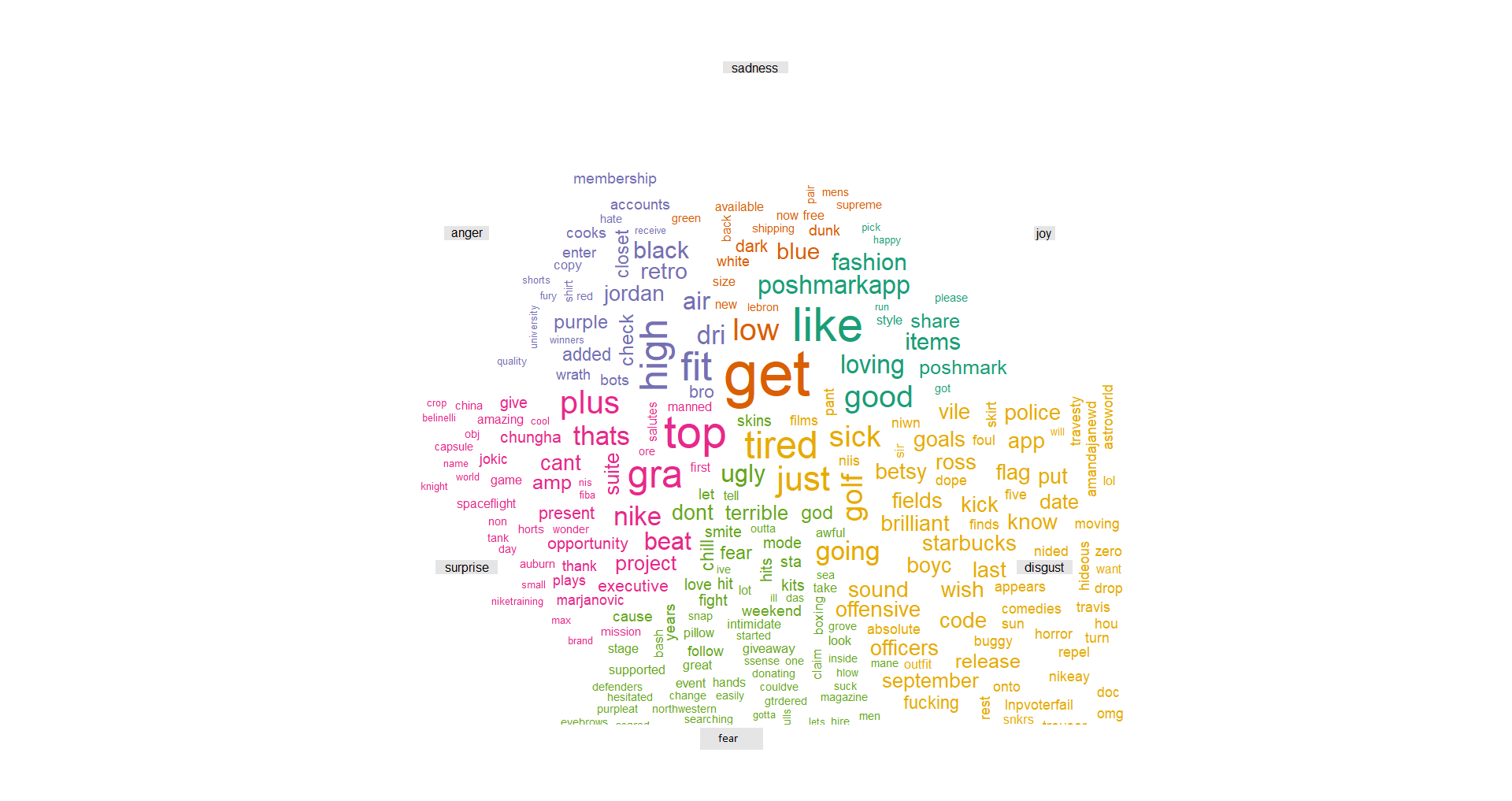}}&
\subfloat[Cristiano Ronaldo]{\includegraphics[width = 2in]{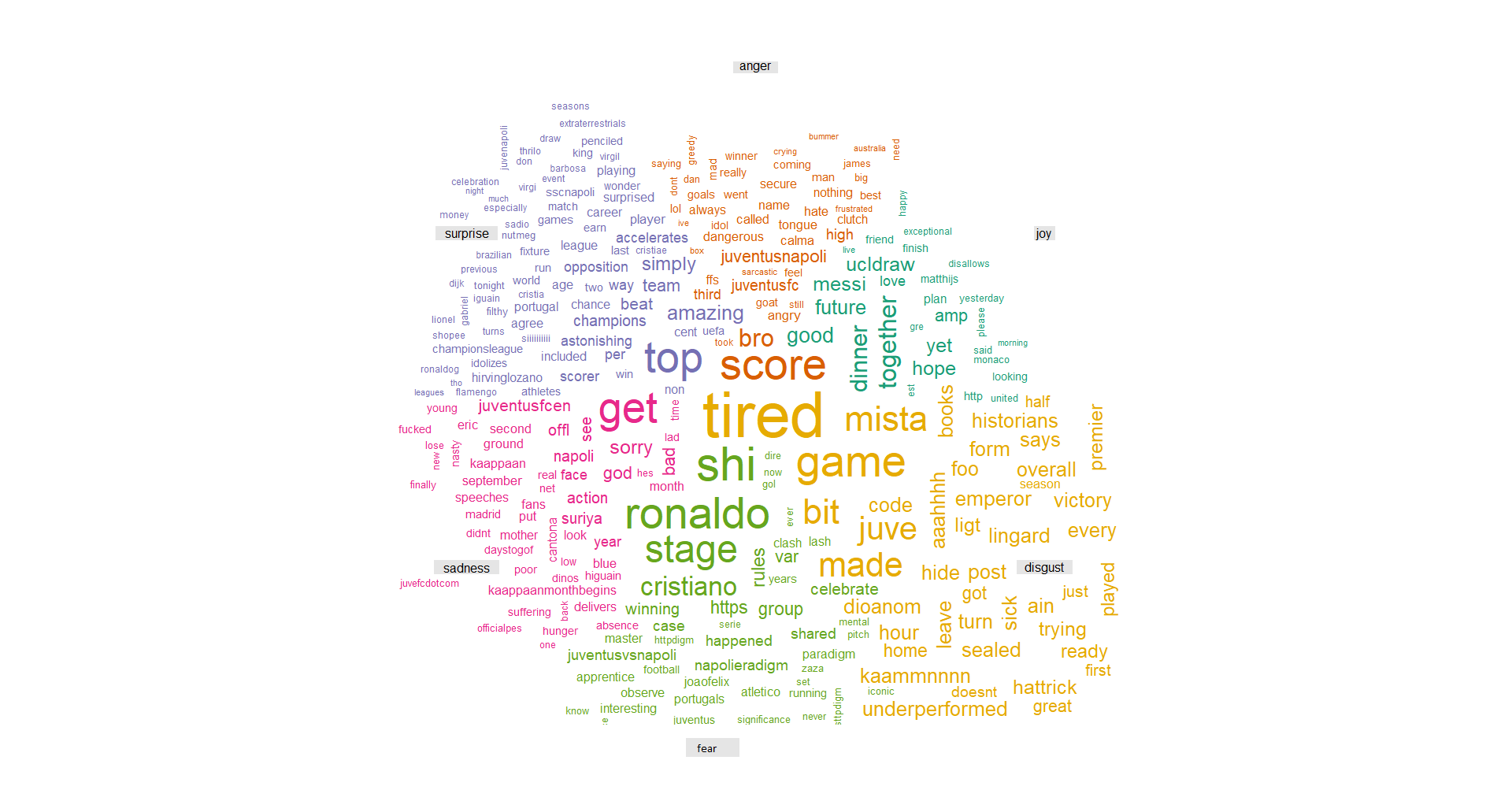}}&
\subfloat[Neymar]{\includegraphics[width = 2in]{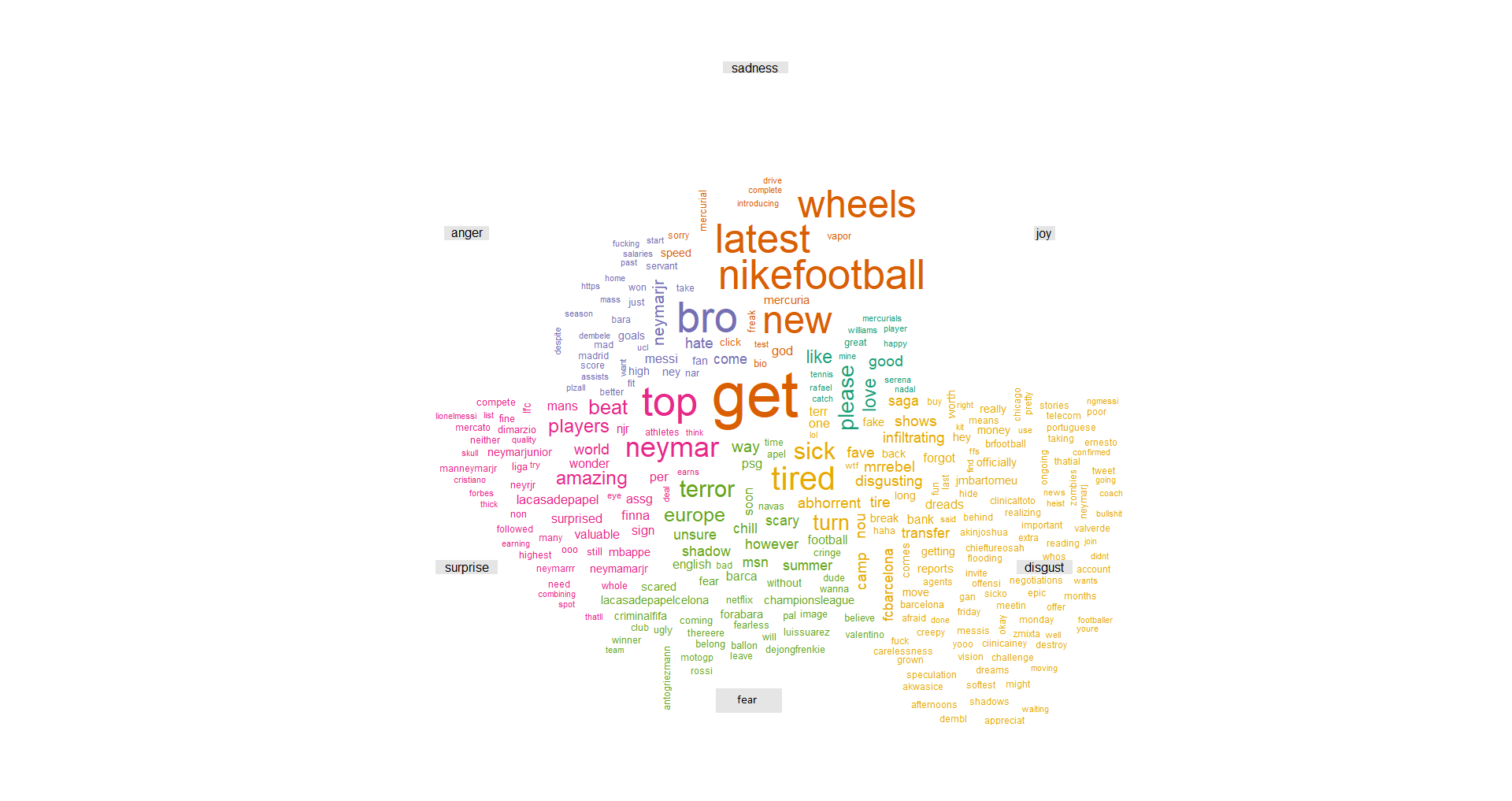}}

\end{tabular}
\caption{WordClouds according to emotions for all brands of interest}
\end{figure}

%
%
%

As mentioned in the figure 7 the WordCloud outline the most important words in tweets as a matter of the number of the repeat.\\

The first method solely applied the Naïve algorithm as a lexicon-based analysis technique. Since the results were demonstrated earlier, one cannot conclude Coca Cola to be closer to Messi based on the sentiment as compared to Beyoncé. Therefore, the second method was employed, which was a Naïve Bayes-based machine learning technique that analyzes results by classifying the emotions into six basic emotion sets and demonstrates each character’s polarity. Eventually, each celebrity’s word cloud, their polarities, and tweets were drawn based on the six basic emotions. Finally, the results of each method are provided in the following tables:

\FloatBarrier
\begin{table}[H]
\caption{{Method 1 (Lexicon Based -Naïve) ((More than 0.2 => \Smiley{})  (Less than 0.2 => \Frowny{}))} }
\label{tw-9dff510fd0a5}
\def\arraystretch{1}
\ignorespaces 
\centering 
\begin{tabulary}{\linewidth}{p{\dimexpr.25\linewidth-2\tabcolsep}p{\dimexpr.25\linewidth-2\tabcolsep}p{\dimexpr.2431\linewidth-2\tabcolsep}p{\dimexpr.2569\linewidth-2\tabcolsep}}
\tbltoprule Brands & Celebrities & AnswerToQ1 & AnswerToQ2(Recommendation) \\

\tblmidrule 
Pepsi \LARGE\Smiley{}& Messi  &\LARGE\Smiley{} &Messi \\
	& Beyoncé &\LARGE\Frowny{} &  \\
Nike \LARGE\Smiley{}&	C.Ronaldo  & \LARGE\Smiley{} &C.Ronaldo \\
	&Neymar&\LARGE\Frowny{} & \\
Burger King \LARGE\Frowny{}	&Snoop Dogg  & \LARGE\Smiley{} &Snoop Dogg\\
	&ConorMcgregor &  \LARGE\Frowny{} &\\
Coca Cola \LARGE\Smiley{}&Taylor Swift  &\LARGE\Smiley{} &Taylor Swift\\
	&SelenaGomez& \LARGE\Frowny{} &\\
Gillete \LARGE\Frowny{}&	Roger Federer  & \LARGE\Smiley{} &Both\\
	&JLO  &\LARGE\Smiley{} &\\

\tblbottomrule 

\end{tabulary}\par 
\end{table}

Table 8 categorizes the sentiments were classified based on their mean sentiment scores into two classes and labels them with emojis. In addition, the two questions were answered.

\FloatBarrier
\begin{table}[H]
\caption{{Method 2 (Machine Learning -Naïve Bayes)} }
\label{tw-9dff510fd0a5}
\def\arraystretch{1}
\ignorespaces 
\centering 
\begin{tabulary}{\linewidth}{p{\dimexpr.25\linewidth-2\tabcolsep}p{\dimexpr.25\linewidth-2\tabcolsep}p{\dimexpr.2431\linewidth-2\tabcolsep}p{\dimexpr.2569\linewidth-2\tabcolsep}}
\tbltoprule Brands & Celebrities & AnswerToQ1 (SA\_PA\_WC) & AnswerToQ2(Recommendation) \\

\tblmidrule 
Pepsi & Messi & 2  \_  2  \_   2 &Beyoncé \\
	& Beyoncé & 1  \_  1  \_  1 &  \\
Nike&	C.Ronaldo& 1  \_  2  \_   2 &Neymar \\
	&Neymar&2  \_  1  \_   1 & \\
Burger King	&Snoop Dogg& 1  \_  1  \_  1 &Snoop Dogg\\
	&ConorMcgregor & 2  \_  2  \_   2 &\\
Coca Cola	&Taylor Swift& 1  \_  1  \_  1 &Taylor Swift\\
	&SelenaGomez& 2  \_  2  \_   2 &\\
Gillete&	Roger Federer& 1  \_  2  \_   2 &JLO\\
	&JLO&2  \_  1  \_   1 &\\

\tblbottomrule 
\end{tabulary}\par 
\end{table}

Table 9 compares the brand’s two related celebrities based on three ranks, namely sentiment analysis (SA), polarity analysis (PA), and word cloud (WC), and highlights the ultimate recommendations while answering the second question. 
\subsection{Managerial Implications}
We found seven major implications that can help managers in many ways:\\
1- Decision Making And Market Research are so vital to businesses and managers. One of the ways that could help them significantly are the approaches of the current study. Another help as \cite{He2017} mentioned in their results, would be companies exploiting the big data existing in social media to build a clear connection with strategic planning and competitive positioning.\\ 
2- We are completely agree with \cite{Ibrahim2019} that an detailed concentration on the consumer-generated content could specially help web based retailing brands manage customer
relationships effectively and efficiently. Thoughts and encounters shared 
by clients via social media platforms permit organizations to associate with them and remain side by side of what is happening on Twitter.\\
3- If we consider the suggested Business SMA framework by \cite{Holsapple2018}, we can get to the four overarching end goals mentioned -Intelligence
Fetching, Making Sense out of Data, Insight Producing, and Decision Making- through the current research.\\
4- One may use the WordCloud approach we used as an easier way to get to related texts instead of more complex ways like lexical chain generation algorithm to cluster texts from an account suggested by \cite{Chua2019}.\\
5- As \cite{Chen2018} mentioned in their article, link prediction is a significant approach in recognizing relations in business and social media. To anticipate the possible relations in multi-relational social networks like twitter and other systems, one should consider the importance and effect between different kinds of relations, for instance by taking advantage of social media CGCs. So for the sake of this important subject, managers should use a handful of tools and algorithms like the ones we used in this article such as: Sentiment Analysis, Polarity Analysis, etc.\\
6- \cite{FronzettiColladon2019} Showed in their article that getting the most out of SMM or SMA and big data can be so useful in predicting the future behaviors of the consumers, and in their specific case "Anticipating the international tourist arrivals". Like that we presented the fact that managers can Improve Decision Making, Celebrity Branding And Market Research: Using Social Media Mining (SMM). In fact we can probably come to this conclusion that data mining in any aspects like our case of interest; Social Media, could be helpful and their outputs as metrics to us, prove their value in increasing the forecasting accuracy of different models in any industry.\\
7- Those who take advantage of the approaches mentioned in the research in hand can easily map the insights obtained to the actor roles mentioned in \cite{Hacker2017} findings and get a more holistic output.\\

\section{Discussion}

\FloatBarrier
\begin{figure}[H]
\centering
\includegraphics[max size={\textwidth}{\textheight}, keepaspectratio]{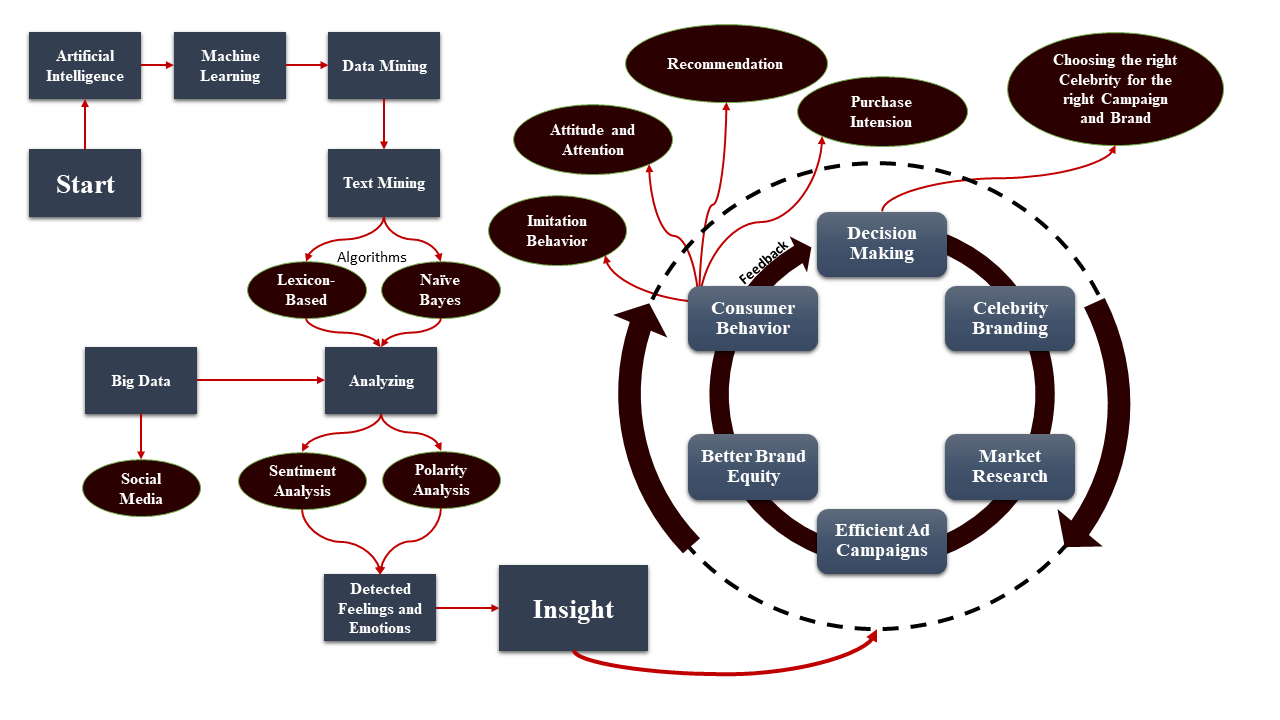}
\caption{{Final inferred Model of the Research(Compiled By The Author)}}
\label{caption.conceptual model}
\end{figure}    

According to \cite{jain2016}, \cite{majeed2017}, \cite{Awasthi2015}, \cite{zhu2019}, \cite{mogaji2017}, \cite{sailunaz2019} ,\cite{ambroise2019} we can clearly see that there is a connection between consumers’ feelings and celebrities’ icon and consumer behavior and decision making according to the perceived sentiment so it is really important to examine all the accessible resources specially big data contents because people are technology oriented these days and express their feelings more directly and more often using social media. Next according to the figure 8 which is our final model and output we do all these process to optimize the process of decision making about celebrity branding and consequently about market research and all the process till the end. And all these steps give us valuable insights and competitive advantage. Furthermore by using the attained insights we can have optimized campaigns which can have much better outputs and are able to save us time and money. This method is going to get us to a stage that we act with less risks, more benefits and optimum usage of resources and makes us powerful enough to be free of error as a matter of putting up our new campaigns during the time. \\


\subsection{Limitations and Future Studies}
There are some ways that future researchers could help and extend the research:\\
1- Using another social media platforms contents rather than twitter.\\
2- Exploiting qualitative research side by side of the data mining methodology like mixed method approaches.\\
3- Trying to extend the scope of the CGCs gathered, in both geography and industry.\\
4- Changing the objectives of the research and do it all again and to see if they can get better or at least acceptable results.\\
5- Using various ML algorithms and seeking better and more optimized insights.\\
6- Trying to check another ways of analytics to the same CGCs.\\
7- One can use semantic lexical chaining methods to clustering obtained data, such as NMF and GSDMM which have offered by \cite{Chua2019}\\

	
	\bibliographystyle{APA}
\bibliography{SMM-Full}

%

\textbf{}\\
\textbf{Funding: }\\
This research article did not receive any specific grant from funding agencies in the public, commercial, or not-for-profit sectors.\\
\textbf{CRediT authorship contribution statement: }\\
All authors contributed equally to this research article as primary authors.\\
\textbf{Declaration of interests: }\\
The authors declare that they have no known competing financial interests or personal relationships that could have appeared to influence the work reported in this paper.\\

\textbf{ORCID iD: }\\
Ali Nikseresht : \url{https://orcid.org/0000-0002-6107-7699}\\
Mohammad Hosein Raeisi : \url{https://orcid.org/0000-0001-5252-3485}\\
Hossein Abbasian Mohammadi : \url{https://orcid.org/0000-0002-6836-4850}

\end{document}